\documentclass[11pt,a4paper,final]{article}
\usepackage[utf8]{inputenc}
\usepackage[english]{babel}
\usepackage{amsmath}
\usepackage{amsfonts}
\usepackage{amssymb}
\usepackage{appendix}
\usepackage{faktor}
\usepackage{yfonts}
\usepackage{upgreek}
\usepackage{lipsum}
\usepackage[hidelinks]{hyperref}
\hypersetup{
    colorlinks=true,
    urlcolor=blue,
    linkcolor=blue,
    citecolor=red,
    }
\usepackage{amsthm}
\usepackage{bbold}
\usepackage{mathrsfs}
\usepackage{wasysym}
\usepackage{mathtools}
\usepackage{bm}
\usepackage{cancel}
\usepackage{color}
\usepackage{enumitem,multicol}
\usepackage{multicol}
\usepackage{tikz}
\usepackage{tikz-cd}
\usepackage[Glenn]{fncychap}
\usepackage{subfig}
\usepackage{accents}
\usepackage{slashed}
\usepackage{psfrag}
\usepackage{todonotes}

\usepackage{stackengine}
	
\usetikzlibrary{babel}
\makeatletter
\newcommand*\owedge{\mathpalette\@owedge\relax}
\newcommand*\@owedge[1]{%
	\mathbin{%
		\ooalign{%
			$#1\m@th\bigcirc$\cr
			\hidewidth$#1\m@th\wedge$\hidewidth\cr
		}%
	}%
}
\makeatother

\usepackage{cjhebrew}

\newcounter{mnotecount}

\newcommand{\mnote}[1]
{\protect{\stepcounter{mnotecount}}$^{\mbox{\tiny
			$\,\bullet$\themnotecount}}$ \marginpar{
		\raggedright\tiny\em
		$\,\bullet$\themnotecount: #1} }

\newtheorem{teo}{Theorem}[section]
\newtheorem{cor}[teo]{Corollary}
\newtheorem{prop}[teo]{Proposition}
\newtheorem{lema}[teo]{Lemma}
\newtheorem{defi}[teo]{Definition}
\newtheorem{eje}[teo]{Example}

\newtheorem{rmk}[teo]{Remark}
\newtheorem{nota}[teo]{Notation}

\usepackage{makeidx}
\usepackage{graphicx}
\usepackage[left=2.60cm, right=2.60cm, top=2.45cm, bottom=2.60cm]{geometry}

\usepackage{scalerel}
\usepackage{stackengine,wasysym}

\newcommand{\uwh}[1]{%
	\mathpalette\douwidehat{#1}%
}

\makeatletter
\newcommand{\douwidehat}[2]{%
	\sbox0{$\m@th#1\widehat{\hphantom{#2}}$}%
	\sbox2{$\m@th#1x$}
	\sbox4{$\m@th#1#2$}
	\dimen0=\ht0
	\advance\dimen0 -.8\ht2
	\dimen2=\dp4
	\rlap{%
		\raisebox{\dimexpr\dimen0-\dimen2}{%
			\scalebox{1}[-1]{\box0}%
		}%
	}%
	{#2}%
}
\makeatother

\usepackage{color}
    \makeatletter
\renewcommand\part{%
	\if@openright
	\cleardoublepage
	\else
	\clearpage
	\fi
	\thispagestyle{empty}%
	\if@twocolumn
	\onecolumn
	\@tempswatrue
	\else
	\@tempswafalse
	\fi
	\null\vfil
	\secdef\@part\@spart}
\makeatother

\newcommand{\wt}{\widetilde}
\newcommand{\wh}{\widehat}

\newcommand{\elltwo}{\ell^{(2)}}

\newcommand{\nablacero}{\accentset{\circ}{\nabla}}

\newcommand{\bY}{\textup{\textbf Y}}
\newcommand{\Y}{\textup{Y}}

\newcommand{\bcr}{\mathbb r}
\newcommand{\bck}{\mathbb k}
\renewcommand{\k}{\mathscr{K}}
\newcommand{\bT}{\textup{\textbf T}}
\newcommand{\T}{\textup{T}}

\newcommand{\bZ}{\textup{\textbf Z}}

\newcommand{\bz}{\textup{\textbf z}}
\newcommand{\z}{\textup{z}}
\newcommand{\Z}{\textup{Z}}
\newcommand{\bU}{\textup{\textbf U}}
\newcommand{\U}{\textup{U}}

\newcommand{\bF}{\textup{\textbf F}}
\newcommand{\F}{\textup{F}}
\newcommand{\br}{\textup{\textbf r}}
\renewcommand{\r}{\textup{r}}

\renewcommand{\H}{\mc{H}}

\newcommand{\bs}{\textup{\textbf s}}
\newcommand{\s}{\textup{s}}

\newcommand{\bPi}{\bm\Pi}
\newcommand{\bg}{\bm{\gamma}}

\newcommand{\mf}{\mathfrak}
\newcommand{\real}{\mathbb R}
\newcommand{\tr}{\operatorname{tr}}

\renewcommand{\div}{\operatorname{div}}

\newcommand{\R}{\bm{\mc R}}
\newcommand{\st}{\stackrel}

\newcommand{\Rcero}{\accentset{\circ}{R}}
\newcommand{\sph}{\mathbb S}
\newcommand{\eps}{\varepsilon}
\newcommand{\X}{\mathfrak{X}}
\renewcommand{\d}{\coloneqq}

\newcommand{\riem}{\operatorname{Riem}}
\newcommand{\ric}{\operatorname{Ric}}

\newcommand{\lie}{\mathcal{L}}
\newcommand{\mc}{\mathcal}

\renewcommand{\to}{\longrightarrow}

\renewcommand{\mapsto}{\longmapsto}
\title{\vspace*{-1.35cm}\textbf{Transverse expansion of the metric at null hypersurfaces I. Uniqueness and application to Killing horizons}}
\author{Marc Mars\footnote{\href{mailto:marc@usal.es}{marc@usal.es}}\,\, and Gabriel Sánchez-Pérez\footnote{\href{mailto:gasape21@usal.es}{gasape21@usal.es}} \\
	Departamento de Física Fundamental, Universidad de Salamanca\\
	Plaza de la Merced s/n, 37008 Salamanca, Spain}
\date{\today}

\begin{document}
	\maketitle

\begin{abstract}
This is the first in a series of two papers where we analyze the transverse expansion of the metric on a general null hypersurface. In this paper we obtain general geometric identities relating the transverse derivatives of the ambient Ricci tensor and the transverse expansion of the metric at the null hypersurface. We also explore the case where the hypersurface exhibits a generalized symmetry generator, namely a privileged vector field in the ambient space which, at the hypersurface, is null and tangent (including the possibility of zeroes). This covers the Killing, homothetic, or conformal horizon cases, and, more generally, any situation where detailed information on the deformation tensor of the symmetry generator is available. Our approach is entirely covariant, independent on any field equations, and does not make any assumptions regarding the topology or dimension of the null hypersurface. As an application we prove that the full transverse expansion of the spacetime metric at a non-degenerate Killing horizon (also allowing for bifurcation surfaces) is uniquely determined in terms of abstract data on the horizon and the tower of derivatives of the ambient Ricci tensor at the horizon. In particular, the transverse expansion of the metric in $\Lambda$-vacuum spacetimes admitting a non-degenerate horizon is uniquely determined in terms of abstract data at the horizon. 
\end{abstract}

\section{Introduction}

The initial value problem of the Einstein field equations plays a central role in General Relativity. The fundamental breakthrough of Choquet-Buhat established that by specifying an abstract $n$-dimensional Riemannian manifold $(\Sigma,h)$ along with a symmetric, two covariant tensor field $K$ subject to constraint equations, an $(n+1)$-dimensional spacetime $(\mc M,g)$ solution of the Einstein equations exists where $(\Sigma,h,K)$ is embedded \cite{Choquet,choquet1969global,ringstrom}. Another important initial value problem in general relativity is the characteristic one, where the data is posed on a pair of null hypersurfaces that intersect transversely. Rendall's work \cite{Rendall} demonstrates that proving the existence of solutions to the Einstein equations in a neighborhood of the intersecting surface only requires prescribing the full spacetime metric (but not its transverse derivatives) on the initial hypersurfaces. In \cite{Luk} Luk extends Rendall's result when the intersection is a two-sphere by proving that the spacetime exists on a neighbourhood of the full two null hypersurfaces. In a recent work \cite{chrusciel2023neighborhood} the authors show that the same result holds regardless the dimension and topology of the intersection. Various other approaches to this problem can be found in \cite{friedrich1981,CandP,cabet}. An initial value problem is also known to be well-posed when the data is given on the future null cone of a point \cite{cone}. Recent research \cite{Mio1,Mio2} has also approached the characteristic problem from a more geometric perspective, incorporating transverse derivatives of the metric as initial data (and thus adding the corresponding constraint equations).\\

In this paper we analyze the situation when initial data is posed on a single null hypersurface. From a causal perspective, the absence of a second null hypersurface implies that extra information coming from the past can potentially influence the solution, hence spoiling uniqueness. However, in cases where the solution exhibits some symmetry, such as Killing vectors, initial data on a single null hypersurface may suffice to establish a unique solution to the Einstein equations. This issue has been extensively studied in recent years, for example when the hypersurface is a homothetic or a Killing horizon, as we review next. We emphasize that the following examples are not covered by the characteristic problem presented in the paragraph above.\\

In \cite{fefferman2012ambient} C. Fefferman and R. Graham introduced the so-called \textit{ambient metric}, inspired by the idea that the light-cone of a point in Minkowski spacetime of dimension $n+2$ encodes the conformal structure of the $n$-sphere. In order to generalize this idea to an arbitrary conformal class, their goal was to construct a homothetic horizon from the given conformal class, and embed it into an ambient manifold Ricci flat to infinite order. Fefferman and Graham proved that in the cases where the dimension of the conformal class is odd, the Einstein equations uniquely determine the full transverse expansion of the ambient metric along a transverse null coordinate at the horizon. However, in even dimensions, additional data is required to fully determine the expansion and, in addition there appears a so-called obstruction tensor whose vanishing or not determines whether the ambient metric can be Ricci flat to all orders at the horizon or not. When the obstruction tensor does not vanish, one can still construct an ambient metric which is Ricci flat to infinite order at the horizon by including log terms in the expansion. This gives rise to the so-called ``generalized ambient metric'' which exhibits only finite differentiability at the horizon. In a remarkable work \cite{rodnianski} the authors prove that this generalized ambient metric is not only Ricci flat to infinite order but in fact \textit{exactly} Ricci flat. It is important to note that in both scenarios, the data is not constrained by any equations, unlike in the spacelike and characteristic Cauchy problems.\\


Another example where the curvature conditions of the ambient spacetime fix the transverse expansion is a non-degenerate Killing horizon. In \cite{moncrief1982neighborhoods} Moncrief shows that the transverse expansion of any 3-dimensional, non-degenerate Killing horizon on a Ricci flat spacetime can be described by six functions at the horizon. This analysis was done in a particular coordinate system, which makes it difficult to determine whether two spacetimes are isometric to infinite order at the horizon or not. In a recent work \cite{oliver}, the authors study this problem from a geometric perspective and establish that, for Ricci flat spacetimes, the complete expansion along a null transverse vector field is determined by ``non-degenerate Killing horizon data'', namely a triple $(\H,\bm\sigma,\mc V)$ where $\bm\sigma$ is a Riemannian metric and $\mc V$ is a non-trivial Killing vector field $\mc V$ on $(\H,\bm\sigma)$ with constant norm. Although this analysis characterizes Killing horizons geometrically, it excludes the possibility of $\mc V$ having zeros (and consequently bifurcation surfaces on the horizon) and the field equations are restricted to vacuum with $\Lambda=0$. \\

The last example we want to illustrate is the degenerate Killing horizon case. Given a degenerate Killing horizon, its so-called \textit{near horizon limit} \cite{kunduri2009classification} is characterized by a function $F$, a one-form $\bm\omega$, and a metric $h$ on the cross-sections. In a recent work \cite{katona2024uniqueness} the authors focus on establishing a uniqueness theorem for extremal Schwarzschild-de Sitter spacetime. For a specific value of the mass parameter, this spacetime admits a degenerate Killing horizon with compact, maximally symmetric cross-sections, thus possessing an associated near horizon limit. By staticity and compactness arguments, it can be proven that the one-form $\bm\omega$ vanishes \cite{chrusciel2005non,bahuaud2022static,wylie2023rigidity}, simplifying the data to the function $F$ and the metric $h$. By solving the $\Lambda$-vacuum equations order by order, which translates into elliptic equations for the transverse expansion, the authors show that all the transverse derivatives of the metric at the horizon can be computed from the data $(h,F)$ and that they agree either with those of extremal Schwarzschild-de Sitter or with its near horizon geometry (Nariai). Consequently, in the case of a real analytic spacetime, they can prove that in a neighbourhood of the horizon the spacetime is isometric to one of these two solutions. This analysis has been extended to electrovacuum in \cite{katona2}.\\

Motivated by the previous examples, in this paper we prove general identities that relate the transverse derivatives of the ambient Ricci tensor with the transverse expansion of the metric at an arbitrary null hypersurface. Our analysis is coordinate free, does not require any field equations and holds regardless the signature of the ambient metric\footnote{Provided it admits degenerate hypersurfaces, of course.}, the dimension or the topology of the hypersurface. Furthermore, we analyze the case where the ambient space admits a preferred vector field $\eta$ that is null and tangent to the hypersurface (e.g. a Killing or a homothetic vector field) and whose deformation tensor $\lie_{\eta}g$ is known. Such vectors are called ``symmetry generators''. Moreover, our analysis includes the possibility of $\eta$ vanishing on subsets of $\H$ with empty interior (e.g. a bifurcate Killing horizon).\\

Additional references closely related to the topic of this paper are the classic rigidity theorem by Hawking \cite{HawkingEllis} on existence of a second Killing vector in rotating analytic black holes in equilibrium (see also \cite{moncrief1983symmetries}), the work by Klainerman-Ionescu \cite{ionescu2009uniqueness} and Alexakis-Ionescu-Klainerman \cite{alexakis2010hawking} on the existence of Hawking's Killing field without assuming analiticity, and an analogous result by Petersen \cite{petersen2021extension} and Petersen-Rácz \cite{oliverracz} on compact Cauchy horizons with non-zero surface gravity.\\

To work with abstract null hypersurfaces we employ the so-called \textit{hypersurface data formalism} \cite{Marc3,Marc1,Marc2}, as it allows one to study hypersurfaces in a detached way, irrespective of their causal character. At the core of this formalism lies the concept of \textit{metric hypersurface data}, which comprises an abstract manifold $\mc{H}$ along with a (0,2)-symmetric tensor field $\bg$, a one-form $\bm\ell$, and a scalar function $\elltwo$. When $\mc H$ happens to be embedded in an ambient manifold $(\mc M,g)$, the tensor $\bg$ agrees with the first fundamental form of $\mc H$, while the one-form $\bm\ell$ and the scalar $\elltwo$ capture the transverse-tangent and transverse-transverse components of the ambient metric, respectively. From $\{\bg,\bm\ell,\elltwo\}$ one can also introduce contravariant data on $\H$ that consists of a (2,0) symmetric tensor field $P$, a vector $n$ and a scalar $n^{(2)}$. In addition, metric hypersurface data $\{\H,\bg,\bm\ell,\elltwo\}$ is subject to a gauge freedom that encompasses, at the abstract level, the multiplicity of choices of transverse vector $\xi$ along $\mc H$. Metric hypersurface data is also endowed with a torsion-free connection $\nablacero$ constructed solely in terms of $\bg$, $\bm\ell$ and $\elltwo$. The arbitrary causal character of $\mc H$ allows the analysis of abstract null geometry, even though neither a metric nor its Levi-Civita connection are present in the null case.\\

Let us denote by $2\bY^{(k)}$ the pullback of the $k$-th Lie derivative of $g$ along a transverse vector $\xi$ at $\mc H$. The collection $\{\bY^{(k)}\}_{k\ge 1}$ will be called \textit{transverse} or \textit{asymptotic expansion}. Our approach to determine this expansion at a general null hypersurface using the hypersurface data formalism is as follows. We begin by establishing a general identity that relates the $m$-th Lie derivative of the curvature and Ricci tensors along any vector $\xi$ with the Lie derivatives of a connection along $\xi$. For the Levi-Civita connection this results in an identity linking the $m$-th Lie derivative of the Ricci tensor with the Lie derivatives of the metric. By applying this general identity to a null hypersurface $\H$ with a transverse vector $\xi$ we can explicitly obtain identities for the leading order terms of the (i) completely transverse, (ii) transverse-tangent, and (iii) completely tangent components of the $m$-th Lie derivative of the ambient Ricci tensor along $\xi$ (Corollary \ref{cor_entero}). A key property of the result is its geometric nature, as the identities depend solely on metric hypersurface data and $\{\bY^{(k)}\}_{k\ge 1}$ once the transverse vector $\xi$ is extended off the hypersurface geodesically.\\

The identities (i)-(iii) described in the previous paragraph are of a very distinct nature. Firstly, the completely transversal component of the $m$-th derivative of the Ricci tensor depends algebraically on the trace of the tensor $\bY^{(m+2)}$ w.r.t $P$, that we denote by $\tr_P\bY^{(m+2)}$. Secondly, its transverse-tangent components depend algebraically on the one-form $\bY^{(m+2)}(n,\cdot)$, where $n$ is the null generator of the hypersurface. And finally, the identity for the fully tangential components of the $m$-th derivative of the Ricci depends on $\bY^{(m+1)}$ via a transport equation along the null generator of $\mc H$, but also on the scalar $\tr_P\bY^{(m+1)}$ and the one form $\bY^{(m+1)}(n,\cdot)$ and its derivatives. Since these two last objects depend algebraically on the derivatives of the Ricci tensor, identity (iii) allows us to know the evolution of the transverse expansion on any null hypersurface along the null direction provided the Ricci tensor is known at all orders on $\H$. Observe that, for a fixed $m$, the leading order terms of the identity (iii) carry \textit{at most} $m+1$ transverse derivatives of the metric, whereas the leading order terms in identities (i) and (ii) always depend on $m+2$ transverse derivatives of $g$ at $\H$.\\

When a symmetry generator $\eta$ is present, it is possible to transfer information from its deformation tensor $\lie_{\eta}g$ into the identity (iii). This gives rise to a new identity called ``generalized master equation of order $m$'' (cf. \eqref{lierictang3}) due to its close relationship to the ``generalized master equation'' found in \cite{tesismiguel} for the case $m=1$. Its key property is that its dependence on $\bY^{(m+1)}$ is not via a transport equation anymore. In addition, it also depends on the one-form $\bY^{(m+1)}(n,\cdot)$ and the scalar $\tr_P\bY^{(m+1)}$. This identity will allow us to identify the minimum amount of data at $\H$ in order to determine the full transverse expansion in terms of the Ricci tensor and its derivatives at the hypersurface. Informally, the idea is that, at each order, from the identities (i) and (ii) of order $m-1$ one can obtain algebraically $\tr_P\bY^{(m+1)}$ as well as $\bY^{(m+1)}(n,\cdot)$. By introducing such $\tr_P\bY^{(m+1)}$ and $\bY^{(m+1)}(n,\cdot)$ into the generalized master equation of order $m$, the remaining components of $\bY^{(m+1)}$ can then be obtained algebraically.\\

In the last part of the paper we apply the generalized master equation of order $m$ to the Killing horizon case. For non-degenerate Killing horizons, we are able to show that the full transverse expansion is uniquely determined from abstract data at the horizon as well as on the tower of derivatives of the ambient Ricci tensor at $\H$ (Theorem \ref{teorema1}). This extends the main result of \cite{oliver} in several directions. Firstly because our approach allows for much more general field equations besides vacuum with $\Lambda=0$. Secondly because we are allowing zeroes of $\eta$, which includes the possibility of the horizon having bifurcation surfaces. And finally, because our result extends to arbitrary ambient signature.\\

The collection of transverse derivatives of the ambient Ricci tensor at $\H$ can be thought at least in two different ways. One possibility is to provide such collection as prescribed data on the hypersurface, e.g. given by some external matter field. Another option is to think of the collection as functional relations between metric hypersurface data and the transverse expansion at the horizon. The easiest example that illustrates the second viewpoint are the $\Lambda$-vacuum equations, as in this case the Ricci tensor is proportional to the metric, so its transverse derivatives will depend on the expansion itself. This property is captured in our Definition \ref{defi_hier}, and allows us to prove that when the ambient Ricci tensor fulfills this property, the full expansion at $\H$ is uniquely determined from abstract data at the horizon (Theorem \ref{teorema12}). When applied to the vacuum equations with cosmological constant, our theorem provides a characterization of all possible analytic $\Lambda$-vacuum manifolds in the vicinity of non-degenerate horizons. As a direct consequence, we establish a uniqueness result for non-extremal Schwarzschild-de Sitter spacetime (Proposition \ref{cor_final}), complementing the main result of \cite{katona2024uniqueness}.\\

The paper is organized as follows. In Section \ref{sec_hypersurfacedata}, we provide a self-contained overview of the fundamental concepts of hypersurface data formalism. In particular, we recall the fact that the fully tangential components of the ambient Ricci tensor at a null hypersurface are given in terms of hypersurface data. In Section \ref{sec_fullRicci} we proceed to calculate the remaining components of the Ricci tensor, namely its fully transverse and transverse-tangent components. To this aim we need several identities involving the pull-back of arbitrary covariant tensors and their derivatives on the null hypersurface that we derive in Appendix \ref{appendix}. Once the full Ricci tensor is expressed in terms of metric data and transverse derivatives of the metric on the null hypersurface, in Section \ref{section_higher} we determine the connection between higher order derivatives of the ambient Ricci tensor and higher order derivatives of the metric. Section \ref{section_deformation} is devoted to the analysis of the algebraic identities obtained when a general symmetry generator is present on $\H$. In Section \ref{sec_KH} we focus our attention to non-degenerate Killing horizons and we prove that the asymptotic expansion on any non-degenerate Killing horizon is uniquely determined in terms of abstract data at the horizon. The paper also includes a final section where we discuss the objectives and contents of the sequel of this paper.

\section*{Notation and conventions}

Throughout this paper $(\mc M,g)$ denotes an arbitrary smooth $\mf{n}$-dimensional semi-Riemannian manifold of any signature $(p,q)$ with both $p$ and $q$ different from zero. When we specifically need this signature to be Lorentzian, we will say so explicitly. We employ both index-free and abstract index notation at our convenience. Ambient indices are denoted with Greek letters, abstract indices on a hypersurface are written in lowercase Latin letters, and abstract indices at cross-sections of a hypersurface are expressed in uppercase Latin letters. As usual, square brackets enclosing indices denote antisymmetrization and parenthesis are for symmetrization. The symmetrized tensor product is denoted with $\otimes_s$. By $\mc F(\mc M)$, $\X(\mc M)$ and $\X^{\star}(\mc M)$ we denote, respectively, the set of smooth functions, vector fields and one-forms on $\mc M$. The subset $\mc F^{\star}(\mc M)\subset\mc F(\mc M)$ consists of the nowhere vanishing functions on $\mc M$. The pullback of a function $f$ via a diffeomorphism $\Phi$ will be denoted by $\Phi^{\star}f$ or simply by $f$ depending on the context. Given a diffeomorphism $\Phi$ and a vector field $X$, we define $\Phi^{\star}X\d (\Phi^{-1})_{\star}X$. A $(p,q)$-tensor refers to a tensor field $p$ times contravariant and $q$ times covariant. Given any pair of $(2,0)$ and $(0,2)$ tensors $A^{ab}$ and $B_{cd}$ we denote $\tr_A \bm B \d A^{ab}B_{ab}$. We employ the symbol $\nabla$ for the Levi-Civita connection of $g$. Throughout this paper we use the notation $\lie_X^{(m)}T$ to denote the $m$-th Lie derivative of the tensor $T$ along $X$, and $X^{(m)}(f)$ for the $m$-th directional derivative of the function $f$ along $X$. When $m=1$ we also write $\lie_X T$ and $X(f)$, respectively, and when $m=0$ they are just the identity operators. All manifolds are assumed to be connected and smooth.

\section{Review of hypersurface data formalism}
\label{sec_hypersurfacedata}

This section is devoted to review the basic notions of the so-called \textit{hypersurface data formalism}. Further details can be found in \cite{Marc1,Marc2,Marc3}. The usefulness of this formalism has been recently demonstrated in the context of matching of spacetimes \cite{miguel1,miguel2,miguel4} and in solving the characteristic problem of general relativity from an abstract point of view \cite{Mio1,Mio2}. Let us start by introducing the necessary objects for this paper.

\begin{defi}
	\label{def_hypersurfacedata}
	Let $\mc H$ be a $d$-dimensional manifold, $\bg$ a symmetric (0,2)-tensor field, $\bm\ell$ a one-form and $\elltwo$ a scalar function on $\mc H$. We say that $\{\mc H,\bg,\bm\ell,\elltwo\}$ defines a \textbf{metric hypersurface data} set provided that (0,2) symmetric tensor $\bm{\mc A}|_p$ on $T_p\mc H\times\real$ defined by 
	\begin{equation}
		\label{def_A}
		\mc A|_p\left((W,a),(V,b)\right) \d \bg|_p (W,V) + a\bm\ell|_p(V)+b\bm\ell|_p(W)+ab\ell^{(2)}|_p
	\end{equation}
	is non-degenerate at every $p\in\mc H$. A five-tuple $\{\mc H,\bg,\bm\ell,\elltwo,\bY\}$, where $\bY$ is a (0,2) symmetric tensor field on $\mc H$, is called \textbf{hypersurface data}.
\end{defi}
The non-degeneracy of $\bm{\mc A}$ allows us to introduce its ``inverse'' $\bm{\mc A}^{\sharp}$ by $\bm{\mc A}^{\sharp}\big(\bm{\mc A}((V,a),\cdot),\cdot\big)=(V,a)$ for every $(V,a)\in\X(\mc H)\otimes\mc{F}(\mc H)$. From $\bm{\mc A}^{\sharp}$ one can define a (2,0) symmetric tensor field $P$, a vector $n$ and a scalar $n^{(2)}$ on $\mc H$ by the decomposition 
\begin{equation}
	\label{def_Asharp}
	\mc A^{\sharp}\left((\bm\alpha,a),(\bm\beta,b)\right) = P (\bm\alpha,\bm\beta)+a n(\bm\beta)+bn(\bm\alpha)+ab n^{(2)} \quad\, \forall\, (\bm\alpha,a),(\bm\beta,b)\in\X^{\star}(\mc H)\times F(\mc H),
\end{equation} 
Equivalently, $P$, $n$ and $n^{(2)}$ can be defined by
\begin{multicols}{2}
	\noindent
	\begin{align}
		\gamma_{ab}n^b + n^{(2)}\ell_a&=0,\label{gamman}\\
		\ell_an^a+n^{(2)}\ell^{(2)}&=1.\label{ell(n)}
	\end{align}
	\begin{align}
		P^{ab}\ell_b+\ell^{(2)} n^a&=0,\label{Pell}\\
		P^{ac}\gamma_{cb} + \ell_b n^a &=\delta^a_b.\label{Pgamma}
	\end{align}
\end{multicols}
Despite its name, the notion of hypersurface data does not view $\mc H$ as a hypersurface of another ambient manifold. The connection between Definition \ref{def_hypersurfacedata} and the standard definition of a hypersurface is as follows.
\begin{defi}
	\label{defi_embedded}
	Metric hypersurface data $\{\mc H,\bg,\bm\ell,\ell^{(2)}\}$ is $(\Phi,\xi)$-\textbf{embedded} in a semi-Riemannian manifold $(\mc M,g)$ if there exists an embedding $\Phi:\mc H\hookrightarrow\mc M$ and a vector field $\xi$ along $\Phi(\mc H)$ everywhere transversal to $\Phi(\mc H)$, called rigging, such that
	\begin{equation}
		\label{embedded_equations}
		\Phi^{\star}(g)=\bg, \hspace{0.5cm} \Phi^{\star}\left(g(\xi,\cdot)\right) = \bm\ell,\hspace{0.5cm} \Phi^{\star}\left(g(\xi,\xi)\right) = \ell^{(2)}.
	\end{equation}
	Hypersurface data $\{\mc H,\bg,\bm\ell,\ell^{(2)},\bY\}$ is embedded provided that, in addition, 
	\begin{equation}
		\label{Yembedded}
		\dfrac{1}{2}\Phi^{\star}\left(\lie_{\xi} g\right) = \bY.
	\end{equation}
\end{defi}
In the context of embedded (metric) hypersurface data, $\bg$ being degenerate is equivalent to $\Phi(\mc H)$ being an embedded null hypersurface. It is easy to show \cite{Marc2} that the degeneracy of $\bg$ is equivalent to $n^{(2)}=0$. Hence, (metric) hypersurface data satisfying $n^{(2)}=0$ are called \textbf{null (metric) hypersurface data}. A cross-section (or simply a section) $\mc S$ of $\mc H$ is an embedded hypersurface $\mc S\hookrightarrow\mc H$ with the property that every integral curve of $n$ crosses $\mc S$ exactly once. From now on we restrict ourselves to the null case.\\

Given null metric hypersurface data we define the tensor field 
\begin{equation}
	\label{defK}
	\bU\d \dfrac{1}{2}\lie_n\bg.
\end{equation} 
When the data is embedded $\bU$ coincides with the second fundamental form of $\Phi(\mc H)$ w.r.t the unique normal one-form $\bm{\nu}$ satisfying $\bm{\nu}(\xi)=1$ (see \cite{Marc1}). It is also convenient to introduce the tensors\\
\begin{minipage}{0.5\textwidth}
\begin{equation}
	\label{def_F}
	\bF\d \dfrac{1}{2}d\bm\ell,
\end{equation}
\end{minipage}
\begin{minipage}{0.5\textwidth}
\begin{equation}
	\label{defPi}
	\bPi \d \bY+\bF.
\end{equation}
\end{minipage}
\vspace{0.2cm}

Let $\{\mc H,\bg,\bm\ell,\elltwo\}$ be null metric hypersurface data $(\Phi,\xi)$-embedded in $(\mc M,g)$. By the transversality of $\xi$, given a (local) basis $\{e_a\}$ of $\mc H$, the set $\{\wh{e}_a\d\Phi_{\star}e_a,\xi\}$ is a (local) basis of $\Phi(\mc H)$ with dual basis $\{\bm\theta^a,\bm\nu\}$. Raising the indices we can introduce $\nu\d g^{\sharp}(\bm\nu,\cdot)$ and $\theta^a \d g^{\sharp}(\bm\theta^a,\cdot)$, which are given in terms of $\{\xi, \wh{e}_a\}$ by 
\begin{equation}
	\label{nuthetaa}
	\nu =  n^a \wh{e}_a,\qquad \theta^a = P^{ab}\wh{e}_b + n^a\xi.
\end{equation}
From \eqref{def_Asharp} the inverse metric $g^{\alpha\beta}$ at $\mc H$ can be written in the basis $\{\xi, \wh{e}_a\}$ as
\begin{equation}
	\label{inversemetric}
	g^{\alpha\beta}\st{\mc H}{=}	P^{ab}\wh e_a^{\alpha}\wh e_b^{\beta} + n^{a}\wh e_a^{\alpha}\xi^{\beta} + n^{b}\wh e_b^{\beta}\xi^{\alpha} .
\end{equation}
In the embedded picture the notion of rigging vector is non-unique, since given a rigging $\xi$ any other vector of the form $\xi' = z(\xi+\Phi_{\star}V)$ with $(z,V)\in\mc{F}^{\star}(\mc H)\times\X(\mc H)$ is also transverse to $\Phi(\mc H)$. Translating this into the abstract setting we have the following definition.
\begin{defi}
	\label{defi_gauge}
	Let $\{\mc H,\bg,\bm\ell,\ell^{(2)},\bY\}$ be hypersurface data and $(z,V)\in\mc{F}^{\star}(\mc H)\times\X(\mc H)$. We define the \textbf{gauge} transformed hypersurface data with gauge parameters $(z,V)$ by
	\begin{align}
		\mc{G}_{(z,V)}\left(\bg \right)&\d \bg,\label{transgamma}\\
		\mc{G}_{(z,V)}\left( \bm{\ell}\right)&\d z\left(\bm{\ell}+\bg(V,\cdot)\right),\label{tranfell}\\
		\mc{G}_{(z,V)}\big( \ell^{(2)} \big)&\d z^2\big(\ell^{(2)}+2\bm\ell(V)+\bg(V,V)\big),\label{transell2}\\
		\mc{G}_{(z,V)}\left( \bY\right)&\d z \bY + \bm\ell\otimes_s d z +\dfrac{1}{2}\lie_{zV}\bg.\label{transY}
	\end{align}
\end{defi}
Transformations \eqref{transgamma}-\eqref{transY} induce the corresponding transformations on $P$ and $n$ \cite{Marc1}
\begin{multicols}{2}
	\noindent
	\begin{equation}
		\label{gaugeP}
		\mc{G}_{(z,V)}\left(P \right) = P -2V\otimes_s n,
	\end{equation}
	\begin{equation}
		\label{transn}
		\mc{G}_{(z,V)}\left( n \right)= z^{-1}n.
	\end{equation}
\end{multicols}
The set of gauge transformations defines a group whose composition law and the inverse element are \cite{Marc1}
\begin{multicols}{2}
	\noindent
	\begin{equation}
		\label{group}
		\mc{G}_{(z_1,V_1)}\circ \mc{G}_{(z_2,V_2)} = \mc{G}_{(z_1 z_2,V_2+z_2^{-1}V_1)}
	\end{equation}
	\begin{equation}
		\label{gaugelaw}
		\mc{G}_{(z,V)}^{-1} = \mc{G}_{(z^{-1},-zV)}.
	\end{equation}
\end{multicols}
As expected from the geometric interpretation of $\bU$ as the second fundamental form of $\Phi(\mc H)$ w.r.t $\nu\d\Phi_{\star}n$, its gauge transformation is given by \cite{Marc1}
\begin{equation}
	\label{Ktrans}
	\mc{G}_{(z,V)}\left(\bU \right)= z^{-1}\bU.
\end{equation} 
A consequence of this together with \eqref{gaugeP} and $\bU(n,\cdot)=0$ is
\begin{equation}
	\label{transtrPU}
	\tr_{P'}\bU' = z^{-1}\tr_{P}\bU.
\end{equation}
Given null metric hypersurface data $\{\mc H,\bg,\bm\ell,\elltwo\}$ it is possible to define a torsion-free connection $\nablacero$ on $\mc H$ by means of \cite{Marc2}
\begin{multicols}{2}
	\noindent
	\begin{equation}
		\label{nablagamma}
		\nablacero_a\gamma_{bc} = -\ell_c\U_{ab} - \ell_b\U_{ac},
	\end{equation}
	\begin{equation}
		\label{nablaell}
		\nablacero_a\ell_b  = \F_{ab} - \elltwo\U_{ab}.
	\end{equation}
\end{multicols}
When the data is embedded in $(\mc M,g)$, $\nablacero$ is related with the Levi-Civita connection $\nabla$ of $g$ by 
\begin{equation}
	\label{connections}
	\nabla_{\Phi_{\star}X}\Phi_{\star}Y \st{\mc H}{=} \Phi_{\star}\nablacero_X Y - \bY(X,Y)\nu - \bU(X,Y)\xi \qquad \forall X,Y\in\X(\mc H).
\end{equation}
Unless otherwise indicated, scalar functions related by $\Phi^{\star}$ are denoted with the same symbol. The action of $\nablacero$ on the contravariant data $\{P,n\}$ is given by \cite{Marc2}
\begin{align}
	\nablacero_c n^b & =\s_c n^b + P^{ba}\U_{ca},\label{derivadannull}\\
	\nablacero_c P^{ab} & = -\big(n^aP^{bd}+n^bP^{ad}\big) \F_{cd} - n^an^b (d\elltwo)_c,\label{derivadaP}
\end{align}
where $\bs\d\bF(n,\cdot)$. A direct consequence of \eqref{connections} and \eqref{derivadannull} is 
\begin{equation}
	\nabla_{\nu}\nu \st{\H}{=} -\bY(n,n)\nu.
\end{equation}
It is therefore natural to define the surface gravity of $n$ by $\kappa_n\d -\bY(n,n)$, whose gauge transformation law follows directly from \eqref{transn} and \eqref{transY} and is
\begin{equation}
	\label{transkappa}
	\kappa_n' = z^{-1}\kappa_n - z^{-1} n(\log|z|).
\end{equation}
Applying the Cartan identity $\lie_n = \iota_n \circ d + d\circ \iota_n$ to the one-form $\bm\ell$ and using $2\bF = d\bm\ell$ and $\bm\ell(n)=1$ one has 
\begin{equation}
	\label{lienell}
	\lie_n\bm\ell = 2\bs.
\end{equation}
Another consequence of \eqref{derivadannull} is that for any one-form $\bm\theta$ it holds \cite{miguel3}
\begin{equation}
	\label{nnablacerotheta}
	2n^b \nablacero_{(a}\theta_{b)} = \lie_n\theta_a + \nablacero_a\big(\bm\theta(n)\big) - 2\big(\bm\theta(n)\s_a + P^{bc}\U_{ab}\theta_c\big).
\end{equation}
For future use we need to know the commutator $[P^{ab},\lie_n]$ acting on a 2-covariant, symmetric tensor field $T_{ab}$. 
\begin{lema}
	Let $\T_{ab}$ a (0,2) symmetric tensor field. Then,
	\begin{equation}
		\label{lietrPY}
		P^{ab}\lie_n \T_{ab} = \lie_n(\tr_P\bT) + 4P(\bm t,\bs)+2P^{ac}P^{bd}\U_{cd}T_{ab} + n(\elltwo)\bm t(n),
	\end{equation}
	where $\bm t\d \bT(n,\cdot)$. 
	\begin{proof}
		The result follows at once from the following expression of the Lie derivative of $P$ along $n$ \cite{MarcAbstract} 
		\begin{equation}
			\label{lienP}
			\lie_n P^{ab} = -2 \s_c \left(P^{ac}n^b + P^{bc}n^a\right) -2P^{ac}P^{bd}\U_{cd} - n^an^bn\big(\elltwo\big).
		\end{equation}
	\end{proof}
\end{lema}
%
Later we will also need the $\nabla$-derivative of $\xi$ along tangent directions to $\mc H$ \cite{Marc1}
\begin{equation}
	\label{nablaxi}
	\wh{e}_a^{\mu}\nabla_{\mu}\xi^{\beta}\st{\mc H}{=} (\r-\s)_a\xi^{\beta} + P^{cd}\Pi_{ac}\wh{e}^{\beta}_d + \dfrac{1}{2}\nu^{\beta}\nablacero_a\elltwo,
\end{equation}
where $\br\d \bY(n,\cdot)$. The gauge transformation law of the one-form $\bPi(\cdot,n)=\br-\bs$ is \cite{Mio1}
\begin{equation}
	\label{gauger-s}
	\mc{G}_{(z,V)}\big(\br-\bs\big) =\br-\bs - \bU(V,\cdot) + d\log|z|.
\end{equation}
As proven in \cite{Marc1}, the completely tangential components of the ambient Riemann tensor, as well as its 3-tangential, 1-transverse components can be written in terms of hypersurface data as
\begin{equation}
	\label{ABembedded}
	R_{\alpha\beta\mu\nu}\xi^{\alpha}e^{\beta}_be^{\mu}_ce^{\nu}_d \st{\mc H}{=} A_{bcd}, \hspace{1cm} R_{\alpha\beta\mu\nu} e^{\alpha}_ae^{\beta}_be^{\mu}_ce^{\nu}_d \st{\mc H}{=} B_{abcd},
\end{equation}
where $A$ and $B$ are the tensors on $\mc H$ defined by 
\begin{align}
	A_{bcd}&\d 2\nablacero_{[d}\F_{c]b} + 2\nablacero_{[d}\Y_{c]b} + \U_{b[d}\nablacero_{c]}\elltwo + 2\Y_{b[d}(\r-\s)_{c]},\label{A} \\
	B_{abcd}&\d \gamma_{af}\Rcero^f{}_{bcd} + 2\ell_a\nablacero_{[d}\U_{c]b}+2\U_{a[d}\Y_{c]b} + 2\U_{b[c}\Pi_{d]a},\label{B}
\end{align}
and $\Rcero^f{}_{bcd}$ is the curvature of $\nablacero$.
	It follows from \eqref{ABembedded} that all the tangential components of the ambient Ricci tensor can be written in terms of hypersurface data as \cite{Mio1} $$g^{\alpha\beta} R_{\alpha \mu \beta \nu} e_a^{\mu} e_b^{\nu}\st{\mc H}{=}B_{acbd}P^{cd}- (A_{bac}+A_{abc})n^c.$$ The RHS defines a tensor on $\mc H$ called \textit{constraint tensor} $\R$. Its explicit form is \cite{miguel3}
	\begin{equation}
		\label{constraint}
		\begin{aligned}
			\mc R_{ab}& = \accentset{\circ}{R}_{(ab)} -2\lie_n \Y_{ab} - (2\kappa_n+\tr_P\bU)\Y_{ab} + \nablacero_{(a}\left(\s_{b)}+2\r_{b)}\right)\\
			&\quad -2\r_a\r_b + 4\r_{(a}\s_{b)} - \s_a\s_b - (\tr_P\bY)\U_{ab} + 2P^{cd}\U_{d(a}\left(2\Y_{b)c}+\F_{b)c}\right),
		\end{aligned}
	\end{equation}
	where $\Rcero_{ab}$ is the Ricci tensor of the connection $\nablacero$. The tensor $\R$ is abstract in the sense that it does not require the data to be embedded in any ambient manifold. Note that all the dependence on the tensor $\bY$ in \eqref{constraint} is explicit. Below we shall need the explicit form of the contraction of $\mc{R}_{ab}$ with $n^a$ and $P^{ab}$. The former was obtained in \cite{miguel3} and relies on the following general identity, also derived in \cite{miguel3}
	\begin{equation}
		\label{Rceron}
		\Rcero_{(ab)}n^a = \dfrac{1}{2}\lie_n\s_b - 2P^{ac}\U_{ab}\s_c + P^{ac}\nablacero_c\U_{ab} - \nablacero_b\big(\tr_P\bU\big) + \big(\tr_P\bU\big)\s_b.
	\end{equation}
	The result is
	\begin{equation}
		\label{constraintn}
		\begin{aligned}
			\mc R_{ab}n^a  = -\lie_n(\r_b-\s_b) - \nablacero_b \kappa_n - (\tr_P\bU) (\r_b-\s_b)  - \nablacero_b (\tr_P\bU) + P^{cd}\nablacero_c\U_{bd} - 2P^{cd}\U_{bd}s_c.
		\end{aligned}
	\end{equation}
	Another contraction with $n^b$ gives
	\begin{equation}
		\label{constraintnn}
		\mc R_{ab}n^an^b = -n(\tr_P\bU) + (\tr_P\bU)\kappa_n - P^{ab}P^{cd}\U_{ac}\U_{bd},
	\end{equation}
	which is the abstract version of the Raychaudhuri equation. For the trace of \eqref{constraint} with respect to $P$ we simply need to use \eqref{lietrPY} and get
	\begin{equation}
		\label{trPR}
		\begin{aligned}
			\tr_P\bm{\mc R} &= \tr_P \Rcero -2\lie_n(\tr_P\bY)  - 2\big(\kappa_n+\tr_P\bU\big)\tr_P\bY+ \div_P(\bs + 2\br) \\
			&\quad -2P(\br,\br)-4P(\br,\bs)-P(\bs,\bs)+2\kappa_n n(\elltwo),
		\end{aligned}
	\end{equation}
	where $\div_P\bm{t}\d P^{ab}\nablacero_a t_b$.
\section{Full Riemann tensor at a hypersurface}
\label{sec_fullRicci}

While the components of the form $R_{\alpha\beta\mu\nu}\xi^{\alpha}\wh e^{\beta}_b\wh e^{\mu}_c \wh e^{\nu}_d$ and $R_{\alpha\beta\mu\nu}\wh e^{\alpha}_a\wh e^{\beta}_b\wh e^{\mu}_c \wh e^{\nu}_d$ can be written solely in terms of hypersurface data, the remaining ones, i.e. $R_{\alpha\beta\mu\nu}\xi^{\alpha}\wh e^{\beta}_b\xi^{\mu} \wh e^{\nu}_d$, cannot, because in addition they require second order transverse derivatives of the metric. In order to compute them we use a lemma from \cite{marsperturbations}.
\begin{lema}
	\label{propriemxixi1}
	Let $(\mc M,g)$ be a semi-Riemannian manifold and $\xi,X,Y\in\X(\mc M)$. Then,
	\begin{equation}
		\label{riemxixi}
		\riem(\xi,X,Y,\xi) = \dfrac{1}{2}\big(\lie_{\xi}^{(2)}g\big)(X,Y) - \dfrac{1}{2}\left(\lie_{\nabla_{\xi}\xi} g\right)(X,Y) - g(\nabla_X\xi,\nabla_Y\xi).
	\end{equation}
\end{lema}
This lemma allows us to write the remaining components of the ambient Riemann tensor in terms of second transverse derivatives of the metric and hypersurface data.
\begin{prop}
	\label{propriemxixi}
	Let $\{\mc H,\bg,\bm\ell,\elltwo,\bY\}$ be null hypersurface data $(\Phi,\xi)$-embedded in $(\mc M,g)$ and define $\beta\in\mc F(\mc H)$ and $T\in\X(\mc H)$ by the decomposition $\nabla_{\xi}\xi \st{\mc H}{=} \beta\xi + \Phi_{\star}T$. Then, 
	\begin{equation}
		\label{riemxixiH}
		\Phi^{\star}\left(\riem(\xi,\cdot,\cdot,\xi)\right) = \dfrac{1}{2}\Phi^{\star}\big(\lie_{\xi}^{(2)}g\big) - \beta\bY - d\beta\otimes_s\bm\ell - \dfrac{1}{2}\lie_T\bg - \bPi\cdot\bPi -(\br-\bs)\otimes_s d\elltwo,
	\end{equation}	
	where $\big(\bPi\cdot\bPi\big)(X,Y)\d P(\bPi(X,\cdot),\bPi(Y,\cdot))$.
	\begin{proof}
		Let $\{e_a\}$ be a (local) basis of $\X(\mc H)$ and $\wh{e}_a\d\Phi_{\star}e_a$. From \eqref{nablaxi} together with \eqref{embedded_equations} and recalling that $\nu$ is null, normal to $\Phi(\mc H)$ and satisfies $g(\xi,\nu)=1$, it follows
		\begin{align*}
			g_{\alpha\beta} \nabla_{\wh{e}_a}\xi^{\alpha}\nabla_{\wh{e}_b}\xi^{\beta} & = \elltwo(\r-\s)_a(\r-\s)_b + 2(\r-\s)_{(a|}\big(\ell_dP^{fd}\Pi_{|b)f}+\dfrac{1}{2}\nablacero_{|b)}\elltwo\big) \\
			&\quad\, + P^{cd}P^{fg}\Pi_{ac}\Pi_{bf}\gamma_{dg}\\
			&= P^{cd}\Pi_{ac}\Pi_{bd} + (\r-\s)_{(a}\nablacero_{b)}\elltwo ,
		\end{align*}
		where in the second equality we used $\ell_dP^{fd}=-\elltwo n^f$. Equation \eqref{riemxixiH} follows from \eqref{riemxixi} after using 
		\begin{align}
			\Phi^{\star}\left(\lie_{\nabla_{\xi}\xi}g\right) & = \Phi^{\star}\left(\lie_{\beta\xi}g\right) + \Phi^{\star}\left(\lie_{\Phi_{\star}T}g\right)\nonumber\\
			&=\Phi^{\star}\left(\beta\lie_{\xi}g\right) + \Phi^{\star} \left(2 d\beta\otimes_s g(\xi,\cdot)\right)+ \Phi^{\star}\left(\lie_{\Phi_{\star}T}g\right)\nonumber\\
			&= 2\beta\bY+2d\beta\otimes_s\bm\ell  + \lie_T\bg.\label{prop2aux1}
		\end{align}
	\end{proof}
\end{prop}
Let $X,Y\in\X(\mc H)$. Since $\nabla_{\Phi_{\star}X}\xi$ and $\nabla_{\Phi_{\star}Y}\xi$ only depend on $\xi$ along $\Phi(\mc H)$ and $\riem$ is a tensor, it follows that $\riem(\xi,\Phi_{\star}X,\Phi_{\star}Y,\xi) + g(\nabla_{\Phi_{\star}X}\xi,\nabla_{\Phi_{\star}Y}\xi)$ only depends on $\xi$ along $\Phi(\mc H)$. Thus, by equation \eqref{riemxixi} the tensor $$\dfrac{1}{2}\Phi^{\star}\big(\lie_{\xi}^{(2)}g\big) - \dfrac{1}{2}\Phi^{\star}\left(\lie_{\nabla_{\xi}\xi} g\right)$$ is independent of how one extends $\xi$ off $\Phi(\mc H)$, so by \eqref{prop2aux1} it follows that the tensor field $$\dfrac{1}{2}\Phi^{\star}\big(\lie_{\xi}^{(2)}g\big) - \beta\bY - d\beta\otimes_s\bm\ell - \dfrac{1}{2}\lie_T\bg$$ does not depend on the extension of $\xi$ off $\Phi(\mc H)$. This suggests extending the definition of hypersurface data set as follows.
\begin{defi}
	A septuple $\{\mc H,\bg,\bm\ell,\elltwo,\bY,\bZ^{(2)}\}$ defines an \textbf{extended hypersurface data} set provided $\{\mc H,\bg,\bm\ell,\elltwo,\bY\}$ is hypersurface data and $\bZ^{(2)}$ is a (0,2) symmetric tensor field on $\mc H$. 
\end{defi}
\begin{defi}
	\label{defiZ2}
	An extended hypersurface data set $\{\mc H,\bg,\bm\ell,\elltwo,\bY,\bZ^{(2)}\}$ is said to be $(\Phi,\xi)$-\textbf{embedded} in a semi-Riemannian manifold $(\mc M,g)$ provided $\{\mc H,\bg,\bm\ell,\elltwo,\bY\}$ is $(\Phi,\xi)$-embedded in the sense of Definition \ref{defi_embedded} and, in addition, $$\bZ^{(2)} = \dfrac{1}{2}\Phi^{\star}\big(\lie_{\xi}^{(2)}g\big) - \dfrac{1}{2}\beta\Phi^{\star}\left(\lie_{\xi} g\right) - d\beta\otimes_s\Phi^{\star}(g(\xi,\cdot)) - \dfrac{1}{2}\Phi^{\star}\left(\lie_{\Phi_{\star}T}g\right),$$ where $(\beta,T)\in\mc F(\mc H)\times\X(\mc H)$ are defined by $\nabla_{\xi}\xi \st{\mc H}{=} \beta\xi + \Phi_{\star}T$ and $\xi$ is any extension of the rigging off $\Phi(\mc H)$. 
\end{defi}
In terms of $\bZ^{(2)}$, equation \eqref{riemxixiH} gets rewritten as 
\begin{equation}
	\label{riemxixiH2}
	\Phi^{\star}\left(\riem(\xi,\cdot,\cdot,\xi)\right) = \bZ^{(2)} - \bPi\cdot\bPi - (\br-\bs)\otimes_s d\elltwo.
\end{equation}
Once the full Riemann tensor at $\mc H$ is computable from hypersurface data and $\bZ^{(2)}$, one can write down explicitly the full Ricci tensor on the abstract null hypersurface.
\begin{prop}
	\label{cor_ricxi}
	Let $\mc D = \{\mc H,\bg,\bm\ell,\ell^{(2)},\bY,\bZ^{(2)}\}$ be extended null hypersurface data $(\Phi,\xi)$-embedded in $(\mc M,g)$. Let $\ric$ be the Ricci tensor of $g$ and $\mc{\dot R}\d \Phi^{\star}\big(\ric(\xi,\cdot)\big)$, $\mc{\ddot R}\d  \Phi^{\star}\big(\ric(\xi,\xi)\big)$. Then, 
	\begin{align}		
		\mc{\ddot R} & = -\tr_P\bZ^{(2)} + \tr_P\big(\bPi\cdot\bPi\big) + P(\br-\bs,d\elltwo) ,\label{ricxixi}\\
		\mc{\dot R}_c  & = - P^{ab}A_{abc} + \z^{(2)}_c  - P^{ab}\Pi_{da}\Pi_{cb}n^d +\dfrac{1}{2}\kappa_n \nablacero_c\elltwo -\dfrac{1}{2} n(\elltwo)(\r-\s)_{c} ,\label{ricxiX}
	\end{align}
	where $A_{abc}$ is defined in \eqref{A} and $\bz^{(2)} \d \bZ^{(2)}(n,\cdot)$.
	\begin{proof}
		Let $\{e_a\}$ be a (local) basis of $\X(\mc H)$ and $\wh{e}_a\d\Phi_{\star}e_a$. From the definition of the Ricci tensor, the symmetries of the Riemann tensor and equations \eqref{inversemetric} and \eqref{riemxixiH2} it follows
		\begin{align*}
			\ric(\xi,\xi) & \st{\mc H}{=} \big(P^{ab}\wh e_a^{\alpha}\wh e_b^{\beta} + n^{a}\wh e_a^{\alpha}\xi^{\beta} + n^{b}\wh e_b^{\beta}\xi^{\alpha}\big) R_{\alpha\mu\beta\nu}\xi^{\mu}\xi^{\nu}\\
			&\st{\mc H}{=} -P^{ab}\Z^{(2)}_{ab} +P^{ab} P^{cd}\Pi_{ca}\Pi_{db} + P^{ab}(\r-\s)_a \nablacero_b\elltwo.
		\end{align*}
		In the same way, using the first equation in \eqref{ABembedded}, given $X\in\X(\mc H)$
		\begin{align*}
			\ric(\xi,\Phi_{\star}X) & \st{\mc H}{=} \big(P^{ab}\wh e_a^{\alpha}\wh e_b^{\beta} + n^{a}\wh e_a^{\alpha}\xi^{\beta} + n^{a}\wh e_a^{\beta}\xi^{\alpha}\big) R_{\alpha\mu\beta\nu}\xi^{\mu}(\Phi_{\star}X)^{\nu}\\
			&\st{\mc H}{=} -P^{ab} A_{abc}X^c + \Z^{(2)}_{ab}n^aX^b - P^{cd}\Pi_{ac}\Pi_{bd}n^aX^b - n^aX^b(\r-\s)_{(a}\nablacero_{b)}\elltwo  .
		\end{align*}
	\end{proof}
\end{prop}
\section{Higher order derivatives}
\label{section_higher}

In this section we compute the derivatives $\lie_{\xi}^{(m)}\ric$ on $\mc H$ in terms of transverse derivatives of $g$ on $\mc H$ up to order $m+1$, i.e. making $\lie_{\xi}^{(m+2)}g$ and $\lie_{\xi}^{(m+1)}g$ explicit. In order to simplify the notation let us introduce the tensors $$\bY^{(m)}\d \dfrac{1}{2}\Phi^{\star}\big(\lie_{\xi}^{(m)}g\big), \qquad \br^{(m)}\d \bY^{(m)}(n,\cdot), \qquad \kappa^{(m)}\d -\bY^{(m)}(n,n),$$ 

as well as $$\mc{R}^{(m)}\d \Phi^{\star}\big(\lie_{\xi}^{(m-1)}\ric\big),\quad \dot{\mc R}^{(m)}\d \Phi^{\star}\big(\lie_{\xi}^{(m-1)}\ric(\xi,\cdot)\big),\quad \ddot{\mc R}^{(m)}\d \Phi^{\star}\big(\lie_{\xi}^{(m-1)}\ric(\xi,\xi)\big).$$

 Observe that $\bY^{(1)}$, $\br^{(1)}$ and $\kappa^{(1)}$ agree with $\bY$, $\br$ and $\kappa_n$, respectively. In what follows we refer to the collection of tensors $\{\bY,\bY^{(2)},...\}$ as the \textit{transverse} or \textit{asymptotic expansion}.\\

In the following lemma we recall a well-known identity for derivatives of products of any two objects $S$ and $T$. Note than when $S$ and $T$ are tensors, the expression also holds when contractions are allowed. 
\begin{lema}
	Let $S$ and $T$ be two objects, $S\circledast T$ any product of them and $\mc{D}$ any derivative operator. Then, 
	\begin{equation}
		\label{derivada}
		\mc{D}^{(m)}\big(S\circledast T\big) = \sum_{i=0}^m\binom{m}{i}\big( \mc{D}^{(i)} S\big)\circledast\big( \mc{D}^{(m-i)}T\big).
	\end{equation}
\end{lema}
Given any vector field $\xi$ we introduce the tensor $\Sigma[\xi]\d \lie_{\xi}\nabla$, or in abstract index notation
\begin{equation}
	\label{Sigma}
	\Sigma[\xi]^{\alpha}{}_{\mu\nu} =\dfrac{1}{2}g^{\alpha\beta}\left(\nabla_{\mu}\mc{K}[\xi]_{\nu\beta} + \nabla_{\nu}\mc{K}[\xi]_{\mu\beta} - \nabla_{\beta}\mc{K}[\xi]_{\mu\nu}\right),
\end{equation} 
where $\mc{K}[\xi]$ is the so-called \textit{deformation tensor} of $\xi$, defined by $\mc{K}[\xi]\d \lie_{\xi}g$. In order not to overload the notation in this section we will simply use the symbols $\Sigma$ and $\mc{K}$ for $\Sigma[\xi]$ and $\mc{K}[\xi]$, respectively. In later sections, we will come back to the notation $\mc{K}[\xi]$ because deformation tensors of more than one vector field will occur.\\

To compute $\lie_{\xi}^{(m)}\ric$ up to order $m+1$ we use the following classical result \cite{yano2020theory} relating the Lie derivative of the curvature with the Lie derivative of the connection
\begin{equation}
	\label{Yano}
	\lie_{\xi}R^{\mu}{}_{\alpha\nu\beta} = H^{\gamma\rho}_{\nu\beta}\nabla_{\gamma}\Sigma^{\mu}{}_{\alpha\rho},\qquad  H^{\gamma\rho}_{\nu\beta} \d \delta^{\gamma}_{\nu}\delta^{\rho}_{\beta}-\delta^{\gamma}_{\beta}\delta^{\rho}_{\nu}.
\end{equation}
For future convenience we define 
\begin{equation}
	\label{Sigmadown}
	\uwh\Sigma_{\nu\alpha\beta} \d g_{\mu\nu}\Sigma^{\mu}{}_{\alpha\beta}=\dfrac{1}{2}\left(\nabla_{\alpha}\mc{K}_{\beta\nu} + \nabla_{\beta}\mc{K}_{\alpha\nu} - \nabla_{\nu}\mc{K}_{\alpha\beta}\right) \eqqcolon F^{\rho\lambda\gamma}_{\nu\alpha\beta} \nabla_{\rho} \mc{K}_{\lambda\gamma},
\end{equation} 
where we have introduced the tensor $$F^{\rho\lambda\gamma}_{\nu\alpha\beta} \d \dfrac{1}{2}\left(\delta^{\rho}_{\alpha}\delta^{\lambda}_{\beta}\delta^{\gamma}_{\nu} + \delta^{\rho}_{\beta}\delta^{\lambda}_{\alpha}\delta^{\gamma}_{\nu}-\delta^{\rho}_{\nu}\delta^{\lambda}_{\alpha}\delta^{\gamma}_{\beta}\right).$$ The hat in $\uwh\Sigma_{\nu\alpha\beta}$ is not really necessary because $\uwh\Sigma_{\nu\alpha\beta}$ is just $\Sigma^{\nu}{}_{\alpha\beta}$ with the index lowered. However the distinction will be necessary later for the Lie derivative of $\uwh\Sigma_{\nu\alpha\beta}$ and $\Sigma^{\nu}{}_{\alpha\beta}$.
\begin{rmk}
	The notations $H^{\gamma\rho}_{\nu\beta}$ and $F^{\rho\lambda\gamma}_{\nu\alpha\beta}$ are unambiguous since we shall never lower/raise their indices. We stick to this rule for any tensor written with indices on top of each other.
\end{rmk}
The idea now is to apply the operator $\lie_{\xi}^{(m-1)}$ to \eqref{Yano} and express the result by making $\lie^{(m+1)}_{\xi}g$ and $\lie^{(m+2)}_{\xi}g$ explicit. In order to do that we need to commute $\lie_{\xi}^{(m-1)}$ and $\nabla$ when they act on a $(q,p)$ tensor $A^{\alpha_1\cdots\alpha_q}_{\beta_1\cdots\beta_p}$. We introduce the notation $A^{(m)}\d \lie_{\xi}^{(m-1)}A$, $m\ge 1$. The commutator is found explicitly in the following proposition.
\begin{prop}
	\label{propMarc}
	Let $\xi\in\X(\mc M)$ and $m\ge 1$ be a integer. Then, given any $(p,q)$ tensor $A^{\alpha_1\cdots\alpha_q}_{\beta_1\cdots\beta_p}$ the following identity holds
	\begin{equation*}
		\begin{aligned}
			\lie_{\xi}^{(m)}\nabla_{\gamma}A^{\alpha_1\cdots\alpha_q}_{\beta_1\cdots\beta_p} = \nabla_{\gamma} A^{(m+1)}{}^{\alpha_1\cdots\alpha_q}_{\beta_1\cdots\beta_p} + \sum_{k=0}^{m-1}\binom{m}{k+1}&\left(\sum_{j=1}^{q}A^{(m-k)}{}^{\alpha_1\cdots\alpha_{j-1}\sigma\alpha_{j+1}\cdots\alpha_q}_{\beta_1\cdots\beta_p}\Sigma^{(k+1)}{}^{\alpha_j}{}_{\sigma\gamma}\right. \\
			&\left.- \sum_{i=1}^{p}A^{(m-k)}{}^{\alpha_1\cdots\alpha_q}_{\beta_1\cdots\beta_{i-1}\sigma\beta_{i+1}\cdots\beta_p}\Sigma^{(k+1)}{}^{\sigma}{}_{\beta_i\gamma}\right).
		\end{aligned}
	\end{equation*}
	\begin{proof}
		The case $m=1$ is the classical identity \cite{yano2020theory} $$\lie_{\xi}\nabla_{\gamma}A^{\alpha_1\cdots\alpha_q}_{\beta_1\cdots\beta_p} = \nabla_{\gamma}\lie_{\xi}A^{\alpha_1\cdots\alpha_q}_{\beta_1\cdots\beta_p} + \sum_{j=1}^{q}A^{\alpha_1\cdots\alpha_{j-1}\sigma\alpha_{j+1}\cdots\alpha_q}_{\beta_1\cdots\beta_p}\Sigma^{\alpha_j}{}_{\sigma\gamma} - \sum_{i=1}^{p}A^{\alpha_1\cdots\alpha_q}_{\beta_1\cdots\beta_{i-1}\sigma\beta_{i+1}\cdots\beta_p}\Sigma^{\sigma}{}_{\beta_i\gamma}.$$ We prove the result by induction, so let us assume that the claim is true up to some $m \geq 1$ and show that it is then true for $m+1$ also. We compute, using the induction hypothesis,
		\begin{align*}
			& \lie_{\xi}^{(m+1)} \nabla_{\gamma} A^{\alpha_1\cdots\alpha_q}_{\beta_1\cdots \beta_p}  
			= \lie_{\xi} \left ( \lie^{(m)}_{\xi}
			\nabla_{\gamma} A^{\alpha_1\cdots\alpha_q}_{\beta_1\cdots \beta_p}   \right ) \\
			&  \quad = \lie_{\xi} \left(	\nabla_{\gamma} A^{(m+1)}{}^{\alpha_1\cdots\alpha_q}_{\beta_1\cdots \beta_p} 
			+ \sum_{k=0}^{m-1} \binom{m}{k+1} \left( \sum_{j=1}^q 
			A^{(m-k)}{}^{\alpha_1\cdots\sigma\alpha_q}_{\beta_1\cdots \beta_p} \Sigma^{(k+1)}{}^{\alpha_j}{}_{\sigma \gamma}\right.\right.\\
			&\hfill \hspace{7.5cm} \left.\left. - \sum_{i=1}^p  A^{(m-k)}{}^{\alpha_1\cdots\alpha_q}_{\beta_1\cdots\sigma\cdots \beta_p} \Sigma^{(k+1)}{}^{\sigma}{}_{\beta_i\gamma}\right) \right) \\
			&  \quad = \nabla_{\gamma} \lie_{\xi} A^{(m+1)}{}^{\alpha_1 \cdots \alpha_q}_{\beta_1 \cdots \beta_p}	+ \sum_{j=1}^q A^{(m+1)}{}^{\alpha_1 \cdots\sigma\cdots \alpha_q}_{\beta_1 \cdots \beta_p}\Sigma^{\alpha_j}{}_{\sigma\gamma}	- \sum_{i=1}^p A^{(m+1)}{}^{\alpha_1 \cdots \alpha_q}_{\beta_1\cdots\beta_{i-1}\sigma\beta_{i+1}\cdots\beta_p} \Sigma^{\sigma}{}_{\beta_i\gamma} \\
			& \quad \quad + \sum_{k=0}^{m-1} \binom{m}{k+1} \left ( \sum_{j=1}^q A^{(m-k+1)}{}^{\alpha_1 \cdots\sigma\cdots \alpha_q}_{\beta_1 \cdots \beta_p} \Sigma^{(k+1)}{}^{\alpha_j}{}_{\sigma\gamma} - \sum_{i=1}^p  A^{(m-k+1)}{}^{\alpha_1 \cdots \alpha_q}_{\beta_1\cdots\sigma\cdots\beta_p} \Sigma^{(k+1)}{}^{\sigma}{}_{\beta_i\gamma}	\right )\\
			& \quad \quad + \sum_{k=0}^{m-1} \binom{m}{k+1} \left ( \sum_{j=1}^q A^{(m-k)}{}^{\alpha_1 \cdots\sigma\cdots \alpha_q}_{\beta_1 \cdots \beta_p} \Sigma^{(k+2)}{}^{\alpha_j}{}_{\sigma \gamma}- \sum_{i=1}^p  A^{(m-k)}{}^{\alpha_1 \cdots \alpha_q}_{\beta_1\cdots\sigma\cdots\beta_p} \Sigma^{(k+2)}{}^{\sigma}{}_{\beta_i\gamma}	\right ).
		\end{align*}
		In the last term we rename $k$ as $k-1$ and split the second sum in two parts and the last sum also in two parts,
		\begin{align*}
			& \lie_{\xi}^{(m+1)} \nabla_{\gamma}A^{\alpha_1\cdots\alpha_q}_{\beta_1\cdots \beta_p}    = 	\nabla_{\gamma} \lie_{\xi} A^{(m+1)}{}^{\alpha_1\cdots\alpha_q}_{\beta_1\cdots \beta_p}
			+ \left (1 + \binom{m}{1} \right )	\left ( \sum_{j=1}^q A^{(m+1)}{}^{\alpha_1\cdots\alpha_q}_{\beta_1 \cdots \beta_p}\Sigma^{\alpha_j}{}_{\sigma\gamma} - \sum_{i=1}^p A^{(m+1)}{}^{\alpha_1 \cdots \alpha_q}_{\beta_1\cdots\beta_p} \Sigma^{\sigma}{}_{\beta_i\gamma} 	\right ) \\
			& \quad \quad + \sum_{k=1}^{m-1} \left ( \binom{m}{k+1} + \binom{m}{k} \right )\left ( 	\sum_{j=1}^q A^{(m-k+1)}{}^{\alpha_1\cdots\alpha_q}_{\beta_1 \cdots \beta_p} \Sigma^{(k+1)}{}^{\alpha_j}{}_{\sigma \gamma}
			- \sum_{i=1}^p  A^{(m-k+1)}{}^{\alpha_1\cdots\alpha_q}_{\beta_1 \cdots \beta_p} \Sigma^{(k+1)}{}^{\sigma}{}_{\beta_i\gamma}\right )  \\
			& \quad \quad + \binom{m}{m} \left( \sum_{j=1}^q A^{\alpha_1\cdots\alpha_q}_{\beta_1 \cdots \beta_p} \Sigma^{(m+1)}{}^{\alpha_j}_{\sigma \gamma}- \sum_{i=1}^p  A^{\alpha_1\cdots\alpha_q}_{\beta_1 \cdots \beta_p} \Sigma^{(m+1)}{}^{\sigma}{}_{\beta_i\gamma}\right) .
		\end{align*}
		Using the binomial identity $\binom{m}{i+1} + \binom{m}{i} = \binom{m+1}{i+1}$, the proposition follows by induction.
	\end{proof}
\end{prop}
Once we know how to commute $\lie_{\xi}^{(m-1)}$ and $\nabla$ when acting on an arbitrary tensor field, we can apply the result to equation \eqref{Yano} to compute the explicit expression of the tensor $\lie_{\xi}^{(m)} R^{\mu}{}_{\alpha\nu\beta}$. This will be used below to compute the derivatives $\lie_{\xi}^{(m)}\ric$ on a null hypersurface $\mc H$ in terms of transverse derivatives of $g$ on $\mc H$ up to order $m+1$.
\begin{prop}
	Let $\xi\in\X(\mc M)$ and $m\ge 2$ an integer. Then,
	\begin{equation}
		\label{liemriemann}
		\lie_{\xi}^{(m)} R^{\mu}{}_{\alpha\nu\beta} = H^{\gamma\rho}_{\nu\beta}\left(\nabla_{\gamma}\Sigma^{(m)}{}^{\mu}{}_{\alpha\rho}+ \sum_{k=0}^{m-2}\binom{m}{k+1}\Sigma^{(m-k-1)}{}^{\sigma}{}_{\alpha\rho}\Sigma^{(k+1)}{}^{\mu}{}_{\sigma\gamma}\right)
	\end{equation}
and
\begin{equation}
	\label{liemricci}
	\lie_{\xi}^{(m)} R_{\alpha\beta} = H^{\gamma\rho}_{\mu\beta}\left(\nabla_{\gamma}\Sigma^{(m)}{}^{\mu}{}_{\alpha\rho}+ \sum_{k=0}^{m-2}\binom{m}{k+1}\Sigma^{(m-k-1)}{}^{\sigma}{}_{\alpha\rho}\Sigma^{(k+1)}{}^{\mu}{}_{\sigma\gamma}\right).
\end{equation}
	\begin{proof}
		Applying $\lie_{\xi}^{(m-1)}$ to \eqref{Yano} and using Proposition \ref{propMarc},
		\begin{align*}
			\lie_{\xi}^{(m)} R^{\mu}{}_{\alpha\nu\beta} = H^{\gamma\rho}_{\nu\beta}\Big(\nabla_{\gamma}\Sigma^{(m)}{}^{\mu}{}_{\alpha\rho}+\sum_{k=0}^{m-2}\binom{m-1}{k+1}\Big(&\Sigma^{(m-k-1)}{}^{\sigma}{}_{\alpha\rho}\Sigma^{(k+1)}{}^{\mu}{}_{\sigma\gamma} - \Sigma^{(m-k-1)}{}^{\mu}{}_{\sigma\alpha}\Sigma^{(k+1)}{}^{\sigma}{}_{\rho\gamma}\\
			& -\Sigma^{(m-k-1)}{}^{\mu}{}_{\rho\sigma}\Sigma^{(k+1)}{}^{\sigma}{}_{\alpha\gamma}\Big)\Big).
		\end{align*}
		Since $H^{\gamma\rho}_{\nu\beta}$ is antisymmetric in $\gamma,\rho$ its contraction with the third term vanishes. Renaming $k' = m-2-k$ in the last term the sum simplifies to 
		\begin{align*}
			&H^{\gamma\rho}_{\nu\beta}\sum_{k=0}^{m-2}\binom{m-1}{k+1}\Big(\Sigma^{(m-k-1)}{}^{\sigma}{}_{\rho\alpha}\Sigma^{(k+1)}{}^{\mu}{}_{\sigma\gamma} -\Sigma^{(m-k-1)}{}^{\mu}{}_{\rho\sigma}\Sigma^{(k+1)}{}^{\sigma}{}_{\alpha\gamma}\Big)\\
			&=H^{\gamma\rho}_{\nu\beta}\left(\sum_{k=0}^{m-2}\binom{m-1}{k+1}\Sigma^{(m-k-1)}{}^{\sigma}{}_{\rho\alpha}\Sigma^{(k+1)}{}^{\mu}{}_{\sigma\gamma} -\sum_{k'=0}^{m-2} \binom{m-1}{m-1-k'}\Sigma^{(k'+1)}{}^{\mu}{}_{\rho\sigma}\Sigma^{(m-k'-1)}{}^{\sigma}{}_{\alpha\gamma}\right).
		\end{align*}
		From the antisymmetry of $H^{\gamma\rho}_{\nu\beta}$, the symmetry of $\Sigma^{(k+1)}{}^{\mu}{}_{\rho\sigma}$ and the combinatorial properties $\binom{m-1}{m-1-k'}=\binom{m}{k'}$ and $\binom{m-1}{k+1}+\binom{m-1}{k}=\binom{m}{k+1}$, \eqref{liemriemann} follows. Equation \eqref{liemricci} is immediate since the Lie derivative and the trace commute.
	\end{proof}
\end{prop}

Identity \eqref{liemricci} constitutes the exact relation between the $m$-th Lie derivative of the Ricci tensor with the Lie derivatives of the tensor $\Sigma$. Before restricting it to a null hypersurface $\mc H$ with rigging $\xi$ and computing the leading order terms, we shall establish a property that will play a key role in Section \ref{sec_KH}, namely that when the rigging $\xi$ is extended off $\Phi(\mc H)$ by $\nabla_{\xi}\xi=0$, the tensors $\lie_{\xi}^{(m)}R_{\alpha\beta}$ on $\mc H$ are geometrical in the following sense.

\begin{defi}
	\label{defi_geometrical}
Let $\{\mc H,\bg,\bm\ell,\elltwo\}$ be null metric hypersurface data $(\Phi,\xi)$-embedded in $(\mc M,g)$. Let $\mc{T}$ and $T$ be $(0,p)$ tensor fields on $\mc H$ and $\mc M$, respectively.
\begin{itemize}
	\item We say that $\mc T$ is $\mc H$-geometrical provided that it depends at most on null metric data $\{\bg,\bm\ell,\elltwo\}$ and on the tensors $\{\bY^{(k)}\}_{k\ge 1}$.
	\item We say that $T$ is geometrical provided that the pullback of arbitrary contractions of $T$ with $\xi$ (including no contraction) into $\mc H$ is $\mc H$-geometrical.
\end{itemize}
\end{defi}

In order to prove that $\lie_{\xi}^{(m)} R_{\alpha\beta}$ is geometrical for any natural number $m$ we first explore some general properties of geometrical objects and also establish that several building-block tensors that appear in the argument are indeed geometrical. We start with the latter.

\begin{lema}
	\label{lemanablaxi}
	Let $\{\mc H,\bg,\bm\ell,\elltwo\}$ be null metric hypersurface data $(\Phi,\xi)$-embedded in $(\mc M,g)$ and extend $\xi$ off $\Phi(\H)$ by $\nabla_{\xi}\xi = 0$. Then the tensor $\nabla_{\alpha}\xi_{\beta}$ is geometrical. 
\begin{proof}
Since $\xi^{\alpha}\nabla_{\alpha}\xi_{\beta}=0$ it suffices to check that $\wh e_a^{\alpha} \wh e_b^{\beta}\nabla_{\alpha}\xi_{\beta}$ and $\wh e_a^{\alpha} \xi^{\beta}\nabla_{\alpha}\xi_{\beta}$ only depend on hypersurface data. Using $\wh{e}_a^{\mu}\nabla_{\mu}\xi^{\beta}\st{\mc H}{=} (\r-\s)_a\xi^{\beta} + P^{cd}\Pi_{ac}\wh{e}^{\beta}_d + \dfrac{1}{2}\nu^{\beta}\nablacero_a\elltwo$ (cf. \eqref{nablaxi}) and Definition \ref{defi_embedded} a straightforward computation shows that $$\wh e_a^{\alpha} \wh e_b^{\beta}\nabla_{\alpha}\xi_{\beta} = \Pi_{ab} ,\qquad \wh e_a^{\alpha} \xi^{\beta}\nabla_{\alpha}\xi_{\beta} = \dfrac{1}{2}\nablacero_a\elltwo.$$ Hence, $\nabla_{\alpha}\xi_{\beta}$ is geometrical. 
\end{proof}
\end{lema}

Next we show that when $\nabla_{\xi}\xi = 0$ the tensor $\mc{K}^{(m)}_{\mu\nu}$ is geometrical for all $m\ge 0$. As a preliminary step we first compute in full generality the contraction $\xi^{\alpha}\mc{K}^{(m)}_{\alpha\beta}$ on $\mc H$.
\begin{prop}
	\label{liegxi}
	Let $\mc H$ be a null hypersurface $(\Phi,\xi)$-embedded in $(\mc M,g)$ and let $Z$ be a vector field along $\Phi(\mc H)$. Extend $\xi$ arbitrarily off $\Phi(\mc H)$ and define $a_{\xi}\d\nabla_{\xi}\xi$. Then for every $m\in\mathbb{N}\cup\{0\}$,
	\begin{align}
		\mc{K}^{(m+1)}(\xi,\xi) &\st{\mc H}{=} 2\sum_{i=0}^m\binom{m}{i}\big(\mc{K}^{(m-i)}\big) \big(\lie_{\xi}^{(i)} a_{\xi},\xi\big),\label{liegxixi}\\
		\mc{K}^{(m+1)}(\xi,Z) & \st{\mc H}{=}  \dfrac{1}{2}Z\left(\mc{K}^{(m)}(\xi,\xi)\right) + \sum_{i=0}^m\binom{m}{i}\big(\mc{K}^{(m-i)}\big) \big(\lie_{\xi}^{(i)} a_{\xi},Z\big).\label{liegxiX}
	\end{align}
	\begin{proof}
		We first prove \eqref{liegxiX} and then we show that \eqref{liegxixi} follows from \eqref{liegxiX}. Let us extend $Z$ off $\mc H$ by $\lie_{\xi}Z=0$ (at the end we prove that the result is independent of the extension). First, $$\big(\lie_{\xi}g\big)(\xi,Z) = g(\nabla_{\xi}\xi,Z)+g(\xi,\nabla_Z\xi) = g(a_{\xi},Z) + \dfrac{1}{2}Z\left(g(\xi,\xi)\right).$$ Applying $\lie_{\xi}^{(m)}$ to both sides, $$\big(\lie_{\xi}^{(m+1)}g\big)(\xi,Z) =  \sum_{i=0}^m\binom{m}{i}\big(\lie_{\xi}^{(m-i)}g\big) \big(\lie_{\xi}^{(i)} a_{\xi},Z\big) + \dfrac{1}{2}\xi^{(m)}\left(Z\left(g(\xi,\xi)\right)\right),$$ which becomes \eqref{liegxiX} after commuting $\xi^{(m)}Z=Z\xi^{(m)}$ (since $[\xi,Z]=0$). Obviously equation \eqref{liegxiX} is independent of the extension of $Z$ off $\mc H$. To show \eqref{liegxixi} just use \eqref{liegxiX} with $Z=\xi$.
	\end{proof}
\end{prop}
\begin{cor}
	\label{corkxi}
Let $\{\mc H,\bg,\bm\ell,\elltwo\}$ be null metric hypersurface data $(\Phi,\xi)$-embedded in $(\mc M,g)$ and extend $\xi$ off $\Phi(\mc H)$ by $\nabla_{\xi}\xi=0$. Then $\mc{K}^{(m)}(\xi,\cdot)=0$ for all $m\ge 1$.
\end{cor}

\begin{lema}
\label{cor0}
Let $\{\mc H,\bg,\bm\ell,\elltwo\}$ be null metric hypersurface data $(\Phi,\xi)$-embedded in $(\mc M,g)$ and extend $\xi$ off $\Phi(\H)$ by $\nabla_{\xi}\xi = 0$. Then the tensor $\mc{K}^{(m)}_{\mu\nu}$ is geometrical for all $m\ge 0$.
\begin{proof}
The tensor $\mc{K}^{(0)}_{\mu\nu}\d g_{\mu\nu}$ is clearly geometrical by the definition of embedded metric hypersurface data. By Corollary \ref{corkxi} and recalling $\Phi^{\star}\mc{K}^{(m)}=2\bY^{(m)}$, the tensor $\mc{K}^{(m)}_{\mu\nu}$ for $m\ge 1$ is geometrical as well.
\end{proof}
\end{lema}

\begin{lema}
	\label{lemageo}
	Let $\{\mc H,\bg,\bm\ell,\elltwo\}$ be null metric hypersurface data $(\Phi,\xi)$-embedded in $(\mc M,g)$ and extend $\xi$ off $\Phi(\H)$ by $\nabla_{\xi}\xi = 0$. Let $T$ and $S$ be any two geometrical objects. Then, the following properties are true.
	\begin{enumerate}
		\item $T \otimes S$ is geometrical.
		\item Any trace of $T$ w.r.t $g$ is geometrical.
		\item Any trace of $T$ w.r.t $\lie_{\xi}^{(m)}g^{\mu\nu}$ is geometrical for any $m\ge 1$.
	\end{enumerate} 
	As a consequence, any trace of $T\otimes S$ w.r.t $g$ or $\lie_{\xi}^{(m)}g^{\mu\nu}$ is also geometrical.
	\begin{proof}
		The first property is obvious. The second one follows from $g^{\alpha\beta}\st{\mc H}{=}	P^{ab}\wh e_a^{\alpha}\wh e_b^{\beta} + n^{a}\wh e_a^{\alpha}\xi^{\beta} + n^{b}\wh e_b^{\beta}\xi^{\alpha}$ (cf. \eqref{inversemetric}). We prove the third property by induction. First observe that for $m\ge 1$
		\begin{equation}
			\label{liemg}
			\lie_{\xi}^{(m)}\big(g^{\alpha\mu} g_{\mu\beta}\big) = 0 \qquad \Longrightarrow\qquad \lie_{\xi}^{(m)}g^{\alpha\beta} = - g^{\beta\nu}\sum_{i=1}^{m}\binom{m}{i}(\lie_{\xi}^{(m-i)}g^{\alpha\mu})\lie_{\xi}^{(i)}g_{\mu\nu}.
		\end{equation} 
		Particularizing for $m=1$ gives $\lie_{\xi}g^{\alpha\beta} = - g^{\beta\nu}g^{\alpha\nu}\mc{K}_{\mu\nu}$. By Lemma \ref{cor0} and items 1. and 2. of this lemma, item 3. follows for $m=1$. Assume it is true up to some integer $m-1$. Expression \eqref{liemg} together with Lemma \ref{cor0} then show that it is also true for $m$. 
	\end{proof}
\end{lema}

\begin{lema}
	\label{cor}
	Let $\{\mc H,\bg,\bm\ell,\elltwo\}$ be null metric hypersurface data $(\Phi,\xi)$-embedded in $(\mc M,g)$ and extend $\xi$ off $\Phi(\H)$ by $\nabla_{\xi}\xi = 0$. Then the tensor $\nabla_{\rho}\mc{K}^{(m)}_{\mu\nu}$ is geometrical for all $m\ge 0$.
	\begin{proof}
The argument relies on the following well-known relation between $\nabla$ and $\lie_{\xi}$ acting on a (0,2) symmetric tensor $T$ 
\begin{equation}
	\label{lieandnabla}
	\nabla_{\xi}T_{\lambda\gamma} = \lie_{\xi}T_{\lambda\gamma} - 2 T_{\mu(\lambda}\nabla_{\gamma)}\xi^{\mu}.
\end{equation}	
Particularizing to $T=\mc{K}^{(m)}$ gives $$\xi^{\rho}\nabla_{\rho}\mc{K}^{(m)}_{\mu\nu} = \big(\lie_{\xi}\mc{K}^{(m)}\big)_{\mu\nu} -2 \mc{K}^{(m)}_{\rho(\mu}\nabla_{\nu)}\xi^{\rho} = \mc{K}^{(m+1)}_{\mu\nu} -2 g^{\rho\eps}\mc{K}^{(m)}_{\rho(\mu}\nabla_{\nu)}\xi_{\eps}.$$ The first term is geometrical by Lemma \ref{cor0}, and the second one is also geometrical as a consequence of Lemmas \ref{lemanablaxi}, \ref{cor0} and \ref{lemageo}.
	\end{proof}
\end{lema}

Next we show that the tensor $\uwh\Sigma^{(m)}$ is geometrical.
\begin{prop}
	\label{progeo}
	Let $\{\H,\bg,\bm\ell,\elltwo\}$ be null metric hypersurface data $(\Phi,\xi)$-embedded in $(\mc M,g)$ and extend $\xi$ off $\Phi(\H)$ by $\nabla_{\xi}\xi = 0$. Then, the tensors $\uwh{\Sigma}^{(m)}{}_{\nu\alpha\beta}$ and $\Sigma^{(m)}{}_{\nu\alpha\beta}$ for all $m\ge 1$ are geometrical.
	\begin{proof}
Recalling definition \eqref{Sigmadown}, namely $\uwh\Sigma_{\nu\alpha\beta} \d g_{\nu\mu}\Sigma^{\mu}{}_{\alpha\beta}$, and applying identity \eqref{derivada} gives
	\begin{equation}
		\label{Sigmas4}
		\Sigma^{(m)}{}_{\nu\alpha\beta} = g_{\mu\nu}\Sigma^{(m)}{}^{\mu}{}_{\alpha\beta} = g_{\mu\nu}\lie_{\xi}^{(m-1)}\big(g^{\mu\rho}\uwh\Sigma_{\rho\alpha\beta}\big) = g_{\mu\nu}\sum_{i=0}^{m-1}\binom{m-1}{i}(\lie^{(i)}_{\xi}g^{\mu\rho})\uwh\Sigma^{(m-i)}{}_{\rho\alpha\beta}.
	\end{equation} 
So, by Lemma \ref{lemageo} if $\uwh\Sigma^{(k)}{}_{\nu\alpha\beta}$ is geometrical for all $k\le m$, then $\Sigma^{(m)}{}_{\nu\alpha\beta}$ is geometrical as well. It only remains to prove that $\uwh\Sigma^{(m)}{}_{\nu\alpha\beta}$ is geometrical for all $m\geq 1$. We establish this by an induction argument. From $\uwh\Sigma_{\nu\alpha\beta} = F^{\rho\lambda\gamma}_{\nu\alpha\beta}\nabla_{\rho}\mc{K}_{\lambda\gamma}$ and Lemma \ref{cor} it follows that the tensor $\uwh\Sigma_{\nu\alpha\beta}$ is geometrical. Assume $\uwh{\Sigma}^{(k)}{}_{\nu\alpha\beta}$ (and hence also ${\Sigma}^{(k)}{}_{\nu\alpha\beta}$) is geometrical for all $k\le m$. Applying Proposition \ref{propMarc} to $A=\mc{K}$ and taking into account $\lie_{\xi}F=0$ it follows 
	\begin{align*}
\uwh{\Sigma}^{(m+1)}_{\nu\alpha\beta} = \lie^{(m)}_{\xi}\uwh\Sigma_{\nu\alpha\beta} & = F^{\rho\gamma\lambda}_{\nu\alpha\beta} \nabla_{\rho}\mc{K}^{(m+1)}_{\lambda\gamma} - 2\sum_{k=0}^{m-1}\binom{m}{k+1}F^{\rho\gamma\lambda}_{\nu\alpha\beta} \mc{K}^{(m-k)}_{\sigma(\lambda}\Sigma^{(k+1)}{}^{\sigma}{}_{\gamma)\rho}\\
&= F^{\rho\gamma\lambda}_{\nu\alpha\beta} \nabla_{\rho}\mc{K}^{(m+1)}_{\lambda\gamma}- 2\sum_{k=0}^{m-1}\binom{m}{k+1} F^{\rho\gamma\lambda}_{\nu\alpha\beta} g^{\sigma\eps} \mc{K}^{(m-k)}_{\sigma(\lambda} \Sigma^{(k+1)}{}_{\eps|\gamma)\rho}.
	\end{align*}
Using Lemmas \ref{lemageo} and \ref{cor} and the fact that every ${\Sigma}^{(k)}{}_{\nu\alpha\beta}$ for $k\le m$ is geometrical, we conclude that $\uwh{\Sigma}^{(m+1)}_{\nu\alpha\beta}$ (and thus $\Sigma^{(m+1)}_{\nu\alpha\beta}$) is geometrical as well. 
	\end{proof}
\end{prop}

That $R_{\alpha\beta}$ is geometrical when $\nabla_{\xi}\xi=0$ is immediate from \eqref{constraint}, \eqref{ricxixi} and \eqref{ricxiX} because in this case $\bZ^{(2)}= \bY^{(2)}$ (by Definition \ref{defiZ2}). In order to prove that $\lie_{\xi}^{(m)} R_{\alpha\beta}$ is geometrical for every $m\ge 1$ when $\nabla_{\xi}\xi=0$ it is convenient to first rewrite \eqref{liemricci} making the tensor $H$ explicit, namely 
\begin{equation}
	\label{liemricci2}
	\begin{aligned}
		\lie_{\xi}^{(m)} R_{\alpha\beta} &= \nabla_{\mu}\Sigma^{(m)}{}^{\mu}{}_{\alpha\beta} - \nabla_{\beta}\Sigma^{(m)}{}^{\mu}{}_{\alpha\mu} + \sum_{k=0}^{m-2}\binom{m}{k+1}g^{\sigma\rho}g^{\mu\nu}\Sigma^{(m-k-1)}{}_{\rho\alpha\beta}\Sigma^{(k+1)}{}_{\nu\sigma\mu}\\
		&\qquad\, - \sum_{k=0}^{m-2}\binom{m}{k+1}g^{\sigma\rho}g^{\mu\nu}\Sigma^{(m-k-1)}{}_{\rho\alpha\mu}\Sigma^{(k+1)}{}_{\nu\sigma\beta}.
	\end{aligned}
\end{equation}
By Lemma \ref{lemageo} and Proposition \ref{progeo} all the terms in the two sums are geometrical, so it suffices to show that $\nabla_{\mu}\Sigma^{(m)}{}^{\mu}{}_{\alpha\beta}$ and $\nabla_{\beta}\Sigma^{(m)}{}^{\mu}{}_{\alpha\mu}$ are also geometrical. In order to do that we shall contract both tensors with two tangent vectors $\wh e_a^{\alpha}\wh e_b^{\beta}$, one tangent and one transverse $\wh e_a^{\alpha}\xi^{\beta}$, and two transverse vectors $\xi^{\alpha}\xi^{\beta}$ and check that in all cases the result only depends on metric data and the expansion $\{\bY^{(k)}\}$. These computations rely on general expressions for the pullback of ambient tensor fields into an arbitrary null hypersurface $\H$. These are computed in full generality in Appendix \ref{appendix}. The following notation is used.
\begin{nota}
	\label{notationnota}
Given a $(0,p)$ tensor field $T_{\alpha_1\cdots\alpha_p}$ on $\mc M$ we use the standard notation $T_{a_1\cdots a_p}$ to denote its pullback to $\mc H$. Moreover, we introduce the notation ${}^{(i)}T_{\alpha_1\cdots \alpha_{p-1}}$ for $\xi^{\mu}T_{\alpha_1\cdots \alpha_{i-1}\mu\alpha_i\cdots \alpha_{p-1}}$ and ${}^{(i)}T_{a_1\cdots a_{p-1}}$ for the pullback of ${}^{(i)}T_{\alpha_1\cdots \alpha_{p-1}}$ to $\mc H$. In addition, we use ${}^{(i,j)}T_{a_1\cdots a_{p-2}}$ for the pullback to $\mc H$ of the tensor obtained by first the contraction of $T$ with $\xi$ in the j-th slot and then in the i-th slot of the resulting $(0,p-1)$ tensor, i.e. ${}^{(i,j)}T={}^{(i)}\big({}^{(j)}T\big)$. This notation requires care with the order of indices (note that ${}^{(1,1)}T={}^{(1,2)}T$ and ${}^{(1,3)}T={}^{(2,1)}T\neq {}^{(1,2)}T$).
\end{nota}
Let start by analyzing $\nabla_{\mu}\Sigma^{(m)}{}^{\mu}{}_{\alpha\beta}$. Firstly, from Proposition \ref{propdivergencia} and the fact that $\Sigma^{(m)}{}_{\nu\alpha\beta}$ is geometrical, it follows that $$\wh e_a^{\alpha} \wh e_b^{\beta} \nabla_{\mu}\Sigma^{(m)}{}^{\mu}{}_{\alpha\beta} = n^c \big(\lie_{\xi} \Sigma^{(m)}\big)_{cab} + \mc{H}\text{-geometrical terms}  = n^c\Sigma^{(m+1)}{}_{cab} + \mc{H}\text{-geo. terms},$$ and hence by Proposition \ref{progeo} $\wh e_a^{\alpha} \wh e_b^{\beta} \nabla_{\mu}\Sigma^{(m)}{}^{\mu}{}_{\alpha\beta}$ only depends on metric data and $\{\bY^{(k)}\}$. Secondly, its contraction with $\xi^{\alpha}\wh e_b^{\beta}$ yields $$\xi^{\alpha}\wh e_b^{\beta}\nabla_{\mu}\Sigma^{(m)}{}^{\mu}{}_{\alpha\beta} = g^{\mu\nu}\wh e_b^{\beta} \nabla_{\mu} \big({}^{(2)}\Sigma^{(m)}\big)_{\nu\beta} - g^{\mu\nu}g^{\rho\alpha}\wh e_b^{\beta}\Sigma^{(m)}{}_{\nu\alpha\beta}\nabla_{\mu}\xi_{\rho}.$$ By Lemma \ref{lemageo} and Proposition \ref{progeo} the second term is $\mc H$-geometrical. Moreover, by Proposition \ref{propdivergencia}, $$g^{\mu\nu}\wh e_b^{\beta} \nabla_{\mu} \big({}^{(2)}\Sigma^{(m)}\big)_{\nu\beta} = n^c\big(\lie_{\xi}\big({}^{(2)}\Sigma^{(m)}\big)\big)_{cb} + \mc{H}\text{-geo. terms} =  \big({}^{(2)}\Sigma^{(m+1)}\big)_{cb}\, n^c + \mc{H}\text{-geo. terms},$$ where we used $\lie_{\xi}\big({}^{(i)}T^{(m)}\big) = {}^{(i)}T^{(m+1)}$ since obviously $\lie_{\xi}\xi=0$. Again Prop. \ref{progeo} shows that $\xi^{\alpha}\wh e_b^{\beta}\nabla_{\mu}\Sigma^{(m)}{}^{\mu}{}_{\alpha\beta}$ (and by symmetry also $\xi^{\beta}\wh e_b^{\alpha}\nabla_{\mu}\Sigma^{(m)}{}^{\mu}{}_{\alpha\beta}$) only depends on metric data and $\{\bY^{(k)}\}$. Finally, the contraction with $\xi^{\alpha}\xi^{\beta}$ is $$\xi^{\alpha}\xi^{\beta}\nabla_{\mu}\Sigma^{(m)}{}^{\mu}{}_{\alpha\beta} = g^{\mu\nu}\nabla_{\mu} \big({}^{(2,3)}\Sigma^{(m)}\big)_{\nu} - g^{\mu\nu} g^{\beta\rho} \big({}^{(2)}\Sigma^{(m)}\big)_{\nu\beta}\nabla_{\mu}\xi_{\rho}- g^{\mu\nu}g^{\rho\alpha}\big( {}^{(3)}\Sigma^{(m)}\big)_{\nu\alpha}\nabla_{\mu}\xi_{\rho}.$$ By Prop. \ref{propdivergencia} the first term is $$g^{\mu\nu}\nabla_{\mu} \big({}^{(2,3)}\Sigma^{(m)}\big)_{\nu} =  \big({}^{(2,3)}\Sigma^{(m+1)}\big)_{c}n^c + \mc{H}\text{-geo. terms},$$ so by Prop. \ref{progeo} it is $\mc H$-geometrical. The second and third terms are $\mc H$-geometrical as well by Lemmas \ref{lemanablaxi} and \ref{lemageo} and Prop. \ref{progeo}. Hence $\nabla_{\mu}\Sigma^{(m)}{}^{\mu}{}_{\alpha\beta}$ is geometrical. For the tensor $\nabla_{\beta}\Sigma^{(m)}{}^{\mu}{}_{\alpha\mu}$ we introduce $T^{(m)}{}_{\alpha}\d \Sigma^{(m)}{}^{\mu}{}_{\alpha\mu}$, which is geometrical by Lemma \ref{lemageo}. The contractions $\wh e_a^{\alpha}\wh e_b^{\beta}\nabla_{\beta}T_{\alpha}$, $\wh e_a^{\alpha}\xi^{\beta}\nabla_{\beta}T_{\alpha}$ and $\xi^{\alpha}\wh e_b^{\beta}\nabla_{\beta}T_{\alpha}$ are automatically $\mc H$-geometrical after using identities \eqref{identity1}-\eqref{identity3} in Appendix \ref{appendix} and $\lie_{\xi}T^{(m)} = T^{(m+1)}$. Finally, $\xi^{\alpha}\xi^{\beta}\nabla_{\beta}T^{(m)}{}_{\alpha} = \lie_{\xi}\big({}^{(1)}T^{(m)}\big) = {}^{(1)}T^{(m+1)}$ is also $\H$-geometrical, and hence $\nabla_{\beta}\Sigma^{(m)}{}^{\mu}{}_{\alpha\mu}$ is geometrical. Thus, the following result has been proved.
\begin{prop}
	\label{universal0}
	Let $\{\mc H,\bg,\bm\ell,\elltwo\}$ be null metric hypersurface data $(\Phi,\xi)$-embedded in $(\mc M,g)$ and extend $\xi$ off $\Phi(\H)$ by $\nabla_{\xi}\xi = 0$. Then, the tensor $\lie_{\xi}^{(m)} R_{\alpha\beta}$ is geometrical for every $m\ge 0$.
\end{prop}

\subsection{Analysis of the leading order terms}

In this subsection we compute the leading order terms of the tensor $\lie^{(m)}R_{\alpha\beta}$ on an arbitrary null hypersurface $\H$. As we shall see, it turns out that the leading order terms of $\xi^{\alpha}\lie^{(m)}R_{\alpha\beta}$ involve $m+2$ transverse derivatives of the metric, whereas the fully tangential components of $\lie^{(m)}R_{\alpha\beta}$ depend at most on $m+1$ transverse derivatives. In order to write down equalities only to the leading order it is useful introduce the following notation.
\begin{nota}
	\label{nota1}
	Let $(\mc M,g)$ be a semi-Riemannian manifold and $T,S$ be two tensor fields involving $g$ and its derivatives. The notation $T \st{[m]}{=} S$ means that the tensor $T-S$ does not depend on derivatives of $g$ of order $m$ or higher.
\end{nota}
The only terms that have a chance to carry $m+1$ and $m+2$ derivatives of the metric in identity \eqref{liemricci2} are the ones of the form $\nabla_{\gamma}\Sigma^{(m)}{}^{\mu}{}_{\alpha\beta}$. Thus, with the notation above, 
\begin{equation}
	\label{liemricmunu}
	\lie_{\xi}^{(m)} R_{\alpha\beta} \st{[m+1]}{=} \nabla_{\mu}\Sigma^{(m)}{}^{\mu}{}_{\alpha\beta} - \nabla_{\beta}\Sigma^{(m)}{}^{\mu}{}_{\mu\alpha},\qquad m\ge 1.
\end{equation}
Observe that equation \eqref{derivada} together with $\lie_{\xi}g^{\mu\nu}=-\mc{K}^{\mu\nu}$ implies 
\begin{equation}
	\label{Sigmas}
	\Sigma^{(m)}{}^{\mu}{}_{\alpha\beta} = \lie^{(m-1)}_{\xi}\left(g^{\mu\nu}\uwh\Sigma_{\nu\alpha\beta}\right) \st{[m]}{=} g^{\mu\nu}\uwh\Sigma^{(m)}_{\nu\alpha\beta} - (m-1)\mc{K}^{\mu\nu}\uwh\Sigma_{\nu\alpha\beta}^{(m-1)}.
\end{equation}
Lowering the index $\mu$ gives
\begin{equation}
	\label{Sigmas2}
	\Sigma^{(m)}_{\mu\alpha\beta}  \st{[m]}{=} \uwh\Sigma^{(m)}_{\mu\alpha\beta} - (m-1)g^{\nu\rho}\mc{K}_{\mu\nu}\uwh\Sigma_{\rho\alpha\beta}^{(m-1)}, \qquad \Sigma^{(m)}_{\mu\alpha\beta}  \st{[m+1]}{=} \uwh\Sigma^{(m)}_{\mu\alpha\beta},
\end{equation}
and applying $\lie_{\xi}$,
\begin{equation}
	\label{Sigmas3}
	\lie_{\xi}\Sigma^{(m)}_{\mu\alpha\beta}  \st{[m+1]}{=} \uwh\Sigma^{(m+1)}_{\mu\alpha\beta} - (m-1)g^{\nu\rho}\mc{K}_{\mu\nu}\uwh\Sigma_{\rho\alpha\beta}^{(m)},\qquad \lie_{\xi}\Sigma^{(m)}_{\mu\alpha\beta}  \st{[m+2]}{=} \uwh\Sigma^{(m+1)}_{\mu\alpha\beta}.
\end{equation}
\begin{lema}
	Let $\xi\in\X(\mc M)$ and $m\ge 1$ an integer. Then,
	\begin{align}
		\uwh\Sigma_{\nu\alpha\beta}^{(m)} &\st{[m]}{=} F_{\nu\alpha\beta}^{\rho\lambda\gamma}\nabla_{\rho}\mc{K}^{(m)}_{\lambda\gamma} - g^{\sigma\eps} \mc{K}_{\sigma\nu} F_{\eps\alpha\beta}^{\rho\lambda\gamma}\nabla_{\rho}\mc{K}^{(m-1)}_{\lambda\gamma},\label{Sigmam2}\\
		\Sigma^{(m)}{}^{\mu}{}_{\alpha\beta} &\st{[m]}{=} g^{\mu\nu}F^{\rho\lambda\gamma}_{\nu\alpha\beta}\nabla_{\rho}\mc{K}^{(m)}_{\lambda\gamma} - m \mc{K}^{\mu\nu} F^{\rho\lambda\gamma}_{\nu\alpha\beta}\nabla_{\rho}\mc{K}^{(m-1)}_{\lambda\gamma},\label{Sigmam}
	\end{align}
	\begin{proof}
		By Proposition \ref{propMarc} and \eqref{Sigmadown}, together with $\lie_{\xi}F_{\nu\alpha\beta}^{\rho\lambda\gamma}=0$, it follows
		\begin{align*}
			\lie^{(m-1)}_{\xi}\uwh\Sigma_{\nu\alpha\beta}  =F_{\nu\alpha\beta}^{\rho\lambda\gamma}\nabla_{\rho}\mc{K}^{(m)}_{\lambda\gamma} -\sum_{k=0}^{m-2}\binom{m-1}{k+1}&\Big( \mc{K}^{(m-1-k)}_{\sigma(\beta} \Sigma^{(k+1)}{}^{\sigma}{}_{\nu)\alpha} + \mc{K}^{(m-1-k)}_{\sigma(\alpha} \Sigma^{(k+1)}{}^{\sigma}{}_{\nu)\beta} \\
			&- \mc{K}^{(m-1-k)}_{\sigma(\alpha} \Sigma^{(k+1)}{}^{\sigma}{}_{\beta)\nu} \Big),
		\end{align*}
		which simplifies to $$\lie^{(m-1)}_{\xi}\uwh\Sigma_{\nu\alpha\beta}  =F_{\nu\alpha\beta}^{\rho\lambda\gamma}\nabla_{\rho}\mc{K}^{(m)}_{\lambda\gamma} - \sum_{k=0}^{m-2}\binom{m-1}{k+1}\mc{K}^{(m-1-k)}_{\sigma\nu} \Sigma^{(k+1)}{}^{\sigma}{}_{\alpha\beta}$$ because of the symmetries of $\mc{K}^{(i)}_{\alpha\beta}$ and $\Sigma^{(i)}{}^{\sigma}{}_{\mu\nu}$. Hence,
		\begin{multicols}{2}
			\noindent
			\begin{equation}
				\label{Sigma1}
				\hspace{-0.5cm} \uwh\Sigma_{\nu\alpha\beta}^{(m)} \st{[m]}{=} F_{\nu\alpha\beta}^{\rho\lambda\gamma}\nabla_{\rho}\mc{K}^{(m)}_{\lambda\gamma} - \mc{K}_{\sigma\nu} \Sigma^{(m-1)}{}^{\sigma}{}_{\alpha\beta},
			\end{equation}
			\begin{equation}
				\label{Sigma2}
				\uwh\Sigma_{\nu\alpha\beta}^{(m-1)} \st{[m]}{=} F_{\nu\alpha\beta}^{\rho\lambda\gamma}\nabla_{\rho}\mc{K}^{(m-1)}_{\lambda\gamma}.
			\end{equation}
		\end{multicols}
		Combining \eqref{Sigmas} and \eqref{Sigma2} it follows $\Sigma^{(m-1)}{}^{\sigma}{}_{\alpha\beta} \st{[m]}{=} g^{\sigma\nu}\uwh\Sigma^{(m-1)}_{\nu\alpha\beta} \st{[m]}{=} g^{\sigma\eps} F_{\eps\alpha\beta}^{\rho\lambda\gamma}\nabla_{\rho}\mc{K}^{(m-1)}_{\lambda\gamma}$, which inserted into \eqref{Sigma1} gives 
		\begin{equation*}
			\uwh\Sigma_{\nu\alpha\beta}^{(m)} \st{[m]}{=} F_{\nu\alpha\beta}^{\rho\lambda\gamma}\nabla_{\rho}\mc{K}^{(m)}_{\lambda\gamma} - g^{\sigma\eps} \mc{K}_{\sigma\nu} F_{\eps\alpha\beta}^{\rho\lambda\gamma}\nabla_{\rho}\mc{K}^{(m-1)}_{\lambda\gamma}.
		\end{equation*} 
		This establishes \eqref{Sigmam2}. Introducing \eqref{Sigma2} and \eqref{Sigmam2} into \eqref{Sigmas} gives \eqref{Sigmam}.
	\end{proof}
\end{lema}
We quote the following immediate consequence for future reference.

\begin{cor}
	Let $\xi\in\X(\mc M)$ and $m\ge 0$ an integer. Then,
	\begin{multicols}{2}
		\noindent
		\begin{equation}
			\label{Sigmam22}
			\uwh\Sigma_{\nu\alpha\beta}^{(m)} \st{[m+1]}{=} F_{\nu\alpha\beta}^{\rho\lambda\gamma}\nabla_{\rho}\mc{K}^{(m)}_{\lambda\gamma} ,
		\end{equation}
		\begin{equation}
			\label{Sigmam1}
			\Sigma^{(m)}{}^{\mu}{}_{\alpha\beta} \st{[m+1]}{=} g^{\mu\nu}F^{\rho\lambda\gamma}_{\nu\alpha\beta}\nabla_{\rho}\mc{K}^{(m)}_{\lambda\gamma}.
		\end{equation}
	\end{multicols}
\end{cor}

Before computing $\mc{R}^{(m)}$, $\dot{\mc{ R}}^{(m)}$ and $\ddot{\mc R}^{(m)}$ on $\mc H$ it is important to make the following observation based on Proposition \ref{liegxi}.

\begin{rmk}
	\label{remark1}
By Proposition \ref{liegxi} the derivatives $\mc{K}^{(m)}(\xi,\cdot)$ are given in terms of transverse derivatives of $a_{\xi}\d\nabla_{\xi}\xi$ on $\Phi(\mc H)$, metric hypersurface data as well as on the tensors $\{\bY,...,\bY^{(m-1)}\}$. Hence, when computing $\mc{R}^{(m)}$, $\dot{\mc{ R}}^{(m)}$ and $\ddot{\mc R}^{(m)}$, terms of the form $\mc{K}^{(m)}(\xi,\cdot)$ can always be replaced by lower order terms, i.e. terms that depend on metric hypersurface data and $\{\bY,...,\bY^{(m-1)}\}$.
\end{rmk}

\begin{nota}
	\label{nota2}
	Let $(\mc M,g)$ be a semi-Riemannian manifold, $\mc H$ an embedded hypersurface and $T,S$ two tensors involving $g$ and its derivatives. We introduce the notation $T \st{(m)}{=} S$ to denote that the tensor $(T-S)|_{\mc H}$ does not depend on \textbf{transverse} derivatives of $g$ at $\mc H$ of order $m$ or higher.
\end{nota}
In the next lemma and following proposition we compute the pullback of $\uwh\Sigma^{(m)}$ as well as several contractions of the tensors $\Sigma^{(m)}$ and $\uwh\Sigma^{(m)}$ that we shall use below. The computation relies on the general identities for the pullback of ambient tensor fields to arbitrary null hypersurfaces computed in Appendix \ref{appendix}.
\begin{lema}
	Let $\mc H$ be a null hypersurface $(\Phi,\xi)$-embedded in a semi-Riemannian manifold $(\mc M,g)$. Extend $\xi$ arbitrarily off $\Phi(\mc H)$ and let $\{e_a\}$ be a local basis on $\mc H$ and $\wh e_a \d \Phi_{\star}e_a$. Then, \\
	
	\begin{minipage}{0.38\textwidth}
		\noindent
		\begin{equation}
			\label{contraction4}
			\xi^{\lambda} \nabla_{\rho}\mc{K}^{(m)}_{\lambda\gamma} \st{(m+1)}{=} 0,
		\end{equation}
	\end{minipage}
	\begin{minipage}{0.62\textwidth}
		\noindent
		\begin{equation}
			\label{contraction7}
			\hspace{-0.0cm}	\wh e^{\rho}_a\wh e^{\lambda}_b \wh e^{\gamma}_c \nabla_{\rho}\mc{K}^{(m+1)}_{\lambda\gamma} \st{(m+1)}{=} 2\nablacero_a\Y^{(m+1)}_{bc} + 4\Y_{a(b}\r^{(m+1)}_{c)},
		\end{equation}
	\end{minipage}
	\begin{minipage}{0.38\textwidth}
		\noindent
		\begin{equation}
			\label{contraction3}
			\xi^{\gamma}\wh e^{\rho}_a\wh e^{\lambda}_b\nabla_{\rho}\mc{K}^{(m)}_{\lambda\gamma} \st{(m+1)}{=} 0,
		\end{equation}
	\end{minipage}
	\begin{minipage}{0.62\textwidth}
		\begin{equation}
			\label{contraction6}
			\wh e^{\rho}_a\wh e^{\lambda}_b \wh e^{\gamma}_c \nabla_{\rho}\mc{K}^{(m)}_{\lambda\gamma} \st{(m+1)}{=} 0,
		\end{equation}
	\end{minipage}
	\begin{minipage}{0.38\textwidth}
		\noindent
		\begin{equation}
			\label{contraction8}
			\hspace{-0.3cm}	\wh e^{\gamma}_a\wh e^{\lambda}_b \xi^{\rho} \nabla_{\rho}\mc{K}^{(m)}_{\lambda\gamma} \st{(m+1)}{=} 2\Y^{(m+1)}_{ab},		
		\end{equation}
	\end{minipage}
	\begin{minipage}{0.62\textwidth}
		\begin{equation}
			\label{contraction2}
			\wh e_a^{\lambda} \wh e_b^{\gamma} g^{\sigma\rho} \nabla_{\rho}\mc{K}^{(m)}_{\lambda\gamma} \st{(m+1)}{=}  2\nu^{\sigma} \Y^{(m+1)}_{ab},
		\end{equation}
	\end{minipage}
	\begin{minipage}{0.38\textwidth}
		\noindent
		\begin{equation}
			\label{contraction5}
			g^{\lambda\gamma} \wh e_a^{\rho} \nabla_{\rho}\mc{K}^{(m)}_{\lambda\gamma}  \st{(m+1)}{=} 0,	
		\end{equation}
	\end{minipage}
	\begin{minipage}{0.62\textwidth}
		\begin{equation}
			\label{contraction1}
			\wh e_a^{\rho} \wh e_b^{\lambda} g^{\sigma\gamma} \nabla_{\rho}\mc{K}^{(m)}_{\lambda\gamma} \st{(m+1)}{=} 0.
		\end{equation}
	\end{minipage}
	\begin{proof}
Equations \eqref{contraction3} and \eqref{contraction6} are particular cases of \eqref{contraction4} and \eqref{contraction7}, respectively. Moreover, the validity of  \eqref{contraction4} and \eqref{contraction7} implies at once \eqref{contraction5} and \eqref{contraction1} because $g^{\sigma\gamma}$ can be decomposed using \eqref{inversemetric}. Hence it suffices to prove the equations in the first and third lines. In order to prove \eqref{contraction4} we contract $\xi^{\lambda} \nabla_{\rho}\mc{K}^{(m)}_{\lambda\gamma}$ with (i) $\xi^{\rho}\xi^{\gamma}$, (ii) $\xi^{\rho}\wh e^{\gamma}_c$, (iii) $\wh e^{\rho}_c\xi^{\gamma}$ and (iv) $\wh e^{\rho}_c\wh e^{\gamma}_d$ and show that all of them are at most of order $m$. Let us start with (i). Applying \eqref{lieandnabla} for $T= \mc{K}^{(m)}$ gives $$\xi^{\rho}\xi^{\gamma}\xi^{\lambda} \nabla_{\rho}\mc{K}^{(m)}_{\lambda\gamma} = \xi^{\lambda}\xi^{\gamma} \big(\lie_{\xi} \mc{K}^{(m)}_{\lambda\gamma} - 2\mc{K}^{(m)}_{\mu(\lambda}\nabla_{\gamma)}\xi^{\mu}\big) \st{(m+1)}{=} \xi^{\lambda}\xi^{\gamma}  \mc{K}^{(m+1)}_{\lambda\gamma} \st{(m+1)}{=} 0,$$ where in the second equality we used that $\mc{K}^{(m)}$ is of order $m$ and in the last one Remark \ref{remark1}. For (ii) we contract $\xi^{\lambda} \nabla_{\rho}\mc{K}^{(m)}_{\lambda\gamma}$ with $\xi^{\rho}\wh e^{\gamma}_c$ and use again that $\mc{K}^{(m)}$ is of order $m$, $$\wh e^{\gamma}_c\xi^{\rho}  \xi^{\lambda} \nabla_{\rho}\mc{K}^{(m)}_{\lambda\gamma} = \wh e^{\gamma}_c\xi^{\rho} \nabla_{\rho}\big({}^{(1)}\mc{K}^{(m)}\big)_{\gamma}-\wh e^{\gamma}_c\xi^{\rho} \mc{K}^{(m)}_{\lambda\gamma} \nabla_{\rho}\xi^{\lambda} \st{(m+1)}{=} \wh e^{\gamma}_c\xi^{\rho} \nabla_{\rho}\big({}^{(1)}\mc{K}^{(m)}\big)_{\gamma}.$$ By Remark \ref{remark1} the tensor ${}^{(1)}\mc{K}^{(m)}$ is of order $m-1$. Applying \eqref{identity2} for $T={}^{(1)}\mc{K}^{(m)}$ all terms in the right hand side are at most of order $m$, so $\wh e^{\gamma}_c\xi^{\rho}  \xi^{\lambda} \nabla_{\rho}\mc{K}^{(m)}_{\lambda\gamma} \st{(m+1)}{=} 0$. For item (iii) we contract $\xi^{\lambda} \nabla_{\rho}\mc{K}^{(m)}_{\lambda\gamma}$ with $\wh e^{\rho}_c\xi^{\gamma}$, namely $$\wh e^{\rho}_c\xi^{\gamma}\xi^{\lambda} \nabla_{\rho}\mc{K}^{(m)}_{\lambda\gamma} = \wh e^{\rho}_c \nabla_{\rho} \big(\mc{K}^{(m)}(\xi,\xi)\big) - 2\xi^{\gamma}\mc{K}^{(m)}_{\lambda\gamma}\nabla_{\rho} \xi^{\lambda} \st{(m+1)}{=} 0,$$ because ${}^{(1)}\mc{K}^{(m)}$ is at most of order $m-1$. In order to prove (iv), i.e. $\wh e^{\rho}_c\wh e^{\gamma}_d\xi^{\lambda} \nabla_{\rho}\mc{K}^{(m)}_{\lambda\gamma} \st{(m+1)}{=} 0$, we use identity \eqref{identity3} with $T=\mc{K}^{(m)}$. Since all the terms in the right hand side are at most of order $m$, $\wh e^{\rho}_c\wh e^{\gamma}_d\xi^{\lambda} \nabla_{\rho}\mc{K}^{(m)}_{\lambda\gamma} \st{(m+1)}{=} 0$. This establishes \eqref{contraction4}. Equation \eqref{contraction7} follows from \eqref{identity1} with $T=\mc{K}^{(m+1)}$ after recalling that ${}^{(i)}\mc{K}^{(m+1)}$ is at most of order $m$ and that $\Phi^{\star}\mc{K}^{(m+1)} = 2\bY^{(m+1)}$. It only remains to prove \eqref{contraction8} and \eqref{contraction2}. For the first one we insert \eqref{inversemetric} into $\wh e_a^{\lambda} \wh e_b^{\gamma} g^{\sigma\rho} \nabla_{\rho}\mc{K}^{(m)}_{\lambda\gamma}$ and get $$\wh e_a^{\lambda} \wh e_b^{\gamma} g^{\sigma\rho} \nabla_{\rho}\mc{K}^{(m)}_{\lambda\gamma} = \wh e_a^{\lambda} \wh e_b^{\gamma}\big(P^{cd}\wh e_c^{\sigma}\wh e_d^{\rho} + \xi^{\sigma}\wh e_c^{\rho}n^c + \xi^{\rho}\nu^{\sigma}\big)\nabla_{\rho}\mc{K}^{(m)}_{\lambda\gamma} \st{(m+1)}{=} \wh e_a^{\lambda} \wh e_b^{\gamma}\xi^{\rho}\nu^{\sigma}\nabla_{\rho}\mc{K}^{(m)}_{\lambda\gamma},$$ where equation \eqref{contraction6} has been used in the first and second terms. This expression shows that \eqref{contraction2} follows from \eqref{contraction8}. To establish the latter we use \eqref{identity2} with $T=\mc{K}^{(m)}$ because only the first term in the right hand side is of order $m+1$.
	\end{proof}
\end{lema}
\begin{prop}
	\label{contractionsSigma}
	Let $\mc H$ be a null hypersurface $(\Phi,\xi)$-embedded in a semi-Riemannian manifold $(\mc M,g)$ and extend $\xi$ arbitrarily off $\Phi(\mc H)$. Then,
	$$\uwh\Sigma^{(m+1)}_{cab} \st{(m+1)}{=} \nablacero_a\Y^{(m+1)}_{bc} +\nablacero_b\Y^{(m+1)}_{ac} -\nablacero_c\Y^{(m+1)}_{ab} + 2\r^{(m+1)}_c\Y_{ab} + 2\r_{c} \Y^{(m+1)}_{ab},$$
	$$\uwh\Sigma^{(m)}_{abc} \st{(m+1)}{=} 0,\quad \big({}^{(1)}\uwh\Sigma^{(m)}\big)_{ab} \st{(m+1)}{=} -\Y^{(m+1)}_{ab},\quad \big({}^{(2)}\uwh\Sigma^{(m)}\big)_{a b} \st{(m+1)}{=} \Y^{(m+1)}_{ab},\quad \big({}^{(3)}\uwh\Sigma^{(m)}\big)_{ab} \st{(m+1)}{=} \Y^{(m+1)}_{ab}, $$$$\xi^{\alpha}\xi^{\beta}\Sigma^{(m)}{}^{\mu}{}_{\alpha\beta} \st{(m+1)}{=} 0,\qquad \Sigma^{(m)}{}^{\mu}{}_{\mu a} \st{(m+1)}{=} 0,\qquad \xi^{\alpha}\Sigma^{(m)}{}^{\mu}{}_{\mu \alpha} \st{(m+1)}{=} \tr_P\bY^{(m+1)}.$$
	\begin{proof}
		Consider equation \eqref{Sigmam2} for $m+1$ and contract it with $\wh e^{\nu}_c\wh e^{\alpha}_a\wh e^{\beta}_b$, namely $$\uwh\Sigma_{cab}^{(m+1)} \st{[m+1]}{=} \wh e^{\nu}_c\wh e^{\alpha}_a\wh e^{\beta}_b F_{\nu\alpha\beta}^{\rho\lambda\gamma}\nabla_{\rho}\mc{K}^{(m+1)}_{\lambda\gamma} - \wh e^{\nu}_c\wh e^{\alpha}_a\wh e^{\beta}_bg^{\sigma\eps} \mc{K}_{\sigma\nu} F_{\eps\alpha\beta}^{\rho\lambda\gamma}\nabla_{\rho}\mc{K}^{(m)}_{\lambda\gamma}.$$ Defining $F_{cab}^{def} \d \frac{1}{2}\big(\delta^{d}_a\delta^e_b\delta^f_c + \delta^{d}_b\delta^e_a\delta^f_c - \delta^{d}_c\delta^e_a\delta^f_b\big)$ and using equation \eqref{contraction7} the first term takes the form 
		\begin{align*}
			F^{def}_{cab}\big(\nabla\mc{K}^{(m+1)}\big)_{def} &\st{(m+1)}{=} 2 F^{def}_{cab}\big(\nablacero_d\Y^{(m+1)}_{ef} + 2\Y_{d(e}\r^{(m+1)}_{f)}\big)\\
			&\st{(m+1)}{=} \nablacero_a\Y^{(m+1)}_{bc} +\nablacero_b\Y^{(m+1)}_{ac} -\nablacero_c\Y^{(m+1)}_{ab} + 2\r_{c}^{(m+1)} \Y_{ab}.
		\end{align*} 
		Making $F_{\eps\alpha\beta}^{\rho\lambda\gamma}$ explicit the second term is
		\begin{align*}
			\wh e^{\nu}_c\wh e^{\alpha}_a\wh e^{\beta}_bg^{\sigma\eps} \mc{K}_{\sigma\nu} F_{\eps\alpha\beta}^{\rho\lambda\gamma}\nabla_{\rho}\mc{K}^{(m)}_{\lambda\gamma} & = \dfrac{1}{2}\left(\wh e_a^{\rho}\wh e_b^{\lambda} g^{\sigma\gamma} + \wh e_a^{\lambda}\wh e_b^{\rho} g^{\sigma\gamma} - \wh e_a^{\lambda}\wh e_b^{\gamma} g^{\sigma\rho}\right)\mc{K}_{\sigma\nu}\wh e_c^{\nu} \nabla_{\rho}\mc{K}^{(m)}_{\lambda\gamma}.
		\end{align*}
		Applying \eqref{contraction1} to the first and second terms and \eqref{contraction2} in the last term, $$\wh e^{\nu}_c\wh e^{\alpha}_a\wh e^{\beta}_bg^{\sigma\eps} \mc{K}_{\sigma\nu} F_{\eps\alpha\beta}^{\rho\lambda\gamma}\nabla_{\rho}\mc{K}^{(m)}_{\lambda\gamma} \st{(m+1)}{=} -\mc{K}_{\sigma\nu}\wh e_c^{\nu}n^{d}\wh e^{\sigma}_d \Y^{(m+1)}_{ab}  \st{(m+1)}{=} - 2 \r_c \Y^{(m+1)}_{ab},$$ because $\Phi^{\star}\mc K = 2\bY$. Hence the equation of the first line is established. The equations of the second line follow from \eqref{Sigmam22}, namely 
		\begin{equation}
			\label{Sigmam222}
			\uwh\Sigma_{\nu\alpha\beta}^{(m)} \st{[m+1]}{=} F_{\nu\alpha\beta}^{\rho\lambda\gamma}\nabla_{\rho}\mc{K}^{(m)}_{\lambda\gamma}.
		\end{equation} 
		Contracting this with $\wh e^{\nu}_c\wh e^{\alpha}_a\wh e^{\beta}_b$ and using \eqref{contraction6} gives $\uwh\Sigma_{cab}^{(m)} \st{(m+1)}{=} 0$, and contracting it with $\xi^{\nu}e^{\alpha}_a\wh e^{\beta}_b$ gives $$\big({}^{(1)}\uwh\Sigma^{(m)}\big)_{ab} \st{[m+1]}{=} \dfrac{1}{2}\left(\xi^{\gamma}\wh e^{\rho}_a\wh e^{\lambda}_b+\xi^{\gamma}\wh e^{\lambda}_a\wh e^{\gamma}_b-\xi^{\rho}\wh e^{\lambda}_a\wh e^{\gamma}_b\right)\nabla_{\rho}\mc{K}^{(m)}_{\lambda\gamma}.$$ Applying  \eqref{contraction3} to the first and second terms and \eqref{contraction8} to the last one yields $ \big({}^{(1)}\uwh\Sigma^{(m)}\big)_{ab}  \st{(m+1)}{=} - \Y^{(m+1)}_{ab}$. In a similar way, contracting \eqref{Sigmam222} with $\wh e^{\nu}_a\xi^{\alpha}\wh e^{\beta}_b$ gives $$\big({}^{(2)}\uwh\Sigma^{(m)}\big)_{ab} \st{[m+1]}{=}  \dfrac{1}{2}\left(\xi^{\rho}\wh e_a^{\gamma}\wh e_b^{\lambda} + \xi^{\lambda} \wh e_a^{\gamma}\wh e_b^{\rho} - \xi^{\lambda}\wh e_a^{\rho}\wh e_b^{\gamma}\right)\nabla_{\rho}\mc{K}^{(m)}_{\lambda\gamma}.$$ By \eqref{contraction3} the only term that contributes is the first one, which after using \eqref{contraction8} is $\big({}^{(2)}\uwh\Sigma^{(m)}\big)_{a b} \st{(m+1)}{=} \Y^{(m+1)}_{ab}$. By the symmetry of $\uwh\Sigma^{(m)}$, $\big({}^{(3)}\uwh\Sigma^{(m)}\big)_{a b} \st{(m+1)}{=} \Y^{(m+1)}_{ab}$. To prove the third line consider \eqref{Sigmam1}, namely 
		\begin{equation}
			\label{aux0}
			\Sigma^{(m)}{}^{\mu}{}_{\alpha\beta} \st{[m+1]}{=} g^{\mu\nu}F^{\rho\lambda\gamma}_{\nu\alpha\beta}\nabla_{\rho}\mc{K}^{(m)}_{\lambda\gamma}.
		\end{equation} 
		The contraction with $\xi^{\alpha}\xi^{\beta}$,
		\begin{equation}
			\label{aux}
			\xi^{\alpha}\xi^{\beta}\Sigma^{(m)}{}^{\mu}{}_{\alpha\beta} \st{[m+1]}{=} \dfrac{1}{2} \left(2\xi^{\lambda}\xi^{\rho} g^{\mu\gamma} - \xi^{\gamma}\xi^{\lambda} g^{\mu\rho}\right)\nabla_{\rho}\mc{K}^{(m)}_{\lambda\gamma},
		\end{equation} 
		yields $\xi^{\alpha}\xi^{\beta}\Sigma^{(m)}{}^{\mu}{}_{\alpha\beta} \st{(m+1)}{=} 0$ after using \eqref{contraction4}. Taking trace in $\mu,\alpha$ in \eqref{aux0} gives 
		\begin{equation}
			\label{auxaux}
			\Sigma^{(m)}{}^{\mu}{}_{\mu\beta} \st{[m+1]}{=} g^{\mu\nu}F^{\rho\lambda\gamma}_{\nu\mu\beta}\nabla_{\rho}\mc{K}^{(m)}_{\lambda\gamma} = \dfrac{1}{2} \big(g^{\rho\gamma}\delta_{\beta}^{\lambda}+g^{\lambda\gamma}\delta_{\beta}^{\rho}-g^{\lambda\rho}\delta_{\beta}^{\gamma}\big)\nabla_{\rho}\mc{K}^{(m)}_{\lambda\gamma} = \dfrac{1}{2} g^{\lambda\gamma}\nabla_{\beta}\mc{K}^{(m)}_{\lambda\gamma},
		\end{equation} 
		where in the last equality we used the symmetry of $\mc{K}^{(m)}$. Contracting with $\wh e^{\beta}_a$ and using equation \eqref{contraction5} yields $\wh e^{\beta}_a\Sigma^{(m)}{}^{\mu}{}_{\mu\beta} \st{(m+1)}{=} 0$. Finally, from \eqref{lieandnabla} with $T=\mc{K}^{(m)}$,
		\begin{equation}
			\label{lieandnabla2}
			\nabla_{\xi}\mc{K}^{(m)} \st{[m+1]}{=} \mc{K}^{(m+1)},
		\end{equation}
		so contracting \eqref{auxaux} with $\xi^{\beta}$ gives $\xi^{\beta}\Sigma^{(m)}{}^{\mu}{}_{\mu\beta} \st{(m+1)}{=} \frac{1}{2} g^{\lambda\gamma}\mc{K}^{(m+1)}_{\lambda\gamma}$. Using \eqref{inversemetric} and the fact that ${}^{(1)}\mc{K}^{(m+1)}\st{(m+1)}{=}0$, the last equation of the third line follows.
	\end{proof}
\end{prop}
By virtue of \eqref{liemricmunu} the last ingredient to compute the tensors $\mc{R}^{(m)}$, $\dot{\mc{ R}}^{(m)}$ and $\ddot{\mc R}^{(m)}$ on $\mc H$ is being able to calculate the pullback both of a divergence and of a ambient tensor field. These are computed in full generality in Propositions \ref{proppullback} and \ref{propdivergencia}, and as a consequence we have the following three propositions.

\begin{prop}
	\label{propliericxixi}
	Let $\mc H$ be a null hypersurface $(\Phi,\xi)$-embedded in $(\mc M,g)$ and extended $\xi$ arbitrarily off $\Phi(\mc H)$. Then for every $m\ge 0$,
	\begin{equation}
		\label{liericxixi}
		\ddot{\mc R}^{(m+1)} \st{(m+2)}{=} -\tr_P\bY^{(m+2)}.
	\end{equation}
	\begin{proof}
		The case $m=0$ has already been established in \eqref{ricxixi}. In order to prove the identity for $m\ge 1$ we contract the general expression \eqref{liemricmunu} with $\xi^{\alpha}\xi^{\beta}$ so that
		\begin{align}
			\xi^{\alpha}\xi^{\beta}\lie_{\xi}^{(m)}\ric_{\alpha\beta} &\st{[m+2]}{=} \xi^{\alpha}\xi^{\beta}\nabla_{\mu}\Sigma^{(m)}{}^{\mu}{}_{\alpha\beta} - \xi^{\alpha}\xi^{\beta}\nabla_{\beta}\Sigma^{(m)}{}^{\mu}{}_{\mu\alpha}\label{uni1}\\
			&\st{[m+2]}{=} \nabla_{\mu}\big(\xi^{\alpha}\xi^{\beta}\Sigma^{(m)}{}^{\mu}{}_{\alpha\beta}\big) - \nabla_{\xi}\big(\xi^{\alpha}\Sigma^{(m)}{}^{\mu}{}_{\mu\alpha}\big),\label{uni2}
		\end{align}
		where in the second line we used that $\Sigma^{(m)}$ involves up to $m+1$ derivatives of $g$. For the same reason, by \eqref{aux} it follows 
		\begin{align*}
			\nabla_{\mu} \big(\xi^{\alpha}\xi^{\beta}\Sigma^{(m)}{}^{\mu}{}_{\alpha\beta}\big) & \st{[m+2]}{=} \xi^{\lambda}\xi^{\rho} g^{\mu\gamma}\nabla_{\mu}\nabla_{\rho}\mc{K}^{(m)}_{\lambda\gamma} - \dfrac{1}{2}\xi^{\gamma}\xi^{\lambda} g^{\mu\rho}\nabla_{\mu}\nabla_{\rho} \mc{K}^{(m)}_{\lambda\gamma}\\
			& \st{[m+2]}{=} \xi^{\lambda} g^{\mu\gamma} \nabla_{\mu}\nabla_{\xi}\mc{K}^{(m)}_{\lambda\gamma} - \xi^{\lambda} g^{\mu\gamma}\nabla_{\rho}\mc{K}^{(m)}_{\lambda\gamma} \nabla_{\mu}\xi^{\rho} - \dfrac{1}{2}\xi^{\gamma}\xi^{\lambda} g^{\mu\rho}\nabla_{\mu}\nabla_{\rho} \mc{K}^{(m)}_{\lambda\gamma}.
		\end{align*}
		From the expression of $g^{\mu\rho}$ in \eqref{inversemetric} and the fact that $\nabla\mc{K}^{(m)}$ is at most order $m+1$, the only term that has a chance of carrying $m+2$ transverse derivatives of $g$ is the first one. Using \eqref{lieandnabla2} and recalling \eqref{contraction4}, $$\xi^{\lambda} g^{\mu\gamma} \nabla_{\mu}\nabla_{\xi}\mc{K}^{(m)}_{\lambda\gamma}\st{[m+2]}{=} \xi^{\lambda} g^{\mu\gamma} \nabla_{\mu}\mc{K}^{(m+1)}_{\lambda\gamma} \st{(m+2)}{=} 0.$$ Finally from \eqref{auxaux} it follows $$\nabla_{\xi}\big(\xi^{\alpha}\Sigma^{(m)}{}^{\mu}{}_{\mu\alpha}\big) \st{[m+2]}{=} \dfrac{1}{2} g^{\lambda\gamma}\nabla_{\xi}\big(\xi^{\mu}\nabla_{\mu}\mc{K}^{(m)}_{\lambda\gamma}\big) \st{(m+2)}{=}  \dfrac{1}{2} g^{\lambda\gamma} \mc{K}^{(m+2)}_{\lambda\gamma}\st{(m+2)}{=}  \tr_P\bY^{(m+2)},$$ where in the second equality we used $\nabla_{\xi}\mc{K}^{(m)} \st{[m+1]}{=} \mc{K}^{(m+1)}$ and $\nabla_{\xi}\mc{K}^{(m+1)} \st{[m+2]}{=} \mc{K}^{(m+2)}$ (see \eqref{lieandnabla2}) and in the third one we inserted \eqref{inversemetric} and used the fact that ${}^{(1)}\mc{K}^{(m+2)}$ is at most of order $m+1$ (see Remark \ref{remark1}).
	\end{proof}
\end{prop}
\begin{prop}
	\label{propliericxitang}
	Let $\mc H$ be a null hypersurface $(\Phi,\xi)$-embedded in $(\mc M,g)$ and extend $\xi$ arbitrarily off $\Phi(\mc H)$. Then for any $m\ge 0$,
	\begin{equation}
		\label{liericxiX}
		\dot{\mc R}^{(m+1)} \st{(m+2)}{=} \br^{(m+2)}.
	\end{equation}
	\begin{proof}
		The case $m=0$ is \eqref{ricxiX}. From \eqref{liemricmunu} and the fact that $\Sigma^{(m)}$ is at most of order $m+1$,
		\begin{align}
			\wh e^{\alpha}_a\xi^{\beta}\lie_{\xi}^{(m)}R_{\alpha\beta} &\st{[m+1]}{=} \wh e^{\alpha}_a\xi^{\beta}\nabla_{\mu}\Sigma^{(m)}{}^{\mu}{}_{\alpha\beta} - \wh e^{\alpha}_a\xi^{\beta} \nabla_{\beta} \Sigma^{(m)}{}^{\mu}{}_{\mu\alpha}\label{uni3}\\
			&\st{[m+1]}{=} \wh e^{\alpha}_a \nabla_{\mu} \big(\xi^{\beta}\Sigma^{(m)}{}^{\mu}{}_{\alpha\beta}\big) - \wh e^{\alpha}_a\Sigma^{(m)}{}^{\mu}{}_{\alpha\beta} \nabla_{\mu}\xi^{\beta} - \wh e^{\alpha}_a\nabla_{\xi}\big(\Sigma^{(m)}{}^{\mu}{}_{\mu\alpha}\big),\label{uni4}
		\end{align}
		so the only terms capable of containing $m+2$ transverse derivatives are the first and third ones. By Proposition \ref{propdivergencia} applied to ${}^{(3)}\Sigma^{(m)}{}^{\mu}{}_{\alpha}$ and $\lie_{\xi}\xi^{\beta}=0$ it follows $$
		\wh e^{\alpha}_a \nabla_{\mu} \big(\xi^{\beta}\Sigma^{(m)}{}^{\mu}{}_{\alpha\beta}\big) \st{(m+2)}{=} n^b \wh e^{\alpha}_a \wh e^{\mu}_b \xi^{\beta}\lie_{\xi}\Sigma^{(m)}_{\mu\alpha\beta}$$ because the rest of the terms are of (transverse) order $m+1$ or below. Using the second equation in \eqref{Sigmas3} and Proposition \ref{contractionsSigma} we obtain $$
		\wh e^{\alpha}_a \nabla_{\mu} \big(\xi^{\beta}\Sigma^{(m)}{}^{\mu}{}_{\alpha\beta}\big) \st{(m+2)}{=} n^b\wh e^{\alpha}_a\wh e^{\mu}_b\xi^{\beta}\uwh\Sigma^{(m+1)}_{\mu\alpha\beta} \st{(m+2)}{=} \r^{(m+2)}_a.$$ Finally, by \eqref{identity2} with $T=\Sigma^{(m)}{}^{\mu}{}_{\mu\alpha}$ it follows $$\wh e^{\alpha}_a\nabla_{\xi}\big(\Sigma^{(m)}{}^{\mu}{}_{\mu\alpha}\big) \st{(m+2)}{=}\wh e_a^{\alpha}\lie_{\xi}\Sigma^{(m)}{}^{\mu}{}_{\mu\alpha} \st{(m+2)}{=}  \Sigma^{(m+1)}{}^{\mu}{}_{\mu a} \st{(m+2)}{=} 0,$$ where the last equality we used Proposition \ref{contractionsSigma}.
	\end{proof}
\end{prop}
\begin{prop}
	\label{propricxy}
	Let $\mc H$ be a null hypersurface $(\Phi,\xi)$-embedded in $(\mc M,g)$ and extended $\xi$ arbitrarily off $\Phi(\mc H)$. Let $m\ge 1$ be an integer. Then,
	\begin{equation}
		\label{lierictang}
		\begin{aligned}
			\hskip -0.5cm \mc{R}^{(m+1)}_{ab} &\st{(m+1)}{=} -2\lie_n\Y^{(m+1)}_{ab} - \left(2(m+1)\kappa_n + \tr_P\bU\right)\Y^{(m+1)}_{ab}- (\tr_P\bY^{(m+1)})\U_{ab}\\
			&\qquad\,\,\,\, + 4P^{cd}\U_{c(a}\Y^{(m+1)}_{b)d}  +4(\s-\r)_{(a} \r^{(m+1)}_{b)}+ 2\nablacero_{(a}\r^{(m+1)}_{b)}-2\kappa^{(m+1)}\Y_{ab}.
		\end{aligned}
	\end{equation}
	\begin{proof}
		By the general formula \eqref{liemricmunu} we have 
		\begin{equation}
			\label{aux3}
			\mc{R}^{(m+1)}_{ab} \st{[m+1]}{=} \wh e^{\alpha}_a \wh e_b^{\beta}\nabla_{\mu}\Sigma^{(m)}{}^{\mu}{}_{\alpha\beta} - \wh e^{\alpha}_a \wh e_b^{\beta}\nabla_{\beta}\Sigma^{(m)}{}^{\mu}{}_{\mu\alpha},
		\end{equation}
		so we only need to compute each of these terms. For the first one we use Proposition \ref{propdivergencia} applied to $\Sigma^{(m)}$, namely 
		\begin{align*}
			(\div \Sigma^{(m)})_{ab} & \st{(m+1)}{=} P^{bc}\nablacero_b \Sigma^{(m)}_{c ab} +n^c(\lie_{\xi} \Sigma^{(m)})_{cab}  + n^c\nablacero_c \big({}^{(1)}\Sigma^{(m)}\big)_{ab} + \big(2\kappa_n + \tr_P\bU\big) \big({}^{(1)}\Sigma^{(m)}\big)_{ab}\\
			&\quad\, + (\tr_P\bY - n(\elltwo))n^c\Sigma^{(m)}_{c ab} - 2P^{dc}(\r+\s)_d  \Sigma^{(m)}_{c ab}  +  2P^{dc}\Y_{d(a|}\Sigma^{(m)}_{c |b)f}n^f\\
			&\quad\,  +  2P^{dc}\U_{d(a|}\big({}^{(2)}\Sigma^{(m)}\big)_{c |b)}-  2(\r-\s)_{(a|} \big({}^{(2)}\Sigma^{(m)}\big)_{c |b)}n^c -  2V^c{}_{(a|}n^d \Sigma^{(m)}_{d c |b)}+ 2 \r_{(a|} \big({}^{(1)}\Sigma^{(m)}\big)_{c |b)}n^c .
		\end{align*}
		From the second equation in \eqref{Sigmas2}, $\Sigma^{(m)}_{\alpha\beta\gamma} \st{(m+1)}{=}\uwh\Sigma^{(m)}_{\alpha\beta\gamma}$, and thus by Proposition \ref{contractionsSigma} it follows $\Sigma^{(m)}_{abc} \st{(m+1)}{=} 0$, $\big({}^{(1)} \Sigma^{(m)}\big)_{ab} \st{(m+1)}{=} -\Y^{(m+1)}_{ab}$ and $\big({}^{(2)}\Sigma^{(m)}\big)_{ab} \st{(m+1)}{=} \Y^{(m+1)}_{ab}$. Hence the expression for $(\div \Sigma^{(m)})_{ab}$ simplifies to 
		\begin{equation}
			\label{auxdiv}
			\begin{aligned}
				(\div \Sigma^{(m)})_{ab} & \st{(m+1)}{=} n^c(\lie_{\xi} \Sigma^{(m)})_{cab}  - n^c\nablacero_c \Y^{(m+1)}_{ab} - \big(2\kappa_n + \tr_P\bU\big) \Y^{(m+1)}_{ab}\\
				&\quad\,  +  2P^{dc}\U_{d(a|}\Y^{(m+1)}_{c |b)}-  2(\r-\s)_{(a|} \Y^{(m+1)}_{c |b)}n^c - 2 \r_{(a|} \Y^{(m+1)}_{c |b)}n^c .
			\end{aligned}
		\end{equation}
		In order to compute the term $n^c(\lie_{\xi} \Sigma^{(m)})_{cab}$ we contract the first equation in \eqref{Sigmas3} with $\wh e^{\mu}_c\wh e^{\alpha}_a\wh e^{\beta}_b$, which gives
		\begin{equation*}
			(\lie_{\xi} \Sigma^{(m)})_{cab}  \st{[m+1]}{=} \wh e^{\mu}_c\wh e^{\alpha}_a\wh e^{\beta}_b\uwh\Sigma^{(m+1)}_{\mu\alpha\beta} - (m-1)\wh e^{\mu}_c\wh e^{\alpha}_a\wh e^{\beta}_bg^{\nu\rho}\mc{K}_{\mu\nu}\uwh\Sigma_{\rho\alpha\beta}^{(m)}.
		\end{equation*}
		The first term is given by the first line in Proposition \ref{contractionsSigma}, and, for the second one, using \eqref{inversemetric} as well as $\uwh\Sigma^{(m)}_{abc} \st{(m+1)}{=} 0$ and ${}^{(1)} \uwh\Sigma^{(m)}_{ab} \st{(m+1)}{=} -\Y^{(m+1)}_{ab}$ one gets
		\begin{equation*}
			\wh e_c^{\mu}\wh e_a^{\alpha}\wh e_{b}^{\beta} g^{\nu\rho}\mc{K}_{\mu\nu}\uwh\Sigma^{(m)}_{\rho\alpha\beta} \st{(m+1)}{=} -2\r_c\Y^{(m+1)}_{ab}.
		\end{equation*}
		Combining everything it follows $$(\lie_{\xi}\Sigma^{(m)})_{cab}\st{(m+1)}{=} \nablacero_a\Y^{(m+1)}_{bc} +\nablacero_b\Y^{(m+1)}_{ac} -\nablacero_c\Y^{(m+1)}_{ab} + 2\r^{(m+1)}_c\Y_{ab} + 2m \r_c \Y^{(m+1)}_{ab},$$
		and after contracting with $n^c$ equation \eqref{auxdiv} becomes
		\begin{align*}
			(\div \Sigma^{(m)})_{ab} & \st{(m+1)}{=} 2n^c\nablacero_{(a}\Y^{(m+1)}_{b)c} - 2n^c\nablacero_c\Y^{(m+1)}_{ab} - 2\kappa^{(m+1)}\Y_{ab}  - \big(2(m+1)\kappa_n + \tr_P\bU\big) \Y^{(m+1)}_{ab}\\
			&\quad\,  +  2P^{dc}\U_{d(a}\Y^{(m+1)}_{b)c}+ 2(\s-2\r)_{(a} \r^{(m+1)}_{b)}.
		\end{align*}
		Using
		\begin{align*}
			2n^c\nablacero_{(a} \Y^{(m+1)}_{b)c} - 2n^c\nablacero_c\Y^{(m+1)}_{ab}& = 2\nablacero_{(a}\r^{(m+1)}_{b)} - 2\Y^{(m+1)}_{c(a}\nablacero_{b)} n^c - 2\lie_n\Y^{(m+1)}_{ab} + 4 \Y^{(m+1)}_{c(a}\nablacero_{b)} n^c\\
			&=2\nablacero_{(a}\r^{(m+1)}_{b)}  - 2\lie_n\Y^{(m+1)}_{ab} + 2\Y^{(m+1)}_{c(a}\nablacero_{b)} n^c\\
			&= 2\nablacero_{(a}\r^{(m+1)}_{b)}  - 2\lie_n\Y^{(m+1)}_{ab} + 2\Y^{(m+1)}_{c(a}\U_{b)d}P^{cd} + 2\r^{(m+1)}_{(a}\s_{b)},
		\end{align*}
		where the third equality follows from \eqref{derivadannull}, the expression for $(\div \Sigma^{(m)})_{ab}$ is finally
		\begin{equation}
			\label{div}
			\begin{aligned}
				(\div \Sigma^{(m)})_{ab} & \st{(m+1)}{=} - 2\lie_n\Y^{(m+1)}_{ab}- \big(2(m+1)\kappa_n + \tr_P\bU\big) \Y^{(m+1)}_{ab}+4P^{cd}\Y^{(m+1)}_{c(a}\U_{b)d}\\
				&\quad\,   + 4(\s-\r)_{(a} \r^{(m+1)}_{b)} + 	2 \nablacero_{(a}\r^{(m+1)}_{b)}   - 2\kappa^{(m+1)}\Y_{ab}.
			\end{aligned}
		\end{equation}
		To compute the second term in \eqref{aux3} we use equation \eqref{identity1} for $T=\Sigma^{(m)}{}^{\mu}{}_{\mu\alpha}$. Since by Proposition \ref{contractionsSigma} $\Sigma^{(m)}{}^{\mu}{}_{\mu a}\st{(m+1)}{=}0$ and $\xi^{\alpha} \Sigma^{(m)}{}^{\mu}{}_{\mu\alpha}\st{(m+1)}{=} \tr_P\bY^{(m+1)}$, we find that $\wh e^{\alpha}_a \wh e^{\beta}_b\nabla_{\beta} \Sigma^{(m)}{}^{\mu}{}_{\mu\alpha}  = \big(\tr_P\bY^{(m+1)}\big)\U_{ab}$. Inserting this and \eqref{div} into \eqref{aux3} proves the Proposition.
	\end{proof}
\end{prop}
Proposition \ref{propricxy} is interesting because it allows one to determine the evolution of the transverse expansion of the metric along the null generator in any null hypersurface. We next compute two contractions of \eqref{lierictang} as well as its trace w.r.t $P$.

\begin{cor}
	\label{contraction}
	Let $\mc H$ be a null hypersurface $(\Phi,\xi)$-embedded in $(\mc M,g)$ and extend $\xi$ arbitrarily off $\Phi(\mc H)$. Let $\ric$ be the Ricci tensor of $g$. Then,
	\begin{align}
		\mc{R}^{(m+1)}_{ab} n^b &\st{(m+1)}{=} -\lie_n\r^{(m+1)}_a - \big(2m\kappa_n + \tr_P\bU\big)\r^{(m+1)}_a-\nablacero_a\kappa^{(m+1)},\label{lierictangn}\\
		\mc{R}^{(m+1)}_{ab} n^an^b &\st{(m+1)}{=} \big(2m\kappa_n+\tr_P\bU\big) \kappa^{(m+1)},\label{lierictangnn}\\
		P^{ab}\mc{R}^{(m+1)}_{ab} &\st{(m+1)}{=} -2\lie_n\big(\tr_P\bY^{(m+1)}\big) - 2\left((m+1)\kappa_n + \tr_P\bU\right)\tr_P\bY^{(m+1)} \nonumber\\
		&\qquad\,\,\,\, + 2\kappa^{(m+1)}\big(n(\elltwo)-\tr_P\bY\big)-4P\big(\br+\bs,\br^{(m+1)}\big) + 2\div_P\br^{(m+1)}.	\label{Pcontractioneq}
	\end{align}
	\begin{proof}
		To prove \eqref{lierictangn} we contract \eqref{lierictang} with $n^b$ and use
		\begin{align*}
			2n^b\nablacero_{(a}\r^{(m+1)}_{b)} &= -\nablacero_a\kappa^{(m+1)} - \r^{(m+1)}_{b}\nablacero_{a}n^b + n^b\nablacero_b\r^{(m+1)}_a \\
			& = -\nablacero_a\kappa^{(m+1)} - 2\r^{(m+1)}_{b}\nablacero_{a}n^b + \lie_n\r^{(m+1)}_a\\
			&= -\nablacero_a\kappa^{(m+1)} + 2\kappa^{(m+1)}\s_a - 2P^{bc}\U_{ac}\r^{(m+1)}_b + \lie_n\r^{(m+1)}_a,
		\end{align*} 
		where the second equality follows from \eqref{derivadannull}. Contracting \eqref{lierictangn} with $n^a$ \eqref{lierictangnn} follows at once. For the last one apply $P^{ab}$ to \eqref{lierictang} and use \eqref{lietrPY} with $\bT=\bY^{(m+1)}$.
	\end{proof}
\end{cor}

Later on we shall need to use the ``complete'' identities \eqref{liericxixi}, \eqref{liericxiX} and \eqref{lierictang}, i.e. including a term that gathers all the lower order terms. By Proposition \ref{universal0} these terms are $\mc H$-geometrical when $\nabla_{\xi}\xi=0$.

\begin{cor}
	\label{cor_entero}
	Let $\{\mc H,\bg,\bm\ell,\elltwo\}$ be null metric hypersurface data $(\Phi,\xi)$-embedded in $(\mc M,g)$ and extended $\xi$ arbitrarily off $\Phi(\mc H)$. Let $m\ge 1$ be an integer. Then,
	\begin{multicols}{2}
		\noindent
		\begin{equation}
			\label{ddotR}
			\ddot{\mc R}^{(m)} = - \tr_P\bY^{(m+1)} + \mc{O}^{(m)}(\bY^{\le m}),
		\end{equation}
		\begin{equation}
			\label{dotR}
			\dot{\mc R}^{(m)}_a = \r^{(m+1)}_a + \mc{O}^{(m)}_a(\bY^{\le m}),
		\end{equation}
	\end{multicols}
	\vskip -0.4cm
	\begin{equation}
		\label{R}
		\begin{aligned}
			\mc{R}^{(m+1)}_{ab} &= -2\lie_n\Y^{(m+1)}_{ab} - \left(2(m+1)\kappa_n + \tr_P\bU\right)\Y^{(m+1)}_{ab}- (\tr_P\bY^{(m+1)})\U_{ab}\\
			&\quad\,  + 4P^{cd}\U_{c(a}\Y^{(m+1)}_{b)d}  +4(\s-\r)_{(a} \r^{(m+1)}_{b)}+ 2\nablacero_{(a}\r^{(m+1)}_{b)}\\
			&\quad\, -2\kappa^{(m+1)}\Y_{ab} + \mc{O}^{(m)}_{ab}(\bY^{\le m}),
		\end{aligned}
	\end{equation}
	where $\mc{O}^{(m)}$, $\mc{O}^{(m)}_a$ and $\mc{O}^{(m)}_{ab}$ are, respectively, a scalar, a one-form and a (0,2) symmetric tensor on $\mc H$ with the property that when $\nabla_{\xi}\xi = 0$ they only depend on null metric data $\{\bg,\bm\ell,\elltwo\}$ and on the tensors $\{\bY,...,\bY^{(m)}\}$.
\end{cor}

\subsection{Asymptotic uniqueness}

In this subsection we are interested in studying the sufficient conditions for two metric data $\{\H,\bg,\bm\ell,\elltwo\}$ and $\{\mc H',\bg',\bm\ell',\elltwo{}'\}$ to be embedded in ambient manifolds $(\mc M,g)$ and $(\mc M',g')$ \textit{asymptotically isometric at the hypersurface}, i.e. such that the ambient metrics $g$ and $g'$ at the two null hypersurfaces $\Phi(\H)$ and $\Phi'(\H')$ agree at all orders. We begin by introducing some notation.

\begin{nota}
	\label{nota0}
Let $\{\mc H',\bg',\bm\ell',\elltwo{}'\}$ be null metric hypersurface data and $\chi:\mc H\to\mc H'$ a diffeomorphism. We define
\begin{equation}
	\label{defi_pullback0}
\chi^{\star} \{\mc H',\bg',\bm\ell',\elltwo{}'\} \d \{\mc H,\chi^{\star}\bg',\chi^{\star}\bm\ell',\chi^{\star}\elltwo{}'\}.
\end{equation} 
\end{nota}
\begin{rmk}
	\label{lema_diff}
It is immediate to check that if $\{\mc H',\bg',\bm\ell',\elltwo{}'\}$ is null metric hypersurface data and $\chi:\mc H\to\mc H'$ is a diffeomorphism, then $\chi^{\star} \{\mc H',\bg',\bm\ell',\elltwo{}'\}$ is also null metric hypersurface data.
\end{rmk}
%

In order to compare the metrics at $\Phi(\H)$ and $\Phi'(\H')$ we need to construct suitable neighbourhoods around $\Phi(\H)$ and $\Phi'(\H')$ and to map them via a diffeomorphism, as we accomplish next.

\begin{prop}
	\label{prop_diffeo}
	Let $\Phi:\mc H\hookrightarrow\mc M$ and $\Phi':\mc H'\hookrightarrow\mc M'$ be two embedded hypersurfaces in ambient manifolds $(\mc M,g)$ and $(\mc M',g')$ and let $\xi$, $\xi'$ be respectively riggings of $\Phi(\mc H)$, $\Phi'(\mc H')$ extended geodesically. Assume that there exists a diffeomorphism $\chi:\mc H\to\mc H'$. Then, there exist open neighbourhoods $\mc U\subset \mc M$ and $\mc U'\subset\mc M'$ of $\Phi(\mc H)$ and $\Phi'(\mc H')$ and a unique diffeomorphism $\Psi:\mc U\to\mc U'$ satisfying $\Psi_{\star}\xi = \xi'$ and $\Phi'\circ \chi = \Psi\circ \Phi$. 
	\begin{proof}
		Pick a neighbourhood $\mc U\subset \mc M$ of $\Phi(\mc H)$ small enough so that the integral curves of $\xi$ do not intersect each other and intersect $\Phi(\H)$ precisely once. Construct a neighbourhood $\mc U'$ of $\Phi'(\mc H')$ similarly. Then, given a point $q\in\mc U$ there exist a unique $p\in\mc H$ such that the integral curve $\sigma(\tau)$ of $\xi$ through $p$ reaches $q$ at a finite $\tau_q$, i.e. $\sigma(\tau_q)=q$. Now consider the point $p'\d\chi(p)\in\mc H'$ and the integral curve $\sigma'$ of $\xi'$ through $p'$. We define $\Psi(q) \d \sigma'(\tau_q)$. It is clear that $\Psi$ is a diffeomorphism when $\mc U$ and $\mc U'$ are small enough, and by construction it satisfies $\Psi_{\star}\xi = \xi'$ and $\Phi'\circ \chi = \Psi\circ \Phi$.
	\end{proof}
\end{prop}

Finally, once the neighbourhoods around $\Phi(\H)$ and $\Phi'(\H')$ are constructed and mapped one to another, we prove that when the metric data and the asymptotic expansions at $\Phi(\mc H)$ and $\Phi'(\mc H')$ agree, then $(\mc M,g)$ and $(\mc M',g')$ are isometric to infinite order (i.e. satisfy relation \eqref{isometry0} below).

\begin{prop}
	\label{teo_iso0}
	Let $\{\H,\bg,\bm\ell,\elltwo\}$ (respectively $\{\H',\bg',\bm\ell',\elltwo{}'\}$) be null metric hypersurface data $(\Phi,\xi)$-embedded in $(\mc M,g)$ (resp. $(\Phi',\xi')$-embedded in $(\mc M',g')$) with $\xi$ and $\xi'$ extended geodesically. Assume that there exists a diffeomorphism $\chi:\H\to\H'$ such that $\chi^{\star}\{\H',\bg',\bm\ell',\elltwo{}'\} = \{\H,\bg,\bm\ell,\elltwo\}$ and $\chi^{\star}\bY^{(k)}{}' = \bY^{(k)}$ for every $k\ge 1$. Then, there exist neighbourhoods $\mc U\subset \mc M$ and $\mc U'\subset\mc M'$ of $\Phi(\mc H)$ and $\Phi'(\mc H')$ and a diffeomorphism $\Psi:\mc U\to\mc U'$ such that 
	\begin{equation}
		\label{isometry0}
		\Psi^{\star} \lie_{\xi'}^{(i)}g' \st{\mc H}{=} \lie_{\xi}^{(i)}g
	\end{equation} 
	for every $i\in\mathbb{N}\cup\{0\}$.
	\begin{proof}
		By Proposition \ref{prop_diffeo} there exists a unique diffeomorphism $\Psi:\mc U\to\mc U'$ satisfying $\Psi_{\star}\xi = \xi'$ and $\Phi'\circ \chi = \Psi\circ \Phi$. Let us prove that $\Psi^{\star} \lie_{\xi'}^{(i)}g' \st{\mc H}{=} \lie_{\xi}^{(i)}g$ for every $i\ge 0$. The case $i=0$ is immediate because $\bg=\chi^{\star}\bg'$, $\bm\ell=\chi^{\star}\bm\ell'$ and $\elltwo=\chi^{\star}\elltwo{}'$. Proving the case $i\ge 1$ amounts to show $\Psi^{\star}\big(\lie_{\xi'}^{(i)}g'\big)(\xi',\cdot) = \big(\lie_{\xi}^{(i)}g\big)(\xi,\cdot)$ and $\chi^{\star}\bY^{(i)}{}' = \bY^{(i)}$. The latter is part of the hypothesis, and the former is a direct consequence of Proposition \ref{liegxi} because $a_{\xi}=0$, $a_{\xi'}=0$ and $\elltwo=\chi^{\star}\elltwo{}'$.
	\end{proof}
\end{prop}

\section{Deformation tensor and algebraic identities}
\label{section_deformation}

In some situations one has a privileged vector field $\eta$ on $(\mc M,g)$ whose deformation tensor $\mc{K}[\eta]\d\lie_{\eta} g$ is known. This is in general a very valuable information that one may want to incorporate into the identities of Section \ref{section_higher}. For instance, in \cite{tesismiguel} it is shown that the expression of the constraint tensor in \eqref{constraint} can be rewritten so that the dependence on the tensor $\bY$ is algebraic instead of via a transport equation. The corresponding identity was called the \textit{generalized master equation}. The aim of this section is to extend the same idea to the higher order derivatives of the Ricci tensor, i.e. to combine identity \eqref{R} with information on $\mc{K}[\eta]$ so that the dependence on $\bY^{(m+1)}$ becomes algebraic. Let us start by reviewing a result from \cite{tesismiguel}.

\begin{lema}
	Let $\{\mc H,\bg,\bm\ell,\elltwo\}$ be null metric hypersurface data $(\Phi,\xi)$-embedded in $(\mc M,g)$ and let $\eta\in\X(\mc M)$ be such that $\eta|_{\Phi(\mc H)}$ is tangent to $\Phi(\mc H)$. Denote by $\bar\eta$ the vector field on $\mc H$ satisfying $\Phi_{\star}(\bar\eta) = \eta|_{\Phi(\mc H)}$. Then, 
	\begin{equation}
		\label{etaxi}
		[\eta,\xi] \st{\mc H}{=} A_{\eta} \xi + \Phi_{\star}(X_{\eta}),
	\end{equation}
	where 
	\begin{equation}
		\label{Aeta}
		A_{\eta} \d -\mc{K}[\eta](\xi,\nu) + (\lie_{\bar\eta}\bm\ell)(n),	
	\end{equation}
	and the components of $X_{\eta}$ in any basis $\{e_a\}$ of $\mc H$ are given by 
	\begin{equation}
		\label{Xeta}
		X_{\eta}^a \d \dfrac{1}{2}\mc{K}[\eta]\big(\xi,n^a\xi-2\theta^a\big) + \dfrac{1}{2}n^a \bar{\eta}(\elltwo) + P^{ab}\lie_{\bar\eta}\ell_b,
	\end{equation}
	where $\theta^a= P^{ab}\Phi_{\star} e_b + n^a\xi$.
\end{lema}

From now on we assume $\eta|_{\mc H}$ to be null and tangent to $\Phi(\mc H)$, so there must exist a function $\alpha\in\mc F(\mc H)$ such that $\eta |_{\mc H} = \alpha\nu$ (in principle we do not assume $\alpha\neq 0$, thus allowing $\eta$ to have zeros). The gauge behaviour of the scalars $\alpha$ and $\mc{K}[{\eta}](\xi,\nu)$ are as follows \cite{tesismiguel}.
\begin{lema}
	\label{lemagauge}
	Let $(z,V)$ be a gauge element and denote with a prime the gauge-transformed objects. Then,
	\begin{equation}
		\alpha' = z\alpha, \qquad \mc K[\eta](\xi',\nu') = \mc K[\eta](\xi,\nu) .
	\end{equation}
\end{lema}
From this it is immediate to check that the transformation of $n(\alpha)$ and that of $A_{\eta}$ are then given by
\begin{multicols}{2}
	\noindent
	\begin{equation}
		\label{transnalfa}
		n'(\alpha')=n(\alpha) +  n(\log |z|)\alpha,
	\end{equation}
	\begin{equation}
		\label{transaeta}
		A_{\eta}' = A_{\eta} +  n(\log |z|)\alpha.
	\end{equation}
\end{multicols}
In particular, $n(\alpha)$ is gauge-invariant at the points where $\alpha=0$.\\

Following \cite{tesismiguel}, let us introduce the scalar function $\kappa$ by means of 
\begin{equation}
	\label{kappasgeneral}
	\kappa \d n(\alpha)+\alpha\kappa_n.
\end{equation}
From equation \eqref{transkappa} and Lemma \ref{lemagauge} it follows that $\kappa$ is gauge invariant. This function extends the standard notion of surface gravity. Indeed, at the points where the vector field $\eta$ does not vanish, its surface gravity $\wt{\kappa}$ is the scalar on $\mc H$ defined by
\begin{equation}
	\label{nablaetageneral}
	\nabla_{\eta}\eta \st{\mc H}{=} \wt{\kappa}\eta.
\end{equation}
Inserting $\eta |_{\mc H} = \alpha\nu$ into \eqref{nablaetageneral} and using \eqref{connections} it follows that $\wt{\kappa} = \kappa$ on the subset of $\mc H$ where $\bar\eta\neq 0$. Note however that $\kappa$ is well defined and smooth everywhere on $\mc H$.\\

Using the identity $\lie_{[X,Y]}= \big[\lie_X,\lie_Y\big]$ applied to the metric $g$ it follows 
\begin{equation}
	\label{aux15}
	\lie_{\eta}\lie_{\xi}g = \lie_{\xi}\lie_{\eta}g - \lie_{[\xi,\eta]}g .
\end{equation} 
Introducing \eqref{etaxi} and pulling back this equation into $\mc H$ gives \cite{tesismiguel}
\begin{equation}
	\label{derivadaY}
	\lie_{\bar\eta}\bY = A_{\eta}\bY +\bm\ell\otimes_s dA_{\eta} + \dfrac{1}{2}\lie_{X_{\eta}}\bg + \dfrac{1}{2}\Phi^{\star}\big(\lie_{\xi}\mc{K}[\eta]\big).
\end{equation}
Inserting $\bar\eta = \alpha n$ and recalling \eqref{lienell} one has $\lie_{\bar\eta}\bm\ell = \lie_{\alpha n}\bm\ell = 2\alpha \bs  + d\alpha$, so equations \eqref{Aeta}, \eqref{Xeta} and \eqref{derivadaY} become
\begin{align}
	A_{\eta} &= -\mc{K}[\eta](\xi,\nu) + n(\alpha),\label{Aeta2}\\
	X_{\eta}^a & = \dfrac{1}{2}\mc{K}[\eta]\big(\xi,n^a\xi-2\theta^a\big) + \dfrac{1}{2}\alpha n(\elltwo) n^a + 2\alpha P^{ab}\s_b + P^{ab}\nablacero_b\alpha,\label{Xeta2}\\
	\alpha	\lie_{n}\bY &= A_{\eta}\bY -2d\alpha\otimes_s\br +\bm\ell\otimes_s dA_{\eta} + \dfrac{1}{2}\lie_{X_{\eta}}\bg + \dfrac{1}{2}\Phi^{\star}\big(\lie_{\xi}\mc{K}[\eta]\big).\label{alfalieY}
\end{align}
The \textit{generalized master equation} \cite{tesismiguel} relates the constraint tensor $\mc{R}_{ab}$, the metric hypersurface data and information on the deformation of $\eta$ codified via the tensorial quantities $\mf{w}, \mf{p}, \mf{q}$ and $\mf{I}$ defined by 
\begin{equation}
	\label{hebrew}
	\mf{w}\d\mc{K}[\eta](\xi,\nu),\qquad \mf{p} \d \mc{K}[\eta](\xi,\xi),\qquad \mf{q} \d \Phi^{\star}\big(\mc{K}[\eta](\xi,\cdot)\big), \qquad \mf{I} \d \dfrac{1}{2}\Phi^{\star}\big(\lie_{\xi}\mc{K}[\eta]\big).
\end{equation} 
The equation is obtained by inserting \eqref{alfalieY} into \eqref{constraint}. The result is \cite{tesismiguel}
\begin{equation}
	\label{constraint22}
	\begin{aligned}
		\alpha \mc R_{ab}& = -\big(2\kappa+\alpha\tr_P\bU-2\mf{w}\big)\Y_{ab} + 2\alpha P^{cd}\U_{d(a}\big(2\Y_{b)c}+\F_{b)c}\big) -2\mf{I}_{ab}\\
		&\quad\, +\big(\mf{p} - \alpha(\tr_P\bY)-\alpha n(\elltwo)\big)\U_{ab}-2\alpha\nablacero_{(a}(\s-\r)_{b)} -4(\s-\r)_{(a}\nablacero_{b)}\alpha      \\
		&\quad\, -2\alpha(\s-\r)_a(\s-\r)_b - 2\nablacero_a\nablacero_b\alpha - \alpha\nablacero_{(a}\s_{b)}+\alpha\s_a\s_b  +2\nablacero_{(a}\mf{q}_{b)} +\alpha \Rcero_{(ab)}.
	\end{aligned}
\end{equation}
As already mentioned before, the dependence on $\bY$ in this relation is purely algebraic. The contraction of \eqref{constraint22} with $n$ is \cite{tesismiguel}
\begin{equation}
	\label{constraintn2}
	\begin{aligned}
		\alpha\mc R_{ab}n^b &= \big(\mf{w}-\alpha\tr_P\bU\big)\r_a -\nablacero_a\kappa -\mf{I}_{ab}n^b + P^{cd}\nablacero_c\big(\alpha\U_{ad} \big) + \alpha(\tr_P\bU)\s_a\\
		&\quad\,- \alpha\nablacero_a\tr_P\bU + \dfrac{1}{2}\left(\nablacero_n\mf{q}_a + \nablacero_a\mf{w} - \mf{w}\s_a - P^{bc}\U_{ca}\mf{q}_b\right),
	\end{aligned}
\end{equation}
and contracting again with $n$,
\begin{equation}
	\label{constraintnn2}
	n(\kappa) = n(\mf{w}) + \kappa_n\mf{w} + \mf{I}(n,n).
\end{equation}

In terms of \eqref{hebrew} the function $A_{\eta}$ and the vector $X_{\eta}$ in \eqref{Aeta2}-\eqref{Xeta2} can be written as 
\begin{equation}
	\label{AX}
A_{\eta} = n(\alpha)-\mf{w},\qquad X_{\eta}^a = \dfrac{1}{2}\big(\alpha n(\elltwo)-\mf{p}\big)n^a + P^{ab}\big(2\alpha\s_b+\nablacero_b\alpha-\mf{q}_b\big).
\end{equation}
Hence, relation \eqref{etaxi} becomes, after using \eqref{kappasgeneral},
\begin{equation}
	\label{liexieta}
	\lie_{\xi}\eta^{\mu} \st{\mc H}{=} \big(\mf{w} +\alpha\kappa_n - \kappa \big)\xi^{\mu} + \dfrac{1}{2}\big(\mf{p} - \alpha n(\elltwo)\big)\nu^{\mu} + P^{ab}\big(\mf{q}_b-2\alpha\s_b-\nablacero_b\alpha\big)\wh{e}_a^{\mu}.
\end{equation}

The idea now is to repeat this process with the identity \eqref{R}, i.e. to replace the derivative $\lie_n\bY^{(m+1)}$ by derivatives of $\mc{K}[\eta]$, $\bY^{(m+1)}$ itself and lower order terms. As usual, let us define $\mc{K}[\eta]^{(m)}\d \lie_{\xi}^{(m-1)}\mc{K}[\eta]$ and 
\begin{equation}
	\label{hebreasm}
	\hspace{-0.2cm} \mf{w}^{(m-1)}\d\mc{K}[\eta]^{(m)}(\xi,\nu),\quad \mf{p}^{(m-1)} \d \mc{K}[\eta]^{(m)}(\xi,\xi),\quad \mf{q}^{(m-1)} \d \Phi^{\star}\big(\mc{K}[\eta]^{(m)}(\xi,\cdot)\big),
\end{equation}
\begin{equation}
	\label{hebreasm2}
	\mf{I}^{(m)} \d \dfrac{1}{2}\Phi^{\star}\big(\mc{K}[\eta]^{(m+1)}\big).
\end{equation}
Observe that $\mf{w}^{(0)} =\mf{w}$, $\mf{p}^{(0)} = \mf{p}$, $\mf{q}^{(0)}=\mf{q}$ and $\mf{I}^{(1)} = \mf{I}$. The rule of thumb is that the number inside the parenthesis denotes the number of transverse derivatives applied to $\mc{K}[\eta]$. Following the notation of Section \ref{section_higher} we introduce the tensor $\Sigma[\eta]\d \lie_{\eta}\nabla$.

\begin{prop}
	\label{prop_lieYm}
	Let $\{\mc H,\bg,\bm\ell,\elltwo\}$ be null metric hypersurface data ($\Phi,\xi$)-embedded in $(\mc M,g)$ and extend $\xi$ off $\mc H$ arbitrarily. Let $\eta$ be a vector field on $(\mc M,g)$ satisfying that $\eta|_{\Phi(\mc H)} = \alpha \nu$ for some $\alpha\in\mc{F}(\mc H)$. Then, for any integer $m\ge 1$
	\begin{equation}
		\label{lieYmexact}
		\lie_{\bar\eta}\bY^{(m)} = \mf{I}^{(m)} + m\big(n(\alpha)-\mf{w}\big)\bY^{(m)} +\mc{P}^{(m)},
	\end{equation}
	where $\mc{P}^{(m)}$ is a tensor that depends on $\{\bY,...,\bY^{(m-1)}\}$ and $\{\lie_{\xi}\eta,...,\lie_{\xi}^{(m)}\eta\}\big|_{\mc H}$ and it is given explicitly by 
	\begin{equation}
		\label{Pm}
		\mc{P}^{(m)}\d  m\lie_{X_{\eta}}\bY^{(m-1)}-\dfrac{1}{2}\sum_{i=2}^m \binom{m}{i}\Phi^{\star}\Big(\lie_{\lie^{(i)}_{\xi}\eta} \lie_{\xi}^{(m-i)}g\Big),
	\end{equation}
and $X_{\eta}^a = \frac{1}{2}\big(\alpha n(\elltwo)-\mf{p}\big)n^a + P^{ab}\big(2\alpha\s_b+\nablacero_b\alpha-\mf{q}_b\big)$. As a consequence, 
	\begin{equation}
		\label{alfalienYmexact}
		\alpha \lie_{n}\bY^{(m)} = \mf{I}^{(m)} +m \big(n(\alpha)-\mf{w}\big) \bY^{(m)} - 2 d\alpha\otimes_s \br^{(m)} +\mc{P}^{(m)}.	
	\end{equation}
	\begin{proof}
		We first show by induction the following relation	
		\begin{equation}
			\label{aux14}
			\lie_{\eta}\lie_{\xi}^{(m)} g = \lie_{\xi}^{(m)}\lie_{\eta}g - \sum_{i=1}^{m}\binom{m}{i}\lie_{\lie^{(i)}_{\xi}\eta} \lie_{\xi}^{(m-i)}g.
		\end{equation} 
		For $m=1$ it holds (cf. \eqref{aux15}). Let us assume \eqref{aux14} is true up to some $m\ge 1$ and show that it is then true for $m+1$ also. We compute
		\begin{align*}
			\lie_{\eta}\lie^{(m+1)}_{\xi} g & = \lie_{\eta}\lie_{\xi}\lie_{\xi}^{(m)} g \\
			&= \lie_{\xi} \lie_{\eta} \lie_{\xi}^{(m)} g + \lie_{[\eta,\xi]}\lie_{\xi}^{(m)} g \\
			&= \lie_{\xi}^{(m+1)}\lie_{\eta} g -\sum_{i=1}^m \binom{m}{i}\lie_{\xi} \lie_{\lie^{(i)}_{\xi}\eta} \lie_{\xi}^{(m-i)}g - \lie_{\lie_{\xi}\eta}\lie_{\xi}^{(m)} g\\
			&= \lie_{\xi}^{(m+1)}\lie_{\eta} g -\sum_{i=1}^m \binom{m}{i} \lie_{\lie^{(i)}_{\xi}\eta} \lie_{\xi}^{(m+1-i)}g-\sum_{i=1}^m \binom{m}{i} \lie_{\lie^{(i+1)}_{\xi}\eta} \lie_{\xi}^{(m-i)}g - \lie_{\lie_{\xi}\eta}\lie_{\xi}^{(m)} g\\
			&= \lie_{\xi}^{(m+1)}\lie_{\eta} g -\sum_{i=1}^m \binom{m}{i} \lie_{\lie^{(i)}_{\xi}\eta} \lie_{\xi}^{(m+1-i)}g-\sum_{i=0}^m \binom{m}{i} \lie_{\lie^{(i+1)}_{\xi}\eta} \lie_{\xi}^{(m-i)}g,
		\end{align*}
		where in the third line we introduced the induction hypothesis. Renaming $i\mapsto i-1$ in the second sum and using the binomial identity $\binom{m}{i-1}+\binom{m}{i} = \binom{m+1}{i}$ it follows 
		\begin{align*}
			\lie_{\eta}\lie^{(m+1)}_{\xi} g &= \lie_{\xi}^{(m+1)}\lie_{\eta} g - \sum_{i=1}^m\binom{m+1}{i} \lie_{\lie^{(i)}_{\xi}\eta} \lie_{\xi}^{(m+1-i)}g - \lie_{\lie_{\xi}^{(m+1)}\eta}g\\
			&= \lie_{\xi}^{(m+1)}\lie_{\eta} g - \sum_{i=1}^{m+1}\binom{m+1}{i} \lie_{\lie^{(i)}_{\xi}\eta} \lie_{\xi}^{(m+1-i)}g.
		\end{align*}
		This establishes \eqref{aux14} for $m\ge 1$. The first term in the sum can be computed by means of \eqref{etaxi}. Pulling \eqref{aux14} back into $\mc H$ and using definition \eqref{hebreasm2} gives $$2\lie_{\bar\eta}\bY^{(m)} = 2\mf{I}^{(m)} + 2mA_{\eta}\bY^{(m)} + 2m\lie_{X_{\eta}}\bY^{(m-1)} -\sum_{i=2}^m \binom{m}{i}\Phi^{\star}\Big(\lie_{\lie^{(i)}_{\xi}\eta} \lie_{\xi}^{(m-i)}g\Big).$$ Relation \eqref{lieYmexact} follows after using \eqref{AX}.
	\end{proof}
\end{prop}

An immediate consequence of the previous Proposition is the following.
\begin{cor}
	\label{cor_lieYm}
	Let $\{\mc H,\bg,\bm\ell,\elltwo\}$ be null metric hypersurface data ($\Phi,\xi$)-embedded in $(\mc M,g)$ and extend $\xi$ off $\mc H$ arbitrarily. Let $\eta$ be a vector field on $(\mc M,g)$ satisfying that $\eta|_{\Phi(\mc H)}=\alpha \nu$ for some function $\alpha\in\mc{F}(\mc H)$. Then, for any integer $m\ge 1$
	\begin{equation}
		\label{lieYm}
		\lie_{\bar\eta}\bY^{(m)} \st{(m)}{=} \mf{I}^{(m)} +m \big(n(\alpha)-\mf{w}\big) \bY^{(m)}.
	\end{equation}
	As a consequence,
	\begin{equation}
		\label{alfalienYm}
		\alpha \lie_{n}\bY^{(m)} \st{(m)}{=} \mf{I}^{(m)} +m \big(n(\alpha)-\mf{w}\big) \bY^{(m)} - 2 d\alpha\otimes_s \br^{(m)}.	
	\end{equation}
\end{cor}

Inserting \eqref{alfalienYmexact} into \eqref{R} we arrive at the main result of this section, namely the \textit{generalized master equation of order} $m>1$,
\begin{equation}
	\label{lierictang3}
	\begin{aligned}
		\alpha \mc{R}^{(m+1)}_{ab} &= -\left(2(m+1)(\kappa-\mf{w})+\alpha\tr_P\bU\right)\Y^{(m+1)}_{ab}  - 2\mf{I}^{(m+1)}_{ab}\\
		&\quad\, -\alpha\big(\tr_P\bY^{(m+1)}\big)\U_{ab} + 4\alpha P^{cd}\U_{c(a}\Y^{(m+1)}_{b)d}  +4\alpha (\s-\r)_{(a} \r^{(m+1)}_{b)} \\
		&\quad\, +4\r^{(m+1)}_{(a}\nablacero_{b)}\alpha + 2\alpha\nablacero_{(a}\r^{(m+1)}_{b)}-2\alpha\kappa^{(m+1)}\Y_{ab} + \alpha\mc{O}^{(m)}_{ab} + \mc{P}_{ab}^{(m)},
	\end{aligned}
\end{equation}
where recall that when $\nabla_{\xi}\xi=0$ the tensor $\mc{O}^{(m)}$ depends \textit{only} on metric hypersurface data and $\{\bY,...,\bY^{(m)}\}$ (see Corollary \ref{cor_entero}). The tensor $\mc{P}^{(m)}$ also depends on $\{\bY,...,\bY^{(m)}\}$ and in addition on the vectors $\lie_{\xi}\eta,...,\lie_{\xi}^{(m+1)}\eta$ on $\Phi(\mc H)$. Its key property is that it vanishes when $X_{\eta}=0$ and $\lie_{\xi}^{(i)}\eta \st{\mc H}{=} 0$ for all $i=2,...,m+1$. Finally we prove an interesting property of the vector field $\lie_{\xi}\eta$ that will play a key role in the next section.

	\begin{lema}
		\label{lemalieeta}
Let $\{\mc H,\bg,\bm\ell,\elltwo\}$ be null metric hypersurface data $(\Phi,\xi)$-embedded in $(\mc M,g)$ and extend $\xi$ off $\Phi(\mc H)$ by $\nabla_{\xi}\xi=0$. Let $\eta\in\X(\mc M)$ be such that $\eta|_{\Phi(\mc H)}$ is null and tangent to $\Phi(\mc H)$ and such that its deformation tensor vanishes to all orders on $\Phi(\mc H)$. If $\lie_{\xi}\eta\st{\mc H}{=}h \xi$ for some function $h\in\mc{F}(\mc H)$, then $\lie_{\xi}^{(k)}\eta \st{\mc H}{=} 0$ for any integer $k\ge 2$.
		\begin{proof}
			Let $m\ge 1$. By Proposition \ref{propMarc} it follows
			\begin{equation}
				\label{aux2}
				\lie_{\xi}^{(m)}\nabla_{\gamma}\xi^{\alpha} = \nabla_{\gamma}\lie_{\xi}^{(m)}\xi^{\beta} + \sum_{k=0}^{m-1}\binom{m}{k+1}(\lie_{\xi}^{(m-k-1)}\xi^{\sigma})\Sigma[\xi]^{(k+1)}{}^{\alpha}{}_{\sigma\gamma}=\xi^{\sigma}\Sigma[\xi]^{(m)}{}^{\alpha}{}_{\sigma\gamma}.
			\end{equation}
			Moreover, equation \eqref{derivada} and the fact that $\nabla_{\xi}\xi = 0$ imply
			\begin{equation}
				\label{aux5}
				0 = \lie_{\xi}^{(m)}\nabla_{\xi}\xi^{\alpha} = \xi^{\gamma}\lie_{\xi}^{(m)}\nabla_{\gamma}\xi^{\alpha}.
			\end{equation}
			Combining \eqref{aux2} and \eqref{aux5} gives
			\begin{equation}
				\label{aux7}
				\xi^{\gamma}\xi^{\sigma}\Sigma[\xi]^{(i)}{}^{\alpha}{}_{\gamma\sigma} \st{\mc H}{=} 0 \qquad\forall i\ge 0.
			\end{equation}
			The definition of $\Sigma[\eta]$ entails the general identity
			\begin{equation}
				\label{entails}
				-\nabla_{\xi}\lie_{\eta}\xi = \Sigma[\eta](\xi,\xi) + \nabla_{\lie_{\eta}\xi}\xi - \lie_{\eta}\big(\nabla_{\xi}\xi\big).
			\end{equation} 
			This together with $\nabla_{\xi}\xi =0$ gives
			\begin{equation}
				\label{lie2eta2}
				\lie_{\xi}^{(2)}\eta = -\lie_{\xi}\lie_{\eta}\xi = -\nabla_{\xi}\lie_{\eta}\xi + \nabla_{\lie_{\eta}\xi}\xi = \Sigma[\eta](\xi,\xi) - 2\nabla_{\lie_{\xi}\eta}\xi .
			\end{equation}
			Since $\lie_{\xi}\eta \st{\mc H}{=}h\xi$ and $\Sigma[\eta]$ vanishes on $\Phi(\mc H)$, it follows $\lie_{\xi}^{(2)}\eta \st{\mc H}{=} 0$. Applying $\lie_{\xi}^{(m)}$ to \eqref{lie2eta2} and using \eqref{derivada} and \eqref{aux2} gives
			\begin{equation}
				\label{aux6}
				\lie_{\xi}^{(m+2)}\eta = \Sigma[\eta]^{(m+1)}(\xi,\xi) -2 \sum_{k=0}^{m}\binom{m}{k}(\lie_{\xi}^{(k+1)}\eta^{\gamma})\xi^{\sigma}\Sigma[\xi]^{(m-k)}{}_{\sigma\gamma}.
			\end{equation}
			We prove the statement by induction. Let $m\ge 1$, assume $\lie^{(i)}_{\xi}\eta \st{\mc H}{=} 0$ for all $i=2,...,m+1$ and let us prove that $\lie^{(m+2)}_{\xi}\eta \st{\mc H}{=} 0$. Since $\Sigma[\eta]^{(k)}(\xi,\xi)\st{\mc H}{=} 0$ for all $k\ge 1$ equation \eqref{aux6} on $\Phi(\mc H)$ reads $$\lie_{\xi}^{(m+2)}\eta \st{\mc H}{=} -2 (\lie_{\xi}\eta^{\gamma})\xi^{\sigma}\Sigma[\xi]^{(m)}{}_{\sigma\gamma}\st{\mc H}{=} -2h \xi^{\gamma}\xi^{\sigma} \Sigma[\xi]^{(m)}{}_{\sigma\gamma}\st{\mc H}{=} 0,$$ where in the last equality we used \eqref{aux7}.
		\end{proof}
	\end{lema}
	
	\begin{rmk}
It is remarkable that Lemma \ref{lemalieeta} is insensitive to the extension of $h$ off $\Phi(\mc H)$. At first sight one could think that this is not possible and that counterexamples are easy to construct. For instance, assume that $\lie_{\xi} \eta$ has the form $h\xi$ not just on $\mc H$ but everywhere. Clearly the lemma can only hold true if $\xi (h)=0$. So, the conditions of the lemma must somehow ensure that this property holds. And indeed this can be proved, as we show next. Assume $\lie_{\xi}\eta = h\xi$ everywhere, $\nabla_{\xi}\xi=0$ and that the deformation tensor of $\eta$ vanishes. Then, equation \eqref{entails} reads $$-\nabla_{\xi}\lie_{\eta}\xi = \Sigma[\eta](\xi,\xi) + \nabla_{\lie_{\eta}\xi}\xi - \lie_{\eta}\big(\nabla_{\xi}\xi\big) = -h\nabla_{\xi}\xi = 0 \qquad \Longrightarrow  \qquad  \xi(h) \xi =0.$$ Since $\xi$ is a rigging of $\Phi(\mc H)$ there exists a neighbourhood $\mc U\subset\mc M$ of $\Phi(\mc H)$ in which $\xi\neq 0$, and hence $\xi(h)=0$ on $\mc U$.
	\end{rmk}

When the deformation tensor of $\eta$ vanishes in a neighbourhood of $\Phi(\H)$ (and not only to infinite order on $\Phi(\H)$) and $h$ is assumed to be constant along $\xi$, then in fact $\lie_{\xi}\eta = h\xi$ holds in a neighbourhood of $\Phi(\H)$, as we prove next.

\begin{lema}
\label{lemaMarc}
Let $(\mc M,g)$ admit a Killing vector $\eta$ and consider an embedded hypersurface $\Phi:\H\hookrightarrow\mc M$ with rigging $\xi$. Extend $\xi$ off $\Phi(\H)$ by means of $\nabla_{\xi}\xi=0$ and assume $\lie_{\xi}\eta \st{\mc H}{=} h\xi$ with $\xi(h)=0$. Then, $\lie_{\xi}\eta = h\xi$ in a neighbourhood of $\Phi(\H)$.
\begin{proof}
Define $\zeta \d \lie_{\xi}\eta - c\xi$. Equation \eqref{entails} together with the fact that $\xi$ is geodesic yields $$\nabla_{\xi}\zeta + \nabla_{\zeta}\xi - \Sigma[\eta](\xi,\xi) = 0.$$ Since $\eta$ is a Killing the last term vanishes and then we have a linear homogeneous transport equation for $\zeta$. Since $\zeta\st{\H}{=}0$ we conclude $\zeta=0$ in a neighbourhood of $\Phi(\H)$.
\end{proof}
\end{lema}
\section{Application to non-degenerate Killing horizons}
\label{sec_KH}

In this section we study the case when $(\mc M,g)$ admits a Killing vector $\eta$ (and therefore $\mc K[\eta]=\lie_{\eta}g=0$). First, we review some well known properties of Killing horizons and particularize the identities of Section \ref{section_deformation} to the present case. This will lead to a natural definition of ``abstract Killing horizon data'' as well as its embedded counterpart. We will prove that, fixing the extension of the rigging to being geodesic, the transverse expansion at any non-degenerate Killing horizon is uniquely determined in terms of its abstract Killing horizon data and the ambient Ricci tensor to infinite order on the horizon. Moreover, when the Ricci tensor fulfills a so-called hierarchical dependence, the transverse expansion only depends on abstract Killing horizon data. Finally, we apply this result to characterize $\Lambda$-vacuum manifolds near non-degenerate horizons.\\


A Killing horizon of $\eta$ is an embedded hypersurface $\Phi:\mc H\hookrightarrow\mc M$ to which $\eta$ is tangent, null and nowhere zero. We say that the Killing horizon is non-degenerate when its surface gravity satisfies $\wt\kappa\neq 0$ at some point, and we say it is degenerate when $\wt\kappa=0$ everywhere. In the literature it is common to define the notion of ``non-degenerate'' by the requirement that $\wt\kappa \neq 0$ everywhere.  We prefer the weaker definition above because then a Killing horizon is either degenerate or non-degenerate. In many relevant cases of interest (e.g. $\Lambda$-vacuum spacetimes) both definitions turn out to be equivalent.\\

A well-known property of Killing horizons is that they are totally geodesic, i.e. $\bU=0$, which follows at once from $\Phi^{\star}\mc K[\eta] = 2\alpha\bU$. As a consequence, given any $X\in\X(\mc H)$, equations \eqref{connections} and \eqref{derivadannull} give
\begin{equation}
	\label{connectionskilling}
	\nabla_{\Phi_{\star}X}\eta = \nabla_{\Phi_{\star}X}(\alpha \nu) = \alpha\nabla_{\Phi_{\star}X} \nu + X(\alpha)\nu =  \big(\alpha (\bs-\br) + d\alpha\big)(X) \nu .
\end{equation}
Moreover equations \eqref{constraint}-\eqref{constraintnn} simplify to
\begin{align}
	\mc R_{ab}& = \accentset{\circ}{R}_{(ab)} -2\lie_n \Y_{ab} - 2\kappa_n\Y_{ab} + \nablacero_{(a}\left(\s_{b)}+2\r_{b)}\right) -2\r_a\r_b + 4\r_{(a}\s_{b)} - \s_a\s_b,\label{khconstraint}\\
	\mc R_{ab}n^a & = \lie_n(\s_b-\r_b) - \nablacero_b \kappa_n,\label{khconstraintn}\\
	\mc R_{ab}n^an^b &=0,\label{khconstraintnn}
\end{align}
while, using that the tensors \eqref{hebrew} all vanish, equations \eqref{constraint22}-\eqref{constraintn2} become
\begin{align}
	\alpha\mc R_{ab} &= -2\kappa \Y_{ab} -2\alpha\nablacero_{(a}(\s-\r)_{b)} -4(\s-\r)_{(a}\nablacero_{b)}\alpha  -2\alpha(\s-\r)_a(\s-\r)_b \nonumber\\
	&\quad\,  - 2\nablacero_a\nablacero_b\alpha - \alpha\nablacero_{(a}\s_{b)}+\alpha\s_a\s_b  +\alpha \Rcero_{(ab)},\label{khconstraint2}\\
	\alpha\mc R_{ab}n^b &= -\nablacero_a\kappa.\label{khconstraintn2}
\end{align}
Equations \eqref{khconstraint}-\eqref{khconstraintn2} have been obtained and discussed in \cite{tesismiguel} where among other things the following results are obtained. Equation \eqref{constraintnn2} gives $n(\kappa)=0$, so the surface gravity is constant along the null generator of the horizon. Hence, $\lie_n d\kappa = d(n(\kappa)) = 0$ as well, so $d\kappa$ is also Lie-constant along the null generators. For bifurcate horizons \eqref{khconstraintn2} implies $d\kappa = 0$ on the bifurcation surface (and hence everywhere), which recovers the well-known fact that $\kappa$ is constant for bifurcate horizons \cite{kay1991theorems}. One can also recover easily the result that when $d\kappa\neq 0$ geodesics terminate in a curvature singularity \cite{racz1992extensions}.\\


Particularizing \eqref{lierictang3} for $m>1$ to the case $\bU=0$ and $\mc{K}[\eta]=0$,
\begin{equation}
	\label{lierictangkh2}
	\begin{aligned}
		\alpha \mc{R}^{(m+1)}_{ab} & =  -2(m+1)\kappa\Y^{(m+1)}_{ab}  +4\alpha (\s-\r)_{(a} \r^{(m+1)}_{b)}+4\r^{(m+1)}_{(a}\nablacero_{b)}\alpha \\
		&\quad\,  + 2\alpha\nablacero_{(a}\r^{(m+1)}_{b)}-2\alpha\kappa^{(m+1)}\Y_{ab} + \alpha\mc{O}^{(m)}_{ab} + \mc{P}_{ab}^{(m)}.
	\end{aligned}
\end{equation}

In view of relation \eqref{connectionskilling} it is useful to introduce the one-form
\begin{equation}
	\label{def_omega}
	\bm\tau\d \alpha(\bs-\br)+d\alpha
\end{equation} 
which will play an important role below. Let us discuss some of its basic properties. Note first that for any tangent vector $X \in \X(\mc H)$ it holds
\begin{equation}
	\label{nablaeta}
	\nabla_{\Phi_{\star} X}   \eta = \bm\tau (X) \nu.
\end{equation}
This one-form extends the commonly used one-form $\bm\upvarpi$ defined by $\nabla_{\Phi_{\star}X}\eta = \bm\upvarpi(X)\eta$ (see for instance \cite{ashtekar2002geometry,frolov2012black}). Indeed, at the points where $\alpha\neq 0$ the one-form $\bm\tau$ agrees with $\alpha \bm\upvarpi$. Note however that $\bm\tau$ is well-defined and smooth everywhere on $\H$.

\begin{prop}
	\label{properties_omega}
	Let $\bm\tau$ be as in \eqref{def_omega}. Then,\\
	\begin{minipage}{0.3\textwidth}
		\noindent
		\begin{enumerate}
			\item $\mc G_{(z,V)}\bm\tau=z\bm\tau$,
		\end{enumerate}
	\end{minipage}
	\begin{minipage}{0.3\textwidth}
		\noindent
		\begin{enumerate}
			\item[2.] $\bm\tau(n)=\kappa$,
		\end{enumerate}
	\end{minipage}
	\begin{minipage}{0.4\textwidth}
		\noindent
		\begin{enumerate}
			\item[3.] $\alpha\lie_n\bm\tau = n(\alpha)\bm\tau - \bm\tau(n) d\alpha$.
		\end{enumerate}
	\end{minipage}
	\begin{proof}
		The gauge transformation of $\bm\tau$ follows from \eqref{gauger-s} and Lemma \ref{lemagauge}. Using \eqref{kappasgeneral} and $\br(n)=-\kappa_n$ the second property is immediate. For the third one we use that $\eta$ is a Killing vector, so $\lie_{\eta}\nabla=0$ and thus $0=\lie_{\eta}\nabla_X\eta -\nabla_X\lie_{\eta}\eta -\nabla_{\lie_{\eta}X}\eta$ for every $X\in\X(\mc M)$. When $X$ is tangent to $\mc H$ the vector $\lie_{\eta}X$ on $\mc H$ is also tangent (because $\eta$ is tangent as well). Then, $$0=\lie_{\eta}\nabla_X\eta  -\nabla_{\lie_{\eta}X}\eta = \lie_{\bar\eta}\big(\bm\tau(X)\big) \nu + \bm\tau(X)\lie_{\eta}\nu - \bm\tau\big(\lie_{\bar\eta}X\big) \nu=\big(\lie_{\bar\eta}\bm\tau-n(\alpha)\bm\tau\big)(X)\nu,$$ where we used $\lie_{\eta}\nu = \lie_{\alpha\nu}\nu = -\nu(\alpha)\nu = -n(\alpha)\nu$. Then $\lie_{\bar\eta}\bm\tau = n(\alpha)\bm\tau$, so item 3. follows after using $\lie_{\bar\eta}\bm\tau = \lie_{\alpha n}\bm\tau = \alpha\lie_n\bm\tau + \bm\tau(n) d\alpha$.
	\end{proof}
\end{prop}
In order to construct the abstract notion of ``Killing horizon data'' we want to find the minimum amount of data on the horizon that allows for the determination of the full transverse expansion. As we will see below, when $\kappa\neq 0$ everywhere, the required information involves the null metric hypersurface data, the one-form $\bm\tau$ and the function $\alpha$. The abstract conditions we need to incorporate into the definition must be such that, once the data is embedded, the corresponding horizon satisfies (i) $\Phi^{\star}\big(\lie_{\eta}g\big)=0$, (ii) the one-form $\bm\tau$ satisfies item 3. of Proposition \ref{properties_omega}, (iii) the set of zeros of $\alpha$ has empty interior and (iv) that one-form $\alpha^{-1}(\bm\tau-d\alpha)$ extends smoothly to all $\mc H$. Item (iii) is necessary because Killing vectors that vanish on any hypersurface are necessarily identically zero. In terms of $\eta$ this means that $\alpha$ vanishing on any open subset of $\mc H$ would only be compatible with  $\eta$ being identically vanishing. Item (iv) is necessary because for actual Killing horizons, the following equality holds (cf. \eqref{def_omega}) $$\alpha^{-1} ( \bm\tau - d \alpha) = \bs - \br,$$ and the right hand side is smooth everywhere on $\mc H$. Note that the condition that the zeroes of $\alpha$ have empty interior means, in particular, that the extension of $\alpha^{-1} ( \bm\tau - d \alpha)$ is necessarily unique. Note also than when $\alpha$ has no zeroes,  condition (iv) is automatically satisfied. As we explained at the beginning of this section, condition (i) gives $\bU=0$. This discussion motivates the following definition.
\begin{defi}
	\label{defi_AKH}
	We say $\{\mc H,\bg,\bm\ell,\elltwo,\bm\tau,\alpha\}$ is \textbf{abstract Killing horizon data} (AKH data) provided that (i) $\{\mc H,\bg,\bm\ell,\elltwo\}$ is null metric hypersurface data satisfying $\bU=0$, (ii) $\alpha$ is a smooth function such that the set $\{\alpha=0\}$ has empty interior and (iii) $\bm\tau$ is a one-form such that $\alpha\lie_n\bm\tau = n(\alpha)\bm\tau - \bm\tau(n) d\alpha$ and the one-form $\alpha^{-1}(\bm\tau-d\alpha)$ extends smoothly to all $\mc H$.
\end{defi}

%

\begin{rmk}
	It is worth comparing Definition \ref{defi_AKH} with the notions of abstract Killing horizons of order zero (AKH$_0$) and one (AKH$_1$) introduced in \cite{manzano2024embedded}. The main difference is that the definitions in \cite{manzano2024embedded} involve {\it full} hypersurface data (i.e. involve the tensor $\bY$) while the definition above makes no reference to $\bY$. In fact, we want to use Definition \ref{defi_AKH} in combination with the field equations to {\it construct} $\bY$ in such a way that the data corresponds to a Killing horizon. This is why we have added the term ``data'' in the definition of  ``AKH data''.
\end{rmk}
Next we extend the notion of gauge transformation to the context of AKH data motivated by Lemma \ref{lemagauge} and Proposition \ref{properties_omega}.
\begin{defi}
	\label{gaugeAKH}
	Let $\k=\{\mc H,\bg,\bm\ell,\elltwo,\bm\tau,\alpha\}$ be AKH data and $(z,V)$ gauge parameters. We define the gauge-transformed data by $\mc{G}_{(z,V)}\k\d \{\mc H,\bg',\bm\ell',\elltwo{}',z\bm\tau,z\alpha\}$, where $\{\bg',\bm\ell',\elltwo{}'\}$ are given by \eqref{transgamma}-\eqref{transell2}.
\end{defi}

The condition $\bU=0$ of Definition \ref{defi_AKH} only guarantees that the pullback of the deformation tensor vanishes on $\mc H$. To capture the full information about the deformation tensor we need to restrict ourselves to the embedded case.
\begin{defi}
	\label{EKHdata}
	Let $\k=\{\mc H,\bg,\bm\ell,\elltwo,\bm\tau,\alpha\}$ be AKH data and define $\bar\eta\d \alpha n$. We say that $\k$ is $(\Phi,\xi)$-\textbf{embedded} in $(\mc M,g)$ if (i) $\{\mc H,\bg,\bm\ell,\elltwo\}$ is $(\Phi,\xi)$-embedded in $(\mc M,g)$ as in Def. \ref{defi_embedded} and (ii) $\nabla_{\Phi_{\star}X}\Phi_{\star}\bar\eta = \bm\tau(X)\nu$ for every $X\in\X(\mc H)$. Moreover, we say that $\k$ is an $(\Phi,\xi)$-\textbf{embedded Killing horizon data} (EKH data) if, additionally, (iii) there exist an extension $\eta$ of $\Phi_{\star}\bar\eta$ such that its deformation tensor $\mc K[\eta]\d\lie_{\eta} g$ vanishes to all orders at $\Phi(\mc H)$. 
\end{defi}

\begin{rmk}
	\label{remarkextension}
It is easy to check that if $\k=\{\mc H,\bg,\bm\ell,\elltwo,\bm\tau,\alpha\}$ is $(\Phi,\xi)$-embedded in $(\mc M,g)$, then $\mc{G}_{(z,V)}\k$ is $(\Phi,\xi')$-embedded in $(\mc M,g)$ with $\xi' = z(\xi+\Phi_{\star}V)$. Moreover, since the property of $\mc{K}[\eta]$ vanishing to all orders on $\Phi(\mc H)$ is independent of $\xi|_{\Phi(\mc H)}$ and its extension off $\Phi(\mc H)$, it follows that if $\k$ is $(\Phi,\xi)$-EKH data, then $\mc{G}_{(z,V)}\k$ is $(\Phi,\xi')$-EKH data.	
\end{rmk}

\begin{rmk}
	\label{notation}
The definition of AKH data allows us to define a smooth one-form $\bcr\d \bs - \alpha^{-1}(\bm\tau-d\alpha)$ and a scalar $\bck_n\d \alpha^{-1}(\bm\tau(n)-n(\alpha))$. When the data happens to be embedded, the one-form $\bcr$ and the function $\bck_n$ agree with $\br$ and $\kappa_n$, respectively. This is because from equation \eqref{connectionskilling} and condition (ii) of Definition \ref{EKHdata} it follows $\alpha(\bcr-\br)=0$, and since the interior of the zeroes of $\alpha$ is empty, then $\bcr=\br$. Hence in the embedded case we shall not distinguish between $\bcr$ and $\br$ anymore. Observe also that this fact together with \eqref{kappasgeneral} imply that the surface gravity $\kappa$ of the hypersurface is given by $\kappa = n(\alpha) + \alpha \kappa_n = \bm\tau(n)$.
\end{rmk}

\begin{eje}
	\label{example}
Consider the $d$-dimensional Schwarzschild-de Sitter spacetime $(\mc M,g)$. In ingoing Eddington-Finkelstein coordinates $\{v,r\}$ the metric $g$ is $$g = -\left(1-\dfrac{2M}{r^{d-3}}-\dfrac{\Lambda}{d-1}r^2\right)dv^2 + 2dv dr + r^2 \bg_{\sph^{d-2}},$$ where $\bg_{\sph^{d-2}}$ is the $d-2$ dimensional spherical metric. When $M$ and $\Lambda$ are both positive, and $M$ sufficiently small, the polynomial $G(r)\d -r^{d-3}g_{vv} = r^{d-3}-2M-\frac{\Lambda}{d-1}r^{d-1}$ admits precisely two positive roots $r_0^{+}\ge R\d  \sqrt{\frac{d-3}{\Lambda}}\ge  r_0^{-}$. Since $\partial_r g_{vv}\big|_{r=r_0^{\pm}}= -\frac{2M(d-3)}{(r_0^{\pm})^{d-2}} + \frac{2\Lambda}{d-1}r_0^{\pm} = \Lambda r_0^{\pm} - \frac{d-3}{r_0^{\pm}}$ it follows that when $r_0^{+}> R>  r_0^{-}$ the two null hypersurfaces $\mc{H}^{\pm}\d\{r=r_0^{\pm}\}$ are non-degenerate Killing horizons with Killing $\eta = \partial_v$. $\H^+$ is called cosmological horizon, and $\H^-$ is called event horizon. When $r_0^+=r_0^-=R$, $\H^+=\H^-$ is a degenerate Killing horizon. Let us compute the induced hypersurface data of any of the two horizons (we use $r_0$ to denote at once $r_0^+$ and $r_0^-$). Choosing $\xi=\partial_r$ as the rigging (observe $\nabla_{\xi}\xi=0$) it follows $$\bg = r_0^2 \bg_{\sph^{d-2}}, \qquad \bm\ell = dv,\qquad \elltwo = 0,\qquad \bY = \left( - \dfrac{(d-3)M}{r_0^{d-2}}+\dfrac{\Lambda r_0}{d-1}\right) dv^2 + r_0\bg_{\sph^{d-2}}.$$ It follows at once that $\bm\ell(\eta)=1$, $\bs=\frac{1}{2}d\bm\ell(\eta,\cdot)=0$, so by \eqref{def_omega} with $\alpha=1$ $$\bm\tau = -\br = -\bY(\eta,\cdot) = \Big(\frac{(d-3)M}{r_0^{d-2}}-\frac{\Lambda r_0}{d-1}\Big)dv$$ and thus $\kappa = \bm\tau(\eta) = \frac{(d-3)M}{r_0^{d-2}}-\frac{\Lambda r_0}{d-1} = \frac{d-3}{2r_0}-\frac{\Lambda r_0}{2}$. When $r_0\neq R$ it follows $\kappa\neq 0$. In this case $\k_{SdS}\d \{\mc H,\bg,\bm\ell,\elltwo,\bm\tau,\alpha=1\}$ is EKH in Schwarzschild-de Sitter spacetime with $\kappa\neq 0$ everywhere.
\end{eje}

In Proposition \ref{necesary} we find necessary conditions for two AKH data to be embeddable in isometric semi-Riemannian manifolds. The idea is to use this information to then define a notion of isometry at the AKH data level. First we extend Notation \ref{nota0} to the AKH case.
\begin{nota}
	Let $\k'=\{\mc H',\bg',\bm\ell',\elltwo{}',\bm\omega',\alpha'\}$ be AKH data and $\chi:\mc H\to\mc H'$ a diffeomorphism. We define
	\begin{equation}
		\label{defi_pullback}
\chi^{\star}\k'=\chi^{\star} \{\mc H',\bg',\bm\ell',\elltwo{}',\bm\tau',\alpha'\} \d \{\mc H,\chi^{\star}\bg',\chi^{\star}\bm\ell',\chi^{\star}\elltwo{}',\chi^{\star}\bm\tau',\chi^{\star}\alpha'\}.
	\end{equation} 
\end{nota}
\begin{lema}
	\label{well}
Let $\k'=\{\mc H',\bg',\bm\ell',\elltwo{}',\bm\omega',\alpha'\}$ be AKH data and $\chi:\mc H\to\mc H'$ a diffeomorphism. Then $\chi^{\star}\k'$ is AKH data.
\begin{proof}
Firstly, from Remark \ref{lema_diff} it follows that $\{\H,\bg,\bm\ell,\elltwo\}\d \chi^{\star}\{\H',\bg',\bm\ell',\elltwo{}'\}$ is null metric hypersurface data. Since $\chi^{\star}\lie_{\chi_{\star}n}\bg' = \lie_{n}\bg$ it also follows $\bU=0$. Moreover, since $\chi$ is a diffeomorphism, $\alpha = \chi^{\star}\alpha'$ is smooth and has empty interior. Finally, $$0 = \chi^{\star}\big(\alpha'\lie_{\chi_{{}_{\star}}n}\bm\tau' -(\chi_{\star}n)(\alpha')\bm\tau'+ \bm\tau'(\chi_{\star}n) d\alpha'\big) = \alpha\lie_{n}\bm\tau -n(\alpha)\bm\tau+ \bm\tau(n) d\alpha$$ and $\chi^{\star}\big(\alpha'{}^{-1}(\bm\tau'-d\alpha')\big) = \alpha^{-1}(\bm\tau-d\alpha)$ extends smoothly to all $\mc H$. Hence $\{\H,\bg,\bm\ell,\ell^{(2)}{},\bm\tau,\alpha\}$ is AKH data.
\end{proof}
\end{lema}

\begin{prop}
	\label{necesary}
Let $\k=\{\mc H,\bg,\bm\ell,\elltwo,\bm\omega,\alpha\}$ be AKH data $(\Phi,\xi)$-embedded in $(\mc M,g)$ and let $\k'=\{\mc H',\bg',\bm\ell',\elltwo{}',\bm\omega',\alpha'\}$ be AKH data $(\Phi',\xi')$-embedded in $(\mc M',g')$. Let $\bar\eta\d \alpha n$ and $\bar\eta'\d \alpha' n'$. Assume there exists an isometry $\varphi:(\mc M,g)\to(\mc M',g')$ such that $\varphi(\Phi(\mc H))=\Phi'(\mc H')$ and $\varphi_{\star}\Phi_{\star}\bar\eta = \Phi'_{\star}\bar\eta'$. Then, there exist a diffeomorphism $\chi:\mc H\to\mc H'$ and gauge parameters $(z,V)$ such that
	\begin{equation}
		\label{chi0}
		\chi^{\star} \{\H',\bg',\bm\ell',\elltwo{}',\bm\tau',\alpha'\}= \mc{G}_{(z,V)}\{\H,\bg,\bm\ell,\elltwo,\bm\tau,\alpha\}.
	\end{equation}
	\begin{proof}
First we prove that the vector field $\varphi^{\star}\xi'$ is everywhere transverse to $\Phi(\mc H)$. Observe that since the vector field $\nu'$ is null and normal to $\Phi'(\mc H')$ and $\varphi$ is an isometry, the vector $\varphi^{\star}\nu'$ is also null and normal to $\Phi(\mc H)$, so in particular it is proportional to $\nu$ with non-zero proportionality factor. Then, from $g'(\xi',\nu')=1$ it follows $$1= (\varphi^{\star}g')(\varphi^{\star}\xi',\varphi^{\star}\nu') = g(\varphi^{\star}\xi',\varphi^{\star}\nu') \qquad \Longrightarrow\qquad g(\varphi^{\star}\xi,\nu)\neq 0.$$ This proves that $\varphi^{\star}\xi'$ is everywhere transverse to $\Phi(\mc H)$, so there must exist a function $z\in\mc{F}^{\star}(\mc H)$ and a vector field $V\in\X(\mc H)$ such that $\varphi^{\star}\xi' = z\big(\xi+\Phi_{\star}V\big)$. Since $\Phi(\mc H)$ and $\Phi'(\mc H')$ are diffeomorphic via $\varphi$ and both $\Phi$ and $\Phi'$ are embeddings, there exists a diffeomorphism $\chi$ making the following diagram commutative.
		\begin{center}
			\begin{tikzcd}
				\mc H \arrow[r,"\chi"]\arrow[d,"\Phi"] & \mc H'\arrow[d,"\Phi'"]\\
				\mc M \arrow[r,"\varphi"] & \mc M'
			\end{tikzcd}
		\end{center}
		Then,
\begin{gather*}
\chi^{\star}\bg ' = \chi^{\star}\Phi'{}^{\star} g' = \Phi^{\star}\varphi^{\star}g' = \Phi^{\star}g = \bg,\\
\chi^{\star}\bm\ell ' = \chi^{\star}\Phi'{}^{\star} \big(g'(\xi',\cdot)\big) = \Phi^{\star}\varphi^{\star}\big(g'(\xi',\cdot)\big) = \Phi^{\star}\big(g(z(\xi+\Phi_{\star}V),\cdot)\big) = z\big(\bm\ell + \bg(V,\cdot)\big)\\
\chi^{\star}\ell^{(2)}{}' = \chi^{\star}\Phi'{}^{\star} \big(g'(\xi',\xi')\big) = \Phi^{\star}\varphi^{\star}\big(g'(\xi',\xi')\big) =  \Phi^{\star}\big(g(\varphi^{\star}\xi',\varphi^{\star}\xi')\big) = z^2\big(\elltwo + 2\bm\ell(V) + \bg(V,V)\big).
\end{gather*} 
Hence $\chi^{\star}\{\H',\bg',\bm\ell',\ell^{(2)}{}'\} = \mc{G}_{(z,V)}\{\H,\bg,\bm\ell,\elltwo\}$ (see Def. \ref{defi_embedded}), so $\chi^{\star}\nu' = z^{-1}\nu$ and $\chi^{\star}\alpha' = z\alpha$ because $\varphi_{\star}\Phi_{\star}\bar\eta = \Phi'_{\star}\bar\eta'$. Finally, from condition (ii) in Definition \ref{EKHdata} together with $\varphi_{\star}\Phi_{\star}\bar\eta = \Phi'_{\star}\bar\eta'$ and the fact that $\varphi$ is an isometry, it follows $$\varphi^{\star}\big(\nabla'_{\Phi'_{\star}X'}\Phi'_{\star}\bar\eta'\big) = \nabla_{\Phi_{\star}\chi^{\star}X'}\Phi_{\star}\bar\eta \qquad\Longrightarrow \qquad \chi^{\star}\bm\tau' \otimes\chi^{\star}\nu' = \bm\tau \otimes \nu \qquad\Longleftrightarrow \qquad \chi^{\star}\bm\tau ' = z\bm\tau.$$ Taking into account Definition \ref{gaugeAKH}, \eqref{chi0} is established.
	\end{proof}
\end{prop}

This proposition motivates the following natural notion of isometry in the context of AKH data.

\begin{defi}
	\label{def_isometric}
	Let $\k=\{\mc H,\bg,\bm\ell,\elltwo,\bm\tau,\alpha\}$ and $\k'=\{\mc H',\bg',\bm\ell',\elltwo{}',\bm\tau',\alpha'\}$ be two AKH data. We say $\k$ and $\k'$ are isometric provided that there exists a diffeomorphism $\chi:\mc H\to\mc H'$ and gauge parameters $(z,V)$ such that
	\begin{equation}
		\label{chi}
		\chi^{\star} \{\H',\bg',\bm\ell',\elltwo{}',\bm\tau',\alpha'\}= \mc{G}_{(z,V)}\{\H,\bg,\bm\ell,\elltwo,\bm\tau,\alpha\}.
	\end{equation}
\end{defi}
Lemma \ref{well} guarantees that Definition \ref{def_isometric} is well defined. \\

Our next aim is to show that given EKH data $\k$ satisfying $\bm\tau(n)\neq 0$ everywhere, the full asymptotic expansion $\{\bY^{(k)}\}_{k\ge 1}$ is uniquely determined in terms of $\k$ and the set $\{R^{(m)}_{\alpha\beta}\}_{m\ge 1}$. As we shall see, to prove the uniqueness part of such statement we need to be able to extend the rigging vector $\xi$ such that the tensor $\lie_{\xi}^{(m)}R_{\alpha\beta}$ is geometrical (in the sense of Definition \ref{defi_geometrical}) for every $m\ge 0$. By identities \eqref{ddotR}, \eqref{dotR} and \eqref{lierictangkh2}, the extension of $\xi$ must be such that the tensors $\mc{O}^{(m)}$, $\mc{O}^{(m)}_a$, $\mc{O}^{(m)}_{ab}$ and $\mc{P}^{(m)}_{ab}$ are $\mc{H}$-geometrical for every $m\ge 0$. In Section \ref{section_higher} we have proved that by extending $\xi$ off $\Phi(\mc H)$ by means of $\nabla_{\xi}\xi=0$ the tensors $\mc{O}^{(m)}$, $\mc{O}^{(m)}_a$ and $\mc{O}^{(m)}_{ab}$ are $\mc H$-geometrical. However, this is not sufficient to guarantee that $\mc{P}^{(m)}_{ab}$ is $\mc H$-geometrical, because this tensor also depends on $X_{\eta}$ and $\lie_{\xi}^{(i)}\eta\big|_{\mc H}$ for $i\ge 2$ (see the comment below equation \eqref{lierictang3}). A natural way to make this dependence disappear is to ensure that $X_{\eta} = 0$ and $\lie_{\xi}^{(i)}\eta \st{\mc H}{=} 0$ for every $i\ge 2$. Our strategy is as follows. In Lemma \ref{lemaguagelieeta} we show that given AKH data embedded on an ambient manifold with rigging $\xi$, one can always choose the gauge such that $\lie_{\xi}\eta$ is proportional to $\xi$ on $\Phi(\mc H)$. With this choice $X_{\eta}$ automatically vanishes (cf. \eqref{etaxi}). By combining this result with Lemma \ref{lemalieeta} we will be able to prove that $\lie_{\xi}^{(i)}\eta \st{\mc H}{=} 0$ for every $i\ge 2$ as well. \\

Particularizing equation \eqref{liexieta} to the Killing horizon case, namely $\mc{K}[\eta]=0$,
\begin{equation}
	\label{liexietaKH}
	\lie_{\xi}\eta^{\mu} \st{\mc H}{=}  (\alpha\kappa_n-\kappa) \xi^{\mu} - \dfrac{\alpha}{2} n(\elltwo)\nu^{\mu} - P^{ab}\big(2\alpha\s_b+\nablacero_b\alpha\big)\wh{e}_a^{\mu}.
\end{equation}
In the following lemma we show that there exists a choice of gauge in which $\lie_{\xi}\eta \st{\mc H}{=}  (\alpha\kappa_n-\kappa) \xi$.
\begin{lema}
	\label{lemaguagelieeta}
Let $\k=\{\mc H,\bg,\bm\ell,\elltwo,\bm\tau,\alpha\}$ be AKH data $(\Phi,\xi)$-embedded in $(\mc M,g)$ satisfying $\bm\tau(n)\neq 0$ everywhere on $\mc H$. Assume there exists an extension $\eta$ of $\bar\eta\d\alpha n$ off $\Phi(\mc H)$ satisfying $\mc{K}[\eta](\xi,\cdot)= 0$ on $\Phi(\mc H)$. Then there exists a family of gauges satisfying $\bm\ell = \kappa^{-1}\bm\tau$ and $\elltwo=0$. Moreover, any element of the family satisfies
	\begin{equation}
		\label{liexietaKH2}
		\lie_{\xi}\eta \st{\mc H}{=} (\alpha\kappa_n-\kappa) \xi ,
	\end{equation}
and the whole family can be generated from any element by the action of the subgroup of transformations $\{\mc G_{(z,0)}\}$ (i.e. this subgroup acts transitively on the family). Any element of this family will be said to be written in an ``\,$\eta$-gauge''.
	\begin{proof}
A vector $V\in\X(\mc H)$ is defined uniquely in terms of the one-form $\bm{w}\d\bg(V,\cdot)$ and the scalar function $f \d\bm\ell(V)$. From \eqref{Pgamma} and \eqref{gamman} it follows $P^{ab}w_a w_b = P^{ab}\gamma_{ac}\gamma_{bd}V^cV^d = \big(\delta^b_c-n^b\ell_c\big)\gamma_{bd}V^cV^d = \gamma_{cd}V^cV^d$, so in terms of $f$ and $\bm w$ the gauge transformations of $\bm\ell$ and $\elltwo$ in \eqref{tranfell}-\eqref{transell2} read
\begin{multicols}{2}
	\noindent
	\begin{equation}
		\label{gaugeell}
\bm\ell' = z(\bm\ell + \bm{w}),
	\end{equation}
	\begin{equation}
		\label{gaugeell2}
\ell^{(2)}{}' = z^2\big(\elltwo + 2f + P(\bm w,\bm w)\big).
	\end{equation}
\end{multicols}
From the transformations of $\bm\tau$ and $\kappa$, namely $\bm\tau' = z\bm\tau$ and $\kappa' = \kappa$, it is straightforward to check that by choosing $\bm w \d \kappa^{-1}\bm\tau-\bm\ell$ and $f\d -\elltwo - \frac{1}{2}\kappa^{-2}P(\bm\tau,\bm\tau)$, the gauge-transformed data satisfies (i) $\bm\ell' = \kappa^{-1}\bm\tau'$ and (ii) $\ell^{(2)}{}' = 0$. Moreover, by the transformations \eqref{gaugeell}-\eqref{gaugeell2} and those of $\bm\tau$ and $\kappa$, it is clear that any additional transformation $\mc{G}_{(z',V')}$ will preserve properties (i) and (ii) if and only if $V'=0$. Thus, the whole family is generated by applying $\mc{G}_{(z,0)}$ (with $z\in\mc{F}^{\star}(\mc H)$ arbitrary) to any element of the family. To prove that in this class of gauges expression \eqref{liexietaKH} simplifies to \eqref{liexietaKH2} it suffices to show that $P^{ab}\big(2\alpha\s_b+\nablacero_b\alpha\big)=0$. Writing item 3. of Proposition \ref{properties_omega} in terms of $\kappa\bm\tau=\bm\ell$ gives $$\alpha\lie_n\bm\tau = \alpha n(\kappa)\bm\ell + \alpha\kappa\lie_n\bm\ell = n(\alpha)\kappa\bm\ell - \kappa d\alpha\qquad\Longrightarrow\qquad \kappa\big(\alpha\lie_n\bm\ell+d\alpha \big) = \big(\kappa n(\alpha)-\alpha n(\kappa)\big)\bm\ell.$$ Using $\lie_{n}\bm\ell=2\bs$ (cf. \eqref{lienell}) it follows that the one-form $2\alpha\bs + d\alpha$ is proportional to $\bm\ell$, and since $\elltwo=0$ then $P^{ab}\big(2\alpha\s_b+\nablacero_b\alpha\big)=0$ as a consequence of \eqref{Pell}.
	\end{proof}
\end{lema}

Observe that the EKH data in Example \ref{example} is written in an $\eta$-gauge. By combining Lemmas \ref{lemalieeta} and \ref{lemaguagelieeta} and Remark \ref{remarkextension} we arrive at the following.

\begin{cor}
	\label{corEnDKHD}
Let $\{\mc H,\bg,\bm\ell,\elltwo,\bm\tau,\alpha\}$ be $(\Phi,\xi)$-EKH in $(\mc M,g)$. Then there exists a choice of $\xi$ on $\Phi(\mc H)$ in which $\lie_{\xi}\eta \st{\mc H}{=} (\alpha\kappa_n-\kappa) \xi$. Such gauge is unique once $n$ is fixed. Moreover, extending $\xi$ off $\Phi(\mc H)$ by $\nabla_{\xi}\xi = 0$, then $\lie_{\xi}^{(k)}\eta \st{\mc H}{=} 0$ for all $k\ge 2$. In particular, all the tensors $\mc{P}^{(m)}$ in \eqref{Pm} vanish.
\end{cor}

We are ready to show one of the main results of this section, namely that the full transverse expansion of an EKH data satisfying $\kappa\neq 0$ everywhere is uniquely determined in terms of AKH data and the collection $\{R^{(m)}_{\alpha\beta}\}_{m\ge 1}$. 

\begin{teo}
	\label{teorema1}
	Let $\k=\{\mc H,\bg,\bm\ell,\elltwo,\bm\tau,\alpha\}$ be $(\Phi,\xi)$-EKH in any $(\mc M,g)$. Assume $\bm\tau(n)\neq 0$ everywhere on $\mc H$. Then, when the data is written in an $\eta$-gauge and $\xi$ is extended geodesically, the full transverse expansion $\{\bY^{(m)}\}_{m\ge 1}$ only depends on $\k$ and $\{R^{(m)}_{\alpha\beta}\}_{m\ge 1}$.
	\begin{proof}
Let us write the data in any of the $\eta$-gauges of Lemma \ref{lemaguagelieeta}, which fixes the vector $\xi$ up to a multiplicative non-vanishing function $\xi\mapsto z\xi$, and extend $\xi$ off $\Phi(\mc H)$ by means of $\nabla_{\xi}\xi = 0$. We want to prove that the tensors $\{\bY^{(k)}\}_{k\ge 1}$ only depend on $\k$ and $\{R^{(m)}_{\alpha\beta}\}_{m\ge 1}$, and thus they are insensitive to the particular manifold they are embedded in, provided their tensors $R^{(m)}_{\alpha\beta}$ agree on $\Phi(\mc H)$. With this choice of gauge and extension of $\xi$ the tensors $\mc{O}^{(m)}$, $\mc{O}^{(m)}_a$ and $\mc{O}^{(m)}_{ab}$ only depend on metric data and $\{\bY,...,\bY^{(m)}\}$ (see Corollary \ref{cor_entero}) and the tensor $\mc{P}^{(m)}$ vanishes for every $m\ge 1$ (see Corollary \ref{corEnDKHD}). These two facts will be used repeatedly throughout the proof. By Remark \ref{notation} we can identify the one-form $\bcr\d \bs - \alpha^{-1}(\bm\tau-d\alpha)$ with $\br\d \bY(n,\cdot)$ and the scalar $\bm\tau(n)$ with $\kappa$. Therefore, equation \eqref{khconstraint2} can be rewritten as
		\begin{equation}
			\label{khconstraint3}
			\begin{aligned}
				\alpha\mc R_{ab} &= -2\kappa \Y_{ab} + C_{ab},
			\end{aligned}
		\end{equation}
where $C_{ab}$ is a tensor that only depends on AKH data $\{\bg,\bm\ell,\elltwo,\bm\tau,\alpha\}$. This proves that when $\kappa\neq 0$ everywhere on $\mc H$, the full tensor $\bY$ is determined in terms of AKH data and the tensor $\mc R_{ab}$. Therefore, since in equations \eqref{ddotR}-\eqref{dotR} for $m=0$ the lower order terms only involve the tensor $\bY$ and metric data, it follows that the scalar $\tr_P\bY^{(2)}$ and the one-form $\br^{(2)}$ only depend on AKH data and the tensor $R_{\alpha\beta}$ on $\mc H$. Hence, equation \eqref{lierictangkh2} for $m=1$ reads
		\begin{equation}
			\label{KHorder2}
			\alpha \mc{R}^{(2)}_{ab} =  -4\kappa\Y^{(2)}_{ab} + C_{ab}^{(2)},
		\end{equation}
where $C^{(2)}_{ab}$ only depends on AKH data and $R_{\alpha\beta}$. When $\kappa\neq 0$ everywhere this shows that the tensor $\Y^{(2)}_{ab}$ is uniquely determined from AKH data and the tensors $R_{\alpha\beta}$ and $\mc{R}^{(2)}_{ab}$ on $\mc H$. Iterating this process by means of equations \eqref{ddotR}, \eqref{dotR} and \eqref{lierictangkh2} one obtains the full transverse expansion $\{\bY^{(k)}\}_{k\ge 1}$, and by Corollaries \ref{corEnDKHD} and \ref{cor_entero} this expansion only depends on AKH data and $\{R^{(m)}_{\alpha\beta}\}_{m\ge 1}$, and not on the particular $(\mc M,g)$ where $\k$ is embedded.
	\end{proof}
\end{teo}

This theorem shows that the asymptotic expansion of an EKH data satisfying $\kappa\neq 0$ everywhere \textbf{only} depends on the abstract data $\k$ and the tensors $\{R^{(m)}_{\alpha\beta}\}_{m\ge 1}$, and thus it is insensitive to the particular $(\mc M,g)$ they may be embedded in. The collection $\{R^{(m)}_{\alpha\beta}\}_{m\ge 1}$ can be thought at least in two different ways. One possibility is to provide each $R^{(m)}_{\alpha\beta}$ as prescribed data on the null hypersurface, e.g. by some external matter field. Another option is to provide $R^{(m)}_{\alpha\beta}$ as a functional relation between the abstract data $\k$ and the transverse expansion $\{\bY^{(1)},...,\bY^{(m)}\}$. The simplest example of the second viewpoint is a $d$-dimensional manifold $(\mc M,g)$ satisfying the $\Lambda$-vacuum equations, where $R^{(m+1)}_{\alpha\beta} = \lambda \mc{K}^{(m)}_{\alpha\beta}$ and $\lambda = \frac{2\Lambda}{d-2}$. Then, $\mc{R}=\lambda\bg$, $\dot{\mc R} = \lambda\bm\ell$, $\ddot{\mc R}=\lambda\elltwo$ and $\mc{R}^{(m+1)} = 2\lambda \bY^{(m)}$, $\dot{\mc R}^{(m+1)} = \lambda \Phi^{\star}\big(\mc{K}^{(m)}(\xi,\cdot)\big)$ and $\ddot{\mc R}^{(m+1)} = \lambda \Phi^{\star}\big(\mc{K}^{(m)}(\xi,\xi)\big)$ for every $m\ge 1$. Since by Remark \ref{remark1} the tensor $\mc{K}^{(m)}(\xi,\cdot)$ on $\Phi(\mc H)$ depends on $\{\bY^{(1)},...,\bY^{(m-1)}\}$, one concludes that the tensor $R^{(m+1)}_{\alpha\beta}$ on $\Phi(\mc H)$ depends at most on AKH data and $\{\bY,...,\bY^{(m)}\}$. In general, when the functional relations are such that each $R^{(m)}_{\alpha\beta}$ depends on low enough transverse derivatives of the metric, i.e. such that the LHS in equations \eqref{ddotR}-\eqref{R} depend on derivatives that we already have under control, the proof of Theorem \ref{teorema1} shows that the transverse expansion only depends on the abstract data. Let us make this property precise.
\begin{defi}
	\label{defi_hier}
Let $\k=\{\H,\bg,\bm\ell,\elltwo,\bm\tau,\alpha\}$ be AKH data $(\Phi,\xi)$-embedded in $(\mc M,g)$, and extend $\xi$ off $\Phi(\mc H)$ by $\nabla_{\xi}\xi = 0$. We say that the Ricci tensor of $g$ is \textbf{hierarchical} on $\Phi(\mc H)$ provided that $\mc{R}^{(1)}$ only depends on $\k$ and, for every $m\ge 1$,
\begin{enumerate}
	\item[(i)] $\mc{R}^{(m+1)}$ depends (at most) on $\k$, $\{\bY,...,\bY^{(m)}\}$, $\tr_P\bY^{(m+1)}$ and $\br^{(m+1)}$.
	\item[(ii)] $\dot{\mc R}{}^{(m)}$ and $\ddot{\mc R}{}^{(m)}$ depend (at most) on $\k$ and $\{\bY,...,\bY^{(m)}\}$.
\end{enumerate}
When a Ricci tensor is hierarchical on $\Phi(\mc H)$ we shall refer to its particular dependence stated in (i) and (ii) by its ``hierarchical dependence''.
\end{defi}

Recall that when $\xi$ is extended geodesically the tensor $R^{(m)}_{\alpha\beta}$ is geometrical for every $m\ge 1$ (cf. Proposition \ref{universal0}). As noted above, the canonical example is the $\Lambda$-vacuum equations, since in this case the tensors $\mc{R}^{(m+1)}$, $\dot{\mc R}^{(m)}$ and $\ddot{\mc R}^{(m)}$ only depend on AKH data and $\{\bY^{(1)},...,\bY^{(m)}\}$. In fact, $\mc{R}^{(1)}_{ab} = \lambda \gamma_{ab}$, $\dot{\mc R}_a = \lambda\ell_a$, $\ddot{\mc R} = \lambda\elltwo$ and $\mc{R}^{(m+1)}=2\lambda\bY^{(m)}$, $\dot{\mc R}^{(m)}_a = 0$ and $\ddot{\mc R}^{(m)} = 0$ for every $m\ge 1$ (see Corollary \ref{corkxi}). An immediate consequence of the proof of Theorem \ref{teorema1} and Definition \ref{defi_hier} is the following.

\begin{teo}
	\label{teorema12}
Let $\k=\{\H,\bg,\bm\ell,\elltwo,\bm\tau,\alpha\}$ be $(\Phi,\xi)$-EKH in any $(\mc M,g)$. Assume $\bm\tau(n)\neq 0$ everywhere on $\mc H$ and that the field equations satisfied by $g$ are such that its Ricci tensor is hierarchical on $\Phi(\mc H)$. Then, when the data is written in an $\eta$-gauge and $\xi$ is extended geodesically, the full transverse expansion $\{\bY^{(m)}\}_{m\ge 1}$ only depends on $\k$.
\end{teo}

This theorem generalizes the recent work \cite{oliver} in several directions. The main theorem in \cite{oliver} proves that for spacetimes admitting a non-degenerate Killing horizon $\H$ and being Ricci flat to infinite order at $\H$, the full asymptotic expansion of the metric along certain privileged transverse vector at the horizon can be determined geometrically from a so-called ``non-degenerate Killing horizon data''. Theorem \ref{teorema12} is more general firstly because we are allowing zeroes of $\eta$ on $\mc H$ (while Killing horizons by definition only include points where $\eta$ is non-zero), secondly because our hierarchical property includes more field equations besides vacuum with vanishing $\Lambda$, and finally because we have extended the result to arbitrary signature (provided it admits degenerate hypersurfaces). We recover the result in \cite{oliver} simply by imposing that $R_{\alpha\beta}$ is zero to all orders on the horizon. As an interesting corollary of Theorem \ref{teorema12} we extend the geometric uniqueness of the transverse expansion to the case of asymptotic $\Lambda$-vacuum spacetimes.

\begin{cor}
	\label{corvacuum}
Let $\k=\{\mc H,\bg,\bm\ell,\elltwo,\bm\tau,\alpha\}$ be $(\Phi,\xi)$-EKH in any $(\mc M,g)$ satisfying the $\Lambda$-vacuum equations to infinite order on $\mc H$. Assume $\bm\tau(n)\neq 0$ at least at one point $p\in\H$ and that $\k$ is written in an $\eta$-gauge and with $\xi$ extended geodesically off $\Phi(\mc H)$. Then the full transverse expansion $\{\bY^{(m)}\}_{m\ge 1}$ is uniquely determined in terms of $\k$. 
\begin{proof}
Since $(\mc M,g)$ satisfies the $\Lambda$-vacuum equations to infinite order on $\mc H$ it follows $\mc{R}_{ab}n^b = 0$, and thus from \eqref{khconstraintn2} one has $d\kappa = 0$, so $\kappa =\bm\tau(n)$ is constant on $\H$. Since $\kappa\neq 0$ at least at one point and $\H$ is assumed to be connected, we conclude $\kappa\neq 0$ everywhere on $\H$, and hence Theorem \ref{teorema12} applies.
\end{proof}
\end{cor}

Let us perform a detailed comparison between our AKH data and the ``non-degenerate Killing horizon data'' in \cite{oliver}, which is a triple $(\mc H,\bm\sigma,\mc V)$ where $\bm\sigma$ is a Riemannian metric, $\mc V$ is a nowhere vanishing Killing vector of $(\H,\bm\sigma)$ with constant (non-zero) norm. In order to compare both objects at the same footing we restrict our AKH data to $\alpha\neq 0$, $\bg$ semi-positive definite, $\bm\tau(n)\neq 0$ at some point and vacuum, which in particular implies $\bm\tau(n)$ must be a nonzero constant (see the proof of Corollary \ref{corvacuum}). Let us see the equivalence between AKH data and ``non-degenerate Killing horizon data''. In one direction, given $\{\mc H,\bg,\bm\ell,\elltwo,\bm\tau,\alpha\}$ the vector $\mc V$ and the metric $\bm\sigma$ can be defined as follows: $$\mc V\d \alpha n,\qquad \bm\sigma \d \bg + \alpha^{-2}\bm\tau\otimes\bm\tau.$$ 

Clearly $\mc V$ is nowhere vanishing and $\bm\sigma(\mc V,\mc V) = \big(\bm\tau(n)\big)^2$ is a non-vanishing constant, so $\bm\sigma$ is a gauge-invariant Riemannian metric on $\mc H$ (see Lemma \ref{lemagauge} and Proposition \ref{properties_omega}). Moreover, since $\lie_{\mc V}\bm\tau = n(\alpha)\bm\tau$, it follows $$\lie_{\mc V}\bm\sigma = 2\alpha\bU + \lie_{\mc V}(\alpha^{-2}) \bm\tau\otimes\bm\tau + 2\alpha^{-2} \lie_{\mc V}\bm\tau\otimes_s\bm\tau = \big(- 2\alpha^{-2}n(\alpha) +2\alpha^{-2}n(\alpha)\big)\bm\tau\otimes\bm\tau=  0.$$

 Conversely, given a ``non-degenerate Killing horizon data'' $(\mc H,\bm\sigma,\mc V)$ as in \cite{oliver} we define $$\alpha\d 1,\qquad \bm\tau \d \dfrac{1}{\sqrt{\bm\sigma(\mc V,\mc V)}}\bm\sigma(\mc V,\cdot),\qquad \bg \d \bm\sigma - \dfrac{1}{\bm\sigma(\mc V,\mc V)}\bm\sigma(\mc V,\cdot)\otimes\bm\sigma(\mc V,\cdot)$$$$\elltwo\d 0,\qquad \bm\ell\d \frac{1}{\bm\sigma(\mc V,\mc V)}\bm\sigma(\mc V,\cdot).$$ 
 
 Since $\bm\ell(\mc V)=1$ and $\bg(\mc V,\cdot)=0$ it follows at once that $\{\bg,\bm\ell,\elltwo\}$ is null metric hypersurface data and $n=\mc V$. Moreover, from $\lie_{\mc V}\bm\sigma=0$ one gets $\bU = \frac{1}{2}\lie_n\bg = 0$ and $\alpha\lie_n\bm\tau=0=n(\alpha)\bm\tau - \bm\tau(n) d\alpha$. In addition $\bm\tau(n)$ is a non-zero constant and the one-form $\alpha^{-1}(\bm\tau-d\alpha)$ extends smoothly to all $\mc H$ (because $\alpha=1$). Hence $\{\mc H,\bg,\bm\ell,\elltwo,\bm\tau,\alpha\}$ fulfills the conditions of Definition \ref{defi_AKH} with $\alpha\neq 0$ and $\bm\tau(n)=const.\neq 0$. \\

In \cite{oliver} the choice of the rigging $\xi$ is such that $\lie_{\xi}\eta = 0$. This is possible only when $\alpha$ is nowhere vanishing. Indeed, when $\alpha\neq 0$ one can exploit the freedom in Lemma \ref{lemaguagelieeta} to set $\alpha=1$, and thus $\kappa=\kappa_n$ (cf. \eqref{kappasgeneral}) and $\lie_{\xi}\eta \st{\mc H}{=} 0$ (see \eqref{liexietaKH2}). Lemma \ref{lemaMarc} then shows that $\lie_{\xi}\eta=0$ whenever $\eta$ is a Killing. However, if $\bar\eta$ admits zeroes on $\H$ this choice of gauge is not possible anymore because the properties $\alpha \neq 0$ or $\alpha=0$ at a point are gauge invariant, so if $\alpha=0$ in some gauge it is impossible to make $\alpha=1$ by a gauge transformation.\\

In Proposition \ref{necesary} we proved that two AKH data $\k$ and $\k'$ embedded in isometric manifolds $(\mc M,g)$ and $(\mc M',g')$ are necessarily isometric (in the sense of Definition \ref{def_isometric}). Our next aim is to prove a kind of converse, namely that isometric EKH data $\k$ and $\k'$ both satisfying $\bm\tau(n)\neq 0$ imply that the respective manifolds they are embedded in are isometric to infinite order. The proof involves two steps. Firstly, we need to construct and map to each other suitable neighbourhoods of $\Phi(\mc H)$ and $\Phi'(\mc H')$, and secondly we prove that within an appropriate gauge and extension of $\xi$, the two asymptotic expansions agree. The first task was accomplished in Proposition \ref{prop_diffeo}, and the second one is a consequence of Proposition \ref{teo_iso0}, as we show next.

\begin{teo}
	\label{teo_isometry}
Let $\k=\{\mc H,\bg,\bm\ell,\elltwo,\bm\tau,\alpha\}$ and $\k'=\{\mc H',\bg',\bm\ell',\elltwo{}',\bm\tau',\alpha'\}$ be two isometric EKH in respective ambient manifolds $(\mc M,g)$ and $(\mc M',g')$ and satisfying $\bm\tau(n)\neq 0$ everywhere on $\mc H$. Assume the field equations on $(\mc M,g)$ and $(\mc M',g')$ are such that their Ricci tensors satisfy the same hierarchical dependence on $\Phi(\mc H)$ and $\Phi'(\mc H')$, respectively. Let $\k'$ be written in an $\eta$-gauge and $\k$ in the gauge in which $\k = \chi^{\star}\k'$ and extend the riggings geodesically. Then, there exist neighbourhoods $\mc U\subset \mc M$ and $\mc U'\subset\mc M'$ of $\Phi(\mc H)$ and $\Phi'(\mc H')$ and a diffeomorphism $\Psi:\mc U\to\mc U'$ such that 
	\begin{equation}
		\label{isometry}
		\Psi^{\star} \lie_{\xi'}^{(i)}g' \st{\mc H}{=} \lie_{\xi}^{(i)}g
	\end{equation} 
	for every $i\in\mathbb{N}\cup\{0\}$.
	\begin{proof}
By Definition \ref{defi_embedded} it follows $\bg=\chi^{\star}\bg'$, $\bm\ell=\chi^{\star}\bm\ell'$ and $\elltwo=\chi^{\star}\elltwo{}'$, so by Proposition \ref{teo_iso0} it suffices to prove $\chi^{\star}\bY^{(i)}{}' = \bY^{(i)}$ for every $i\ge 1$. Observe that since $\k'$ is written in an $\eta$-gauge and $\k=\chi^{\star}\k'$, then $\bm\ell = \chi^{\star}\bm\ell'=\chi^{\star}\big(\kappa'{}^{-1}\bm\tau'\big) = \kappa^{-1}\bm\tau$ and $\elltwo = \chi^{\star}\ell^{(2)}{}' = 0$, so $\k$ is also written in an $\eta$-gauge. Equation \eqref{khconstraint2} and its primed version read $$-2\kappa\Y_{ab} +C_{ab}=\alpha\mc{R}_{ab},\qquad -2\kappa'\Y_{ab}' +C_{ab}=\alpha' \mc{R}'_{ab},$$ where $C_{ab}$ and $C'_{ab}$ only depend on AKH data (see Corollaries \ref{corEnDKHD} and \ref{cor_entero}) and thus agree. Since the dependence of $\mc{R}_{ab}$ on $\k$ is the same as the dependence of $\mc{R}'_{ab}$ on $\k'$, taking the pullback $\chi^{\star}$ of the second equation and subtracting the first one gives $\chi^{\star}\bY' = \bY$. A similar argument applied to \eqref{ddotR}-\eqref{dotR} for $m=0$ proves $\chi^{\star}\big(\tr_{P'}\bY^{(2)}{}'\big) = \tr_P\bY^{(2)}$ and $\chi^{\star}\br^{(2)}{}' = \br^{(2)}$. By iterating this process one shows that $\chi^{\star}\bY^{(i)}{}' = \bY^{(i)}$ for every $i\ge 1$.
	\end{proof}
\end{teo}

An example of two manifolds satisfying the same hierarchical dependence is two spacetimes $(\mc M,g)$ and $(\mc M',g')$ such that $R_{\alpha\beta}=\lambda g_{\alpha\beta}$ and $R'_{\alpha\beta}=\lambda g'_{\alpha\beta}$. This leads to the following immediate corollary. 

\begin{cor}
	\label{cor_iso}
Let $\k=\{\mc H,\bg,\bm\ell,\elltwo,\bm\tau,\alpha\}$ and $\k'=\{\mc H',\bg',\bm\ell',\elltwo{}',\bm\tau',\alpha'\}$ be two isometric EKH in respective $\Lambda$-vacuum ambient manifolds $(\mc M,g)$ and $(\mc M',g')$ with the same cosmological constant. Assume $\bm\tau(n)\neq 0$ everywhere on $\mc H$. Let $\k'$ be written in an $\eta$-gauge and $\k$ in the gauge in which $\k = \chi^{\star}\k'$ and extend the riggings geodesically. Then, there exist neighbourhoods $\mc U\subset \mc M$ and $\mc U'\subset\mc M'$ of $\Phi(\mc H)$ and $\Phi'(\mc H')$ and a diffeomorphism $\Psi:\mc U\to\mc U'$ such that 
\begin{equation}
	\label{isometry2}
	\Psi^{\star} \lie_{\xi'}^{(i)}g' \st{\mc H}{=} \lie_{\xi}^{(i)}g
\end{equation} 
for every $i\in\mathbb{N}\cup\{0\}$.
\end{cor}

A consequence of this corollary is that every analytic $\Lambda$-vacuum manifold admitting a non-degenerate Killing horizon (possibly with bifurcation surfaces) is characterized near the horizon by its AKH data. In a forthcoming paper \cite{Mio4} we show that every AKH data gives rise to a spacetime which is $\Lambda$-vacuum to infinite order, irrespective of any assumption of analiticity neither on the data nor on the constructed spacetime.\\

A direct application of Corollary \ref{cor_iso} is that if a spacetime admits an EKH isometric to the non-extremal Schwarzschild-de Sitter data introduced in Example \ref{example}, then the spacetime is isometric to infinite order (i.e. \eqref{isometry2} holds) to non-extremal Schwarzschild-de Sitter. Moreover, when the spacetime is real analytic it is necessarily isometric to Schwarzschild-de Sitter spacetime (at least in a neighbourhood of the horizon). This result complements the main result of \cite{katona2024uniqueness}, where the extremal case was considered.

\begin{prop}
	\label{cor_final}
Let $(\mc M,g)$ be a $d\ge 4$-dimensional spacetime satisfying the vacuum Einstein equations with cosmological constant $\Lambda>0$ and admitting a non-degenerate Killing horizon $\Phi:\mc H\hookrightarrow\mc M$ with spherical cross-sections. Let $\eta$ be the corresponding Killing vector with surface gravity $\kappa\neq 0$ (necessarily constant). Assume $\Phi^{\star} g = r_0^2\bg_{\sph^{d-2}}$ for some $0<r^2_0\neq \frac{d-3}{\Lambda}$ and $\bm\tau\wedge d\bm\tau = 0$, where $\bm\tau$ is defined in \eqref{def_omega}. Let $M \d \frac{1}{2}r_0^{d-3} - \frac{\Lambda}{2(d-1)} r_0^{d-1}$. 
\begin{enumerate}
	\item If $r_0^2>\frac{d-3}{\Lambda}$, $(\mc M,g)$ is isometric to infinite order to non-extremal Schwarzschild-de Sitter spacetime with mass $M$ at its \textbf{cosmological horizon}.
	\item If $r_0^2<\frac{d-3}{\Lambda}$, $(\mc M,g)$ is isometric to infinite order to non-extremal Schwarzschild-de Sitter spacetime with mass $M$ at its \textbf{event horizon}.
\end{enumerate}
Moreover, if $(\mc M,g)$ is analytic, then it is isometric to Schwarzschild-de Sitter spacetime in a neighbourhood of its cosmological horizon (case 1.) or its event horizon (case 2.).
	\begin{proof}
We denote all the objects referring to Schwarzschild-de Sitter described in Example \ref{example} with the label ``SdS''. Let us scale the Killing vector $\eta$ by the constant $\kappa_{SdS}/\kappa$, where $\kappa_{SdS} \d \frac{d-3}{2r_0}-\frac{\Lambda r_0}{2}\neq 0$. Then, the surface gravity of (the re-scaled) $\eta$ is precisely $\kappa_{SdS}$. Since on the horizon the Killing field is nowhere vanishing we can define $n\d \eta$ and thus $\alpha=1$. From item 3. of Proposition \ref{properties_omega} it follows $\lie_n\bm\tau=0$. Let us pick any section $\mc{S}\subset\H$ and solve the differential equation $n(u)=1$ with initial condition $u|_{\mc S}=0$. This gives a global function $u\in\mc{F}(\H)$ that defines a foliation $\{\mc S_u\}_{u\in\real}$. Let us decompose the cotangent space $T^{\star}\H$ by $T^{\star}\H = \mbox{span}\{du\}\oplus \langle n\rangle^{\bot}$, where as usual $\langle n\rangle^{\bot}$ is the set of covectors that annihilate $n$. Therefore the one-form $\bm\tau$ can be uniquely decomposed as $\bm\tau = A du + \bm{b}$ with $A\in\mc{F}(\H)$ and $\bm b\in T^{\star}\H$ satisfying $\bm b(n)=0$. Condition $\bm\tau(n)=\kappa$ gives $A=\kappa$, and condition $\bm\tau \wedge d\bm\tau = 0$ gives $$\big(\kappa du + \bm b\big) \wedge d\bm b = 0\qquad \Longrightarrow \qquad \kappa du\wedge d\bm b+ \bm{b}\wedge d\bm b = 0.$$ Since $\lie_n\bm\tau = \lie_n\big(\kappa du +\bm b\big)=\lie_n\bm b=0$ then from the Cartan identity $d\bm b(n,\cdot) = \lie_n\bm b- d(\bm b(n)) = 0$. Contracting the equation above with $n$ yields $\kappa d\bm b=0$. Since $\kappa$ is a nonzero constant, the one-form $\bm b$ is closed, and hence exact because $\real\times\sph^{d-2}$ is simply connected for $d\ge 4$. Let $B\in\mc{F}(\H)$ be such that $\bm b = dB$, so $\bm\tau = \kappa d\big(u + \kappa^{-1}B\big)$.\\

Let us define $u' \d u +\kappa^{-1}B$ (hence $\bm\tau = \kappa d u'$). Since $\bm b(n)=0$ then $n(B)=0$, so $n( u')=n(u)=1$. Consider the new foliation defined by $ u'$, $\{\mc S_{ u'}\}_{u'\in\real}$. The remaining gauge freedom can be fixed by requiring $\xi$ to be null and orthogonal to that foliation, which in terms of the abstract data implies $\elltwo=0$ and $\bm\ell = du'$. Then, the tuple $$\k\d \{\H , \bg = r_0^2\bg_{\sph^{d-2}}, \bm\ell = du', \elltwo=0, \bm\tau = \kappa du',\alpha=1\}$$ fulfills all the conditions of Definition \ref{defi_AKH}, so $\k$ is AKH data satisfying $\bm\tau(n)=\kappa\neq 0$ everywhere on $\H$. Moreover, $\k$ is written in an $\eta$-gauge. The last step is to construct the diffeomorphism $\chi:\H \to\H_{SdS}$. Choose any isometry $\phi$ that maps the section $\{u'=0\}$ of $\H$ with the section $\{v=0\}$ of $\H_{SdS}$. Consider any point $p\in\H$ and let $\sigma(u')$ be the integral curve of $n$ through $p$. This curve intersects $\mc{S}_{u'=0}$ at a single point $q$. Now consider the integral curve $\sigma_{SdS}(v)$ of $n_{SdS}$ through $\phi(q)$. We define $\chi(p) \d \sigma_{SdS}(u'(p))$. That $\chi$ is a diffeomorphism follows from the product topology of $\H$ and $\H_{SdS}$ and because $n$ and $n_{SdS}$ are both smooth and globally defined. Since $\phi$ is an isometry and the integral curves of $n$ and $n_{SdS}$ are identified it follows $\chi^{\star}\bg_{SdS} = \bg$. With the definition $M \d \frac{1}{2}r_0^{d-3} - \frac{\Lambda}{2(d-1)} r_0^{d-1}$ it is clear that $\k$ is isometric (as in Definition \ref{def_isometric}) either to $\k^+_{SdS}$ or $\k^-_{SdS}$. In order to distinguish the two cases recall that the equation $1-\frac{2M}{r_0{}^{d-3}}-\frac{\Lambda}{d-1}r_0{}^2=0$ admits exactly two solutions $r_0{}^{+}> \sqrt{\frac{d-3}{\Lambda}}>  r_0{}^{-}$. Therefore, when $r_0^2 > \frac{d-3}{\Lambda}$ then $r_0 = r_0^+$ (and we are in the cosmological horizon case) and when $r_0^2 < \frac{d-3}{\Lambda}$ then $r_0 = r_0^-$ (which is the event horizon case). The Proposition follows after using Corollary \ref{cor_iso}.
	\end{proof}
\end{prop}

The case $r_0^2 =  \frac{d-3}{\Lambda}$ is excluded in Proposition \ref{cor_final} because the horizon described in Example \ref{example} is necessarily degenerate when $r_0^2 =  \frac{d-3}{\Lambda}$. A result analogous to the one in Proposition \ref{cor_final} can be formulated with $\Lambda=0$ and Schwarzschild spacetime, or $\Lambda<0$ and Schwarzschild-anti de Sitter spacetime. It is worth emphasizing that, in contrast to the degenerate case treated in \cite{katona2024uniqueness}, it is not necessary to require neither $\Lambda>0$ nor staticity nor compact cross sections for the uniqueness argument to work. The reason is that, for non-degenerate horizons, the equations that allow us to obtain recursively the asymptotic expansion are all algebraic. In the degenerate (also called extremal) case, once $\br^{(m+1)}$ has been replaced in \eqref{lierictangkh2} using \eqref{dotR}, the leading order in identity \eqref{lierictangkh2} is $\bY^{(m)}$ because $\kappa=0$. The dependence between $\mc{R}^{(m)}$ and $\bY^{(m)}$ is not algebraic anymore, but via a partial differential equation on the cross-sections of $\H$. In order for such equations to determine uniquely the expansion it becomes sufficient to require $\Lambda>0$, staticity and that the cross-sections are maximally symmetric \cite{katona2024uniqueness}.\\

Remarkably, when the $(\mc M,g)$ in Proposition \ref{cor_final} is analytic it necessarily admits a \textbf{static} Killing vector, since Schwarzschild-de Sitter does. It would be interesting to relate this to the recent work \cite{marc2023staticity} where it is shown that for bifurcate Killing horizons the condition $\eta$ being static is equivalent to the torsion one-form being closed on the bifurcation surface.

\section{Conclusions and future work}
\label{conclusions}

In this paper, we have developed a series of general identities that relate the transverse derivatives of the ambient Ricci tensor to the transverse expansion of the metric on a null hypersurface. This analysis has then been combined with the presence of a preferred vector field, such as a Killing or homothetic vector (or, more generally, any vector field of which one has complete knowledge of its deformation tensor). As an application we have established a number of uniqueness results. In particular, we have shown that the full asymptotic expansion at a non-degenerate Killing horizon is uniquely determined in terms of abstract data on the horizon and the tower of derivatives of the ambient Ricci tensor at the horizon.\\

The sequel \cite{Mio4} of this work focuses on establishing existence results of ambient spacetimes provided sufficient data is prescribed on a null hypersurface. Specifically, we will prove the existence of an ambient manifold (in general not analytic) given the full asymptotic expansion at the hypersurface. While the ambient spacetime thus constructed cannot be unique in general, we do have, as shown in the present paper, geometric uniqueness to infinite order at the given hypersurface. When this expansion satisfies at every order a set of constraint equations, we will prove that the ambient manifold satisfies the Einstein equations to infinite order at the hypersurface. Furthermore, when the hypersurface admits a product topology, we will determine the minimum amount of data required on a cross-section to ensure the existence of an ambient space that satisfies the Einstein equations to an infinite order.\\

As an application of these general existence results, we will show that every \textit{abstract Killing horizon} data gives rise to a (generally non-analytic) ambient manifold that satisfies the $\Lambda$-vacuum equations to infinite order, with the given data as Killing horizon. The approach we follow allows for the possibility of the Killing having zeroes at the horizon.
%

\section*{Acknowledgements}
This work has been supported by Projects PID2021-122938NB-I00 (Spanish Ministerio de Ciencia e Innovación and FEDER ``A way of making Europe''). M. Mars acknowledges financial support under projects SA097P24 (JCyL) and RED2022-134301-T funded by MCIN/AEI/10.13039/ 501100011033. G. Sánchez-Pérez also acknowledges support of the PhD. grant FPU20/03751 from Spanish Ministerio de Universidades.
\begin{appendices}
	\section{Some pullbacks into a null hypersurface}
	\label{appendix}
	
	In this appendix we compute the pullback of several derivatives of ambient tensors into a null hypersurface that we shall need in the main text of this paper. Recall from Notation \ref{notationnota} that given a $(0,p)$ tensor field $T_{\alpha_1\cdots\alpha_p}$ on $\mc M$ we use the standard notation $T_{a_1\cdots a_p}$ to denote the pullback of $T$ to $\mc H$. Moreover, we also use the notation ${}^{(i)}T_{\alpha_1\cdots \alpha_{p-1}}$ for $\xi^{\mu}T_{\alpha_1\cdots \alpha_{i-1}\mu\alpha_i\cdots \alpha_{p-1}}$ and ${}^{(i)}T_{a_1\cdots a_{p-1}}$ for the pullback of ${}^{(i)}T_{\alpha_1\cdots \alpha_{p-1}}$ to $\mc H$. In addition, we use ${}^{(i,j)}T_{a_1\cdots a_{p-2}}$ for the pullback to $\mc H$ of the tensor obtained by first the contraction of $T$ with $\xi$ in the j-th slot and then in the i-th slot of the resulting $(0,p-1)$ tensor, i.e. ${}^{(i,j)}T={}^{(i)}\big({}^{(j)}T\big)$.
	\begin{prop}
		\label{proppullback}
		Let $(\mc M,g)$ be a semi-Riemannian manifold and $\Phi:\mc H\hookrightarrow\mc M$ a smooth null hypersurface with rigging $\xi$. Let $T$ be a $(0,p)$-tensor on $\mc M$. Then,
		\begin{align}
\hspace{-0.1cm}			\Phi^{\star}\left(\nabla T\right)_{b a_1\cdots a_p} &= \nablacero_b T_{a_1\cdots a_p} + \sum_{i=1}^p \Y_{ba_i}T_{a_1\cdots a_{i-1} c a_{i+1}\cdots a_p} n^c + \sum_{i=1}^p \U_{ba_i}\big({}^{(i)} T\big)_{a_1\cdots  a_p},\hfill\label{identity1}\\
\hspace{-0.1cm}			\Phi^{\star}\left({}^{(1)}\nabla T\right)_{a_1\cdots a_p} &= (\lie_{\xi}T)_{a_1\cdots a_p} - \sum_{i=1}^p (\r-\s)_{a_i}\big({}^{(i)} T\big)_{a_1\cdots a_p} - \sum_{i=1}^p V^b{}_{a_i} T_{a_1\cdots a_{i-1} b a_{i+1}\cdots a_p},\hfill\label{identity2}\\
\hspace{-0.1cm}			\Phi^{\star}\left({}^{(2)}\nabla T\right)_{ba_1\cdots a_{p-1}} &= \nablacero_{b} \big({}^{(1)} T\big)_{a_1\cdots a_{p-1}} + \sum_{i=1}^{p-1} \Y_{b a_i} \big({}^{(1)} T\big)_{a_1\cdots a_{i-1} c a_{i+1}\cdots a_{p-1}}n^c\hfill \nonumber\\
\hspace{-0.1cm}			&+ \sum_{i=1}^{p-1} \U_{b a_i} \big({}^{(i,1)}T\big)_{a_1\cdots a_{p-1}} - (\r-\s)_{b} \big({}^{(1)} T\big)_{a_1\cdots a_{p-1}}  - V^c{}_{b}T_{ca_1\cdots a_{p-1}},\hfill\label{identity3}
		\end{align}
		where $V^{b}{}_{a} \d P^{cb}\Pi_{a c} + \dfrac{1}{2}(d\elltwo)_{a}n^b$.
		\begin{proof}
			Let $\{e_a\}$ be a local basis of $\mc H$ and $\wh{e}_a \d \Phi_{\star}(e_a)$. To prove the first identity we contract $\nabla_{\beta}T_{\alpha_1\cdots \alpha_p}$ with $\wh e_b^{\beta}\wh e_{a_1}^{\alpha_1}\cdots \wh e_{a_p}^{\alpha_p}$ and use \eqref{connections}, namely $$\nabla_{\wh e_b^{\beta}}\wh e^{\alpha}_a = \big(\nablacero_{e_b} e_a\big)^c \wh e_c^{\alpha} - \Y_{ba} n^c\wh e^{\alpha}_c - \U_{ba}\xi^{\alpha}.$$ For the second one we use the relation between $\lie_{\xi}$ and $\nabla$ $$\nabla_{\xi} T_{\alpha_1\cdots\alpha_p} = \lie_{\xi}T_{\alpha_1\cdots\alpha_p} - \sum_{i=1}^p T_{\alpha_1\cdots\alpha_{i-1}\mu\alpha_{i+1}\cdots\alpha_p}\nabla_{\alpha_i}\xi^{\mu},$$ contract it with $\wh e^{\alpha_1}_{a_1}\cdots \wh{e}{}^{\alpha_p}_{a_p}$ and employ \eqref{nablaxi}. Finally, to prove the last one just contract the relation
			\begin{align*}
				\xi^{\mu}\nabla_{\beta} {}T_{\mu\alpha_1\cdots \alpha_{p-1}} = \nabla_{\beta}\big({}^{(1)} T\big)_{\alpha_1\cdots \alpha_{p-1}} - T_{\mu \alpha_1\cdots \alpha_{p-1}} \nabla_{\beta}\xi^{\mu}.
			\end{align*}
			with $\wh e^{\beta}_b\wh e^{\alpha_1}_{a_1}\cdots \wh e^{\alpha_{p-1}}_{a_{p-1}}$ and use \eqref{identity1} and \eqref{nablaxi}.
		\end{proof}
	\end{prop}
	
	\begin{prop}
		\label{propdivergencia}
		Let $(\mc M,g)$ be a semi-Riemannian manifold and $\Phi:\mc H\hookrightarrow\mc M$ a smooth null hypersurface with rigging $\xi$. Let $T$ be a $(0,p+1)$-tensor on $\mc M$ and denote by $\div T$ the $p$-covariant tensor defined by $\left(\div T\right)_{\alpha_1\cdots\alpha_p}\d g^{\mu\nu}\nabla_{\mu}T_{\nu\alpha_1\cdots\alpha_p}$. Then,
		\begin{equation*}
			\begin{aligned}
				\hspace{-0.3cm}\Phi^{\star}\left(\div T\right)_{a_1\cdots a_p} &= P^{bc}\nablacero_b T_{c a_1\cdots a_p} +n^b(\lie_{\xi} T)_{ba_1\cdots a_p}  + n^c\nablacero_c \big({}^{(1)} T\big)_{a_1\cdots a_p} + \big(2\kappa_n + \tr_P\bU\big) \big({}^{(1)} T\big)_{a_1\cdots a_p}\\
				&\quad\, + (\tr_P\bY - n(\elltwo))n^cT_{c a_1\cdots a_p} - 2P^{ac}(\r+\s)_a  T_{c a_1\cdots a_p} \\
				&\quad\,  + \sum_{i=1}^p P^{bc}\Y_{ba_i}T_{c a_1\cdots a_{i-1} d a_{i+1}\cdots a_p}n^d + \sum_{i=1}^p P^{bc}\U_{ba_i} \big({}^{(i+1)} T\big)_{c a_1\cdots a_{i-1} a_{i+1}\cdots a_p} \\
				&\quad\,  - \sum_{i=1}^p (\r-\s)_{a_i}n^b\, \big({}^{(i+1)} T\big)_{b a_1\cdots a_{i-1} a_{i+1}\cdots a_p}- \sum_{i=1}^p V^c{}_{a_i}n^b T_{b a_1\cdots a_{i-1}c a_{i+1}\cdots a_p}\\
				&\quad\,  + \sum_{i=1}^p \r_{a_i} \big({}^{(1)} T\big)_{a_1\cdots a_{i-1}c a_{i+1}\cdots a_p}n^c .
			\end{aligned}
		\end{equation*}
		\begin{proof}
			From \eqref{inversemetric},
			\begin{align*}
				\wh e^{\alpha_1}_{a_1}\cdots \wh e^{\alpha_p}_{a_p}\nabla_{\mu}T^{\mu}{}_{\alpha_1\cdots\alpha_p} & = \wh e^{\alpha_1}_{a_1}\cdots\wh e^{\alpha_p}_{a_p} g^{\mu\nu}\nabla_{\mu}T_{\nu\alpha_1\cdots\alpha_p}\\
				&= \wh e^{\alpha_1}_{a_1}\cdots \wh e^{\alpha_p}_{a_p}\left(P^{bc}\wh e^{\mu}_b \wh e^{\nu}_c + \xi^{\mu} n^b \wh e_b^{\nu} + \xi^{\nu} n^b \wh e_b^{\mu}\right)\nabla_{\mu}T_{\nu\alpha_1\cdots\alpha_p}.
			\end{align*}
			Using \eqref{identity1}, the first term is 
			\begin{align*}
				P^{bc}\nablacero_b T_{c a_1\cdots a_p} + (\tr_P\bY)T_{d a_1\cdots a_p}n^d + \sum_{i=1}^p P^{bc}\Y_{ba_i}T_{c a_1\cdots a_{i-1} d a_{i+1}\cdots a_p}n^d \\
				+ (\tr_P\bU)\big({}^{(1)} T\big)_{ a_1\cdots a_p}+ \sum_{i=1}^p P^{bc}\U_{ba_i}\big({}^{(i+1)}T\big)_{c a_1\cdots a_{i-1} a_{i+1}\cdots a_p}.
			\end{align*}
			From \eqref{identity2}, $V^c{}_b n^b = P^{ac}(\r+\s)_a + \dfrac{1}{2}n(\elltwo)n^c$ and $\br(n) = -\kappa_n$, $\bs(n)=0$ the second term becomes
			\begin{align*}
				n^b(\lie_{\xi} T)_{ba_1\cdots a_p} + \kappa_n \big({}^{(1)}T\big)_{a_1\cdots a_p} - \sum_{i=1}^p (\r-\s)_{a_i}n^b\, \big({}^{(i+1)} T\big)_{b a_1\cdots a_{i-1} a_{i+1}\cdots a_p} \\
				- \left(P^{ac}(\r+\s)_a + \dfrac{1}{2}n(\elltwo)n^c\right) T_{c a_1\cdots a_p} - \sum_{i=1}^p V^c{}_{a_i}n^b T_{b a_1\cdots a_{i-1}c a_{i+1}\cdots a_p}.
			\end{align*}
			Finally, using \eqref{identity3} and $\bU(n,\cdot)=0$ the last term is 
			\begin{align*}
				n^c\nablacero_c \big({}^{(1)} T\big)_{a_1\cdots a_p} + \sum_{i=1}^p \r_{a_i} \big({}^{(1)} T\big)_{a_1\cdots a_{i-1}c a_{i+1}\cdots a_p}n^c +\kappa_n \big({}^{(1)} T\big)_{a_1\cdots a_p} - \left(P^{ac}(\r+\s)_a + \dfrac{1}{2}n(\elltwo)n^c\right) T_{ca_1\cdots a_p}.
			\end{align*}
			Combining the three the result follows.
		\end{proof}
	\end{prop}
	
	
\end{appendices}

\begingroup
\let\itshape\upshape

\renewcommand{\bibname}{References}
\bibliographystyle{acm}
\bibliography{biblio} 

\end{document}